\documentclass[pra, aps, twocolumn, showpacs, superscriptaddress,10pt,floatfix, nofootinbib]{revtex4-2}

\usepackage{graphicx}
\usepackage{dcolumn}
\usepackage{bm}

\usepackage{hyperref}
\hypersetup{
    colorlinks=true,
    linkcolor=blue,          
    citecolor=green,        
    urlcolor=red          
}
\usepackage{verbatim}
\usepackage{xcolor}
\usepackage{mathrsfs}
\usepackage{physics}
\usepackage[page]{appendix}
\usepackage{tikz}
\usepackage{bbold}
\usepackage{booktabs}
\usepackage{makecell}
\usepackage{multirow}

\definecolor{darkgreen}{rgb}{0.0, 0.5, 0.0}

\usepackage{xspace}

\newcommand*{\algo}{\textsc{XLuminA}\xspace}

\begin{document}
\title{\algo: \\ An Auto-differentiating Discovery Framework for Super-Resolution Microscopy}

\author{Carla Rodríguez}
\email{carla.rodriguez@mpl.mpg.de}
\affiliation{Max Planck Institute for the Science of Light, Erlangen, Germany}
\author{Sören Arlt}
\email{soeren.arlt@mpl.mpg.de}
\affiliation{Max Planck Institute for the Science of Light, Erlangen, Germany}
\author{Leonhard Möckl}
\email{leonhard.moeckl@mpl.mpg.de}
\affiliation{Max Planck Institute for the Science of Light, Erlangen, Germany}
\author{Mario Krenn}
\email{mario.krenn@mpl.mpg.de}
\affiliation{Max Planck Institute for the Science of Light, Erlangen, Germany}

\date{\today}

\begin{abstract}
Driven by human ingenuity and creativity, the discovery of super-resolution techniques, which circumvent the classical diffraction limit of light, represent a leap in optical microscopy. However, the vast space encompassing all possible experimental configurations suggests that some powerful concepts and techniques might have not been discovered yet, and might never be with a human-driven direct design approach. Thus, AI-based exploration techniques could provide enormous benefit, by exploring this space in a fast, unbiased way. We introduce \algo, an open-source computational framework developed using JAX, which offers enhanced computational speed enabled by its accelerated linear algebra compiler (XLA), just-in-time compilation, and its seamlessly integrated automatic vectorization, auto-differentiation capabilities and GPU compatibility. Remarkably, \algo demonstrates a speed-up of 4 orders of magnitude compared to well-established numerical optimization methods. We showcase \algo's potential by re-discovering three foundational experiments in advanced microscopy. Ultimately, \algo identified a novel experimental blueprint featuring sub-diffraction imaging capabilities. This work constitutes an important step in AI-driven scientific discovery of new concepts in optics and advanced microscopy.

\end{abstract}
\maketitle

\section{Introduction}\label{sec_intro}
The space of all possible experimental optical configurations is enormous. For example, if we consider experiments that consist of just 10 optical elements, chosen from 5 different components (such as lasers, lenses, phase shifters, beam splitters and cameras), we already get 10 million possible discrete arrangements. The experimental topology (i.e., how the elements are arranged) will further increase this number greatly. Finally, each of these optical components can have tunable parameters (such as lenses' focal lengths, laser power or splitting ratios of beam splitters) which lead to additional high-dimensional continuous parameter space for each of the previously mentioned discrete possibilities. This vast search space contains all experimental designs possible, including those with exceptional properties. So far, researchers have been exploring this space of possibilities guided by experience, intuition and creativity -- and have uncovered countless exciting experimental configurations and technologies. But due to the complexity of this space, it might be that some powerful concepts and techniques have not been discovered so far, and might never be with a human-driven direct design approach. This is where AI-based exploration techniques could provide enormous benefit, by exploring the space in a fast, unbiased way \cite{wang2023scientific, krenn2022scientific}.

\begin{figure*}[ht!]
  \centering
  \includegraphics[width=1\linewidth]{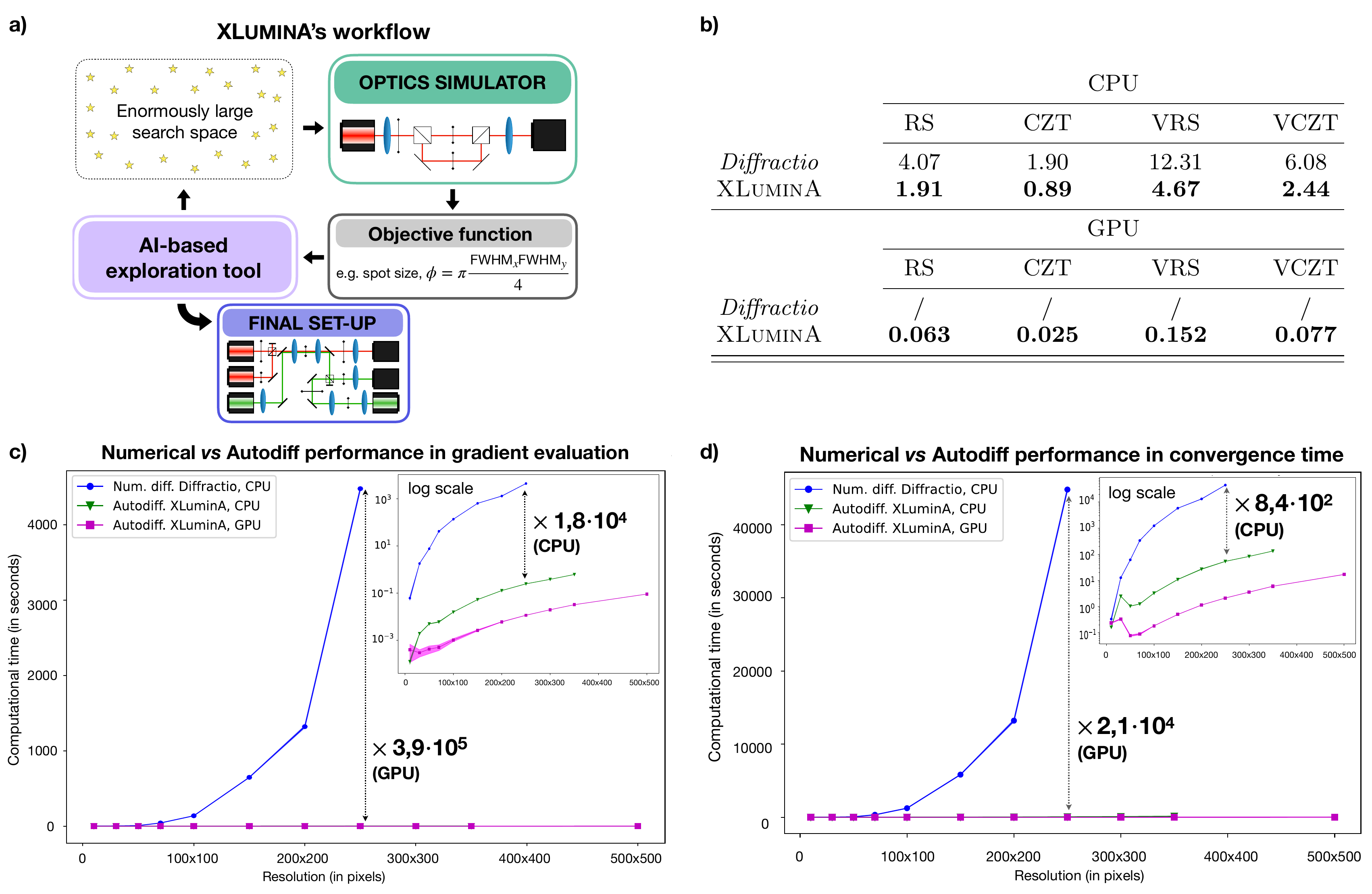}
  \caption{Overview and performance of \algo. (a) Software's workflow, demonstrating the integrated feedback between the AI discovery tool and the optics simulator. Stars depict experimental blueprints with exceptional and useful properties. We start by feeding the system an initial random set of optical parameters, which shape the hardware design on a virtual optical table. The performance of the virtual experiment is computed by the simulator, which leads to detected light (e.g., captured images at the camera). From those simulated outputs, the objective function (for instance, the spot size), is computed. To improve the metric of the cost function, the optimizer adjusts the optical parameters in the initial virtual setup and the cycle is repeated. The whole process is a back-and-forth between the simulator and the optimizer, refining the setup until a convergence is observed. (b) Average execution time (in seconds) over 100 runs, within a resolution of $2048\times 2048$ pixels, for scalar and vectorial field propagation using Rayleigh-Sommerfeld (RS, VRS) and Chirped z-transform (CZT, VCZT) algorithms in \textit{Diffractio} and \algo. Times for \algo correspond to runs with pre-compiled jitted functions. Our methods demonstrate enhanced computational speeds for simulating light diffraction and propagation: a factor of $\times$2 for RS and CZT and about $\times$2.5 for VRS and VCZT using the CPU. With GPU utilization, the speed increases up to two orders of magnitude with factors of $\times$64 for RS, $\times$76 for CZT, $\times$80 for VRS and $\times$78 for VCZT. The single run data is depicted in Extended Data Fig. \ref{fig:extended_data_propagation_runs}. (c) Average time (in seconds) over 5 runs for a single gradient evaluation using numerical differentiation with \textit{Diffractio}'s optical simulator (blue dots) and autodiff methods (green triangles for CPU and magenta squares for GPU) with \algo's optical simulator for different resolutions. The use of \algo with autodiff methods improves the gradient evaluation time by a factor of $\times 3.9 \cdot 10^5$ in the GPU and a factor of $\times 1.8 \cdot 10^4$ on the CPU for resolutions of $250\times250$ pixels. The superior efficiency of autodiff over traditional numerical methods allows for highly efficient optimizations particularly employing the large high resolutions we use (up to 2048$\times$2048 pixels). (d) Average time (in seconds) over 5 runs for convergence time, using numerical differentiation with \textit{Diffractio}'s optical simulator (blue dots) and autodiff methods (green triangles for CPU and magenta squares for GPU) with \algo's optical simulator for different resolutions. Autodiff methods on \algo improves the convergence time with respect to numerical methods by a factor of $\times 2.1 \cdot 10^4$ in the GPU and a factor of $\times 8.4 \cdot 10^2$ in the CPU for a resolution of $250\times250$ pixels. Shaded regions correspond to standard deviation values. The numerical and autodiff methods are computed using BFGS and Adam optimizers, respectively. The speed-up magnitude for different pixel resolutions is depicted in Extended Data Fig. \ref{fig:extended_data_speedup}. Further comparison across different optimizers is presented in Extended Data Fig. \ref{fig:extended_data_optimizers}. All the experiments were run on an Intel CPU Xeon Gold 6130 and Nvidia GPU Quadro RTX 6000.}
  \label{fig:performance}
\end{figure*}

Optical microscopes in today’s sense were invented 300 years ago by Antonj van Leeuwenhoek \cite{Antonj_300years}. Since then, few techniques used in the sciences have seen a similarly rapid development and impact on diverse fields, ranging from material sciences all the way to medicine \cite{AR_microscopy, Sandoghdar_light_microscopy, Bullen_microdrug, PMA_lightmicro}. Arguably, optical microscopy is currently most widely used in biological sciences, where precise labeling of imaging targets enables fluorescence microscopy with exquisite sensitivity and specificity \cite{Grimm_review, Palmer_fluorophores}. In the past two decades, several breakthroughs have broadened the scope of optical microscopy in this area even further. Among them, through the ingenuity and creativity of human researchers, the discovery of super-resolution (SR) methods, which circumvent the classical diffraction limit of light, stand out in particular. Examples for versatile and powerful SR techniques are STED \cite{Hell:94}, PALM/F-PALM \cite{PALM, f-PALM}, (d)STORM \cite{STORM, dSTORM}, SIM \cite{SIM}, and MINFLUX \cite{MINFLUX}, with considerable impact in biology \cite{MOCKL201957, KeXu2013, Yildiz2003}, chemistry \cite{Zhang2015} and material sciences \cite{Mueller2019} for example. Crucially, the motivation of our work goes far beyond small-scale optimization of already known optical techniques. Rather, this work sets out to discover novel, experimentally viable concepts for advanced optical microscopy that are at-present entirely untapped.

We introduce \algo \footnote{GitHub: \href{https://github.com/artificial-scientist-lab/XLuminA}{https://github.com/artificial-scientist-lab/XLuminA}}, an efficient open-source framework developed using JAX \cite{jax2018github}, for the ultimate goal of discovering new optical design principles. \algo offers enhanced computational speed enabled by its accelerated linear algebra compiler (XLA), just-in-time (jit) compilation, and its seamlessly integrated automatic vectorization or batching, auto-differentiation capabilities \cite{autodiffsurvey} and GPU compatibility. We leverage its scope with a specific focus on the area of SR microscopy, which is a set of techniques that has revolutionized biological and biomedical research over the past decade, highlighted by the 2014 Chemistry Nobel Prize \cite{sted_moeckl}. The software's workflow is depicted in Fig. \ref{fig:performance}a. Fundamentally, the simulator is the heart of digital discovery efforts. It translates an experimental design (one point in the vast space of possible designs) to a physical output. The physical output, such as a detector or camera output, can then be used in an objective function to describe the desired design goal. The simulator can either be called directly by gradient-based optimization techniques, or it can be used for generating the training data for deep-learning-based surrogate models. A simulator that can be used for automated design and discovery of new experimental strategies must be (1) fast, (2) reliable, and (3) general. \algo's optical simulator fulfills precisely the aforementioned requirements for advanced microscopy. 

The paper is structured as follows. Upon reviewing previous work, we describe \algo and highlight its efficiency and computational speed advantage over conventional approaches. We demonstrate the applicability of our approach by rediscovering three foundational optical layouts. First, using a data-driven learning methodology, we rediscover an optical configuration commonly used to adjust beam and image sizes. Then, following pure AI-exploratory strategies within a fully continuous framework we rediscover, together with new superior topologies, a beam-shaping technique as employed in STED (stimulated emission depletion) microscopy \cite{Hell:94} and the SR technique exploiting optical vortices \cite{leuchs}. Ultimately, we showcase \algo's capability for genuine discovery, identifying a novel solution that integrates the underlying physical principles present in the two aforementioned SR techniques into a single experimental blueprint, the performance of which exceeds the capabilities of each individual setup. We then discuss the discovered solutions and the applicability of \algo. Finally, we conclude with final remarks and future perspectives.

\subsection{Previous work}
\paragraph{\textbf{Optimization in microscopy}} Our approach is radically different from previous strategies that employ AI for data-driven design of single optical elements \cite{differentiable_microscope, miniscope3D, Wang:19} or data analysis in microscopy, e.g. denoising, contrast enhancement or point-spread-function (PSF) engineering \cite{deepSTORM_PSF, Fu:22, jia2014, Izeddin:12}. While these techniques are influential, they are not meant to change the principle of the experimental approach or the optical layout itself. In contrast, \algo is equipped with tools to simulate, optimize and automatically design new optical setups and concepts from scratch.

\paragraph{\textbf{Discovery in quantum optics}} Numerous works have recently shown how to automatically design new quantum experiments with advanced computational methods \cite{krenn2016automated, knott2016search, ruiz2022digital, valcarce2023automated}, that has led to the discovery of new concepts and numerous blueprints implemented in laboratories \cite{QO_space}. Other simulators such as \textit{Strawberry fields} focus specifically on optimization in photonic quantum computing \cite{killoran2019strawberry}.

\paragraph{\textbf{Design in nanophotonics and photonic materials}} The field of optical \textit{inverse design} focuses on the de-novo design of nano-optical components with practical features\cite{molesky2018inverse, so2020deep}. Examples include on-chip particle accelerators \cite{sapra2020chip}, or wavelength-division multiplexers \cite{su2018inverse}. The main approach is the development of efficient PDE-solvers for Maxwell's equations, including efficient ways to compute the gradients of the vast amount of parameters, usually by a physics-inspired technique called the \textit{adjoint method} \cite{SFan_ID2018, SFan_AD2020}. These techniques are highly computationally expensive \cite{lesina2015convergence} due to their physical targets. We have different physical targets, thus can apply various different approximations in the beam propagation which significantly speeds up our simulator. Interestingly, the adjoint method can be seen as a special case of auto-differentiation (which we use) \cite{SFan_AD2020}.

\paragraph{\textbf{Classical optics simulators}} Several open-source software tools facilitate classical optics phenomena simulations. Some examples are \textit{Diffractio} for light diffraction and interference simulations \cite{diffractio}, \textit{Finesse} for simulating gravitational wave detectors \cite{Finesse}, which do not support auto-differentiation nor GPU compatibility; and \textit{POPPY}, developed as a part of the simulation package of the James Webb Telescope \cite{POPPY2012}, with GPU compatibility but lacking autodiff capabilities. There are also specialized resources like those focusing on the design of Laguerre-Gaussian mode sorters utilizing multi-plane light conversion (MPLC) methods \cite{LGsorter, Labroille:14}, which also do not support GPU computations and autodiff. While these software solutions offer optics simulation capabilities, \algo uniquely integrates simulation with AI-driven automated design powered with JAX's autodiff, just-in-time compilation and automatic GPU compatibility.

\section{Software workflow and performance}\label{sec_software}

\algo allows for the simulation of classical optics hardware configurations and enables the optimization and automated discovery of new setup designs. The software is developed using JAX \cite{jax2018github}, which provides an advantage of enhanced computational speed (enabled by accelerated linear algebra compiler, XLA, with just-in-time compilation, jit) while seamlessly integrating the auto-differentiation framework \cite{autodiffsurvey} and automatic GPU compatibility. It is important to remark that our approach is not restricted to run on CPU (as NumPy-based softwares do): due to JAX-integrated functionalities, by default runs on GPU if available, otherwise automatically falls back to CPU.

The ultimate goal is to discover new concepts and experimental blueprints in optics. Importantly, the most computational expense of an optimization loop comes from running individual optical simulations in each iteration. Thus, it is essential to reduce the computation time by maximizing the speed of optical simulation functions. \algo is equipped with an optics simulator which contains a diverse set of optical manipulation, interaction and measurement technologies. Some specific optical propagation implementations of \algo are inspired by the optics framework \textit{Diffractio} \cite{diffractio}. \textit{Diffractio} is a high-quality, open-source NumPy-based Python module for optics simulation with an active developer community, and is employed in numerous studies in optics and physics in general. We have rewritten and optimized these optical propagation implementations leveraging JAX's jit functionality, which allows for highly efficient code execution, although it imposes some restrictions such as specifying all data structures' dimensions and ensuring their immutability at compile time. On top of that, we developed completely new functions which significantly expand the software capabilities, such as high-resolution propagation methods, and numerous new optical devices which made the current study possible. Further details on the optics simulator can be found in the Methods section. We evaluate the performance of our optimized functions against their counterparts in \textit{Diffractio}. The acquired run-times are shown in Fig. \ref{fig:performance}b. Clearly, our methods significantly enhance computational speeds for simulating light diffraction and propagation. For instance, we observe a speedup of a factor of $\times$2 for RS (Rayleigh-Sommerfeld, a general Fast Fourier Transform-based light propagation algorithm) and CZT (Chirped z-transform, a speed-up version of RS) and about $\times$2.5 for VRS and VCZT (the vectorized versions of RS and CZT, respectively) using the CPU. With GPU utilization, the speedup factors are of $\times$64 for RS, $\times$76 for CZT, $\times$80 for VRS and $\times$78 for VCZT. 

To include the automated discovery feature, \algo's optical simulator and optimizer are tied together by the loss function, as depicted in Fig. \ref{fig:performance}a. The automated discovery tool is designed to explore the vast parameter space encompassing all possible optical designs. When it comes to the nature of the optimizer, it can be either direct (gradient-based) or deep learning-based (surrogate models or deep generative models, e.g., variational autoencoders \cite{VAE}). In this work, we adopt a gradient-based strategy, where the experimental setup's parameters are adjusted iteratively in the steepest descent direction. We first evaluate the time it takes for numerical and analytical (auto-differentiation) methods to compute one gradient evaluation and their convergence times over different resolutions and devices. For this purpose, we use two gradient-descent techniques: the Broyden-Fletcher-Goldfarb-Shanno (BFGS) algorithm \cite{nocedal1999numerical}, which numerically computes the gradients and higher-order derivative approximations and the Adaptive moment estimation (Adam) \cite{kingma2017adam}, an instance of the stochastic-gradient-descent (SGD) method. While BFGS is part of the open-source SciPy Python library \cite{2020SciPy-NMeth} and operates on the CPU, Adam is integrated within the JAX framework and runs in both CPU and GPU. For this last, we take advantage of JAX's built-in autodiff framework and compute analytically the gradients of the loss function. Combined with the jit (just-in-time) compilation functionality, this approach enables the optimizer to efficiently construct an internal gradient function, considerably reducing the computational time per iteration. The acquired results are depicted in Figs. \ref{fig:performance}c and \ref{fig:performance}d. The detailed description of both evaluations is provided in the Methods section. Clearly, autodiff consistently outperforms numerical methods on the gradient evaluation time by up to 4 orders of magnitude on CPU and 5 orders on GPU. In convergence time, autodiff demonstrates superior efficiency up to almost 3 orders of magnitude on CPU and 4 orders on GPU. Given that certain optical elements, such as phase masks, may operate at resolutions as high as $2048\times2048$ pixels, the resulting search space can easily expand to around 8.4 million parameters. This makes the use of autodiff within GPU-accelerated frameworks more appropriate for efficient experimentation. Overall, the computational performance of \algo highlights its suitability for running complex simulations and optimizations with a high level of efficiency.

\begin{figure*}[!htb]
  \centering
  \includegraphics[width=1\linewidth]{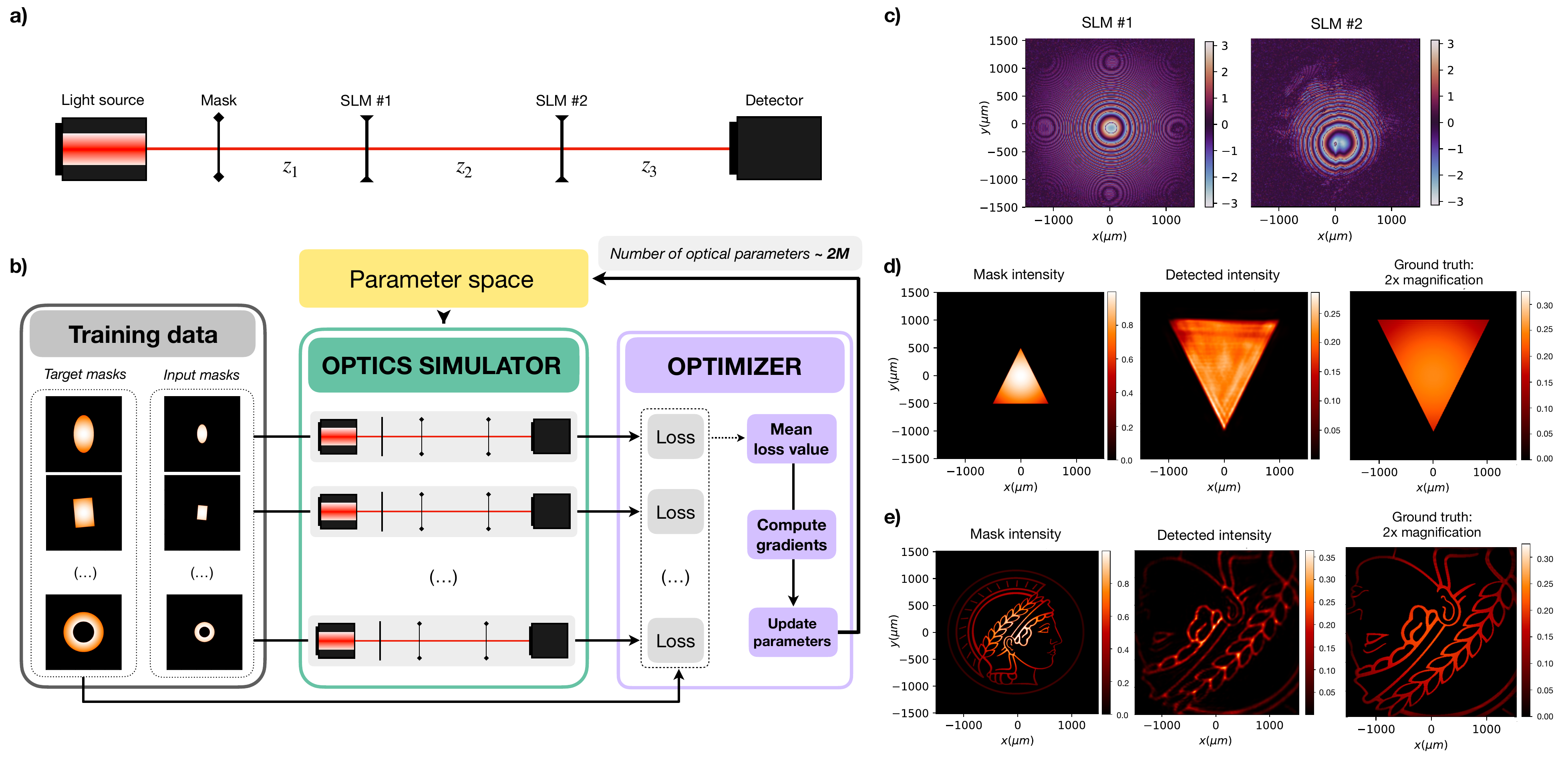}
  \caption{Rediscovery of the optical configuration employed to magnify images. (a) Virtual optical arrangement. It consists of a light source emitting a 650 nm wavelength Gaussian beam. Original lenses are replaced by two spatial light modulators (SLMs) with a resolution of $1024\times1024$ and pixel size of $2.92$ $\mu m$. The parameter space (of $\sim$ 2 million optical parameters) includes the distances, $z_1$, $z_2$ and $z_3$ (in millimeters) and the phase masks (in radians) of the two SLMs. (b) Data-driven discovery scheme. Input-output sample pairs are fed into the optics simulator in batches of 10. The loss function, computed for each virtual optical setup, evaluates the mean squared error between the intensity response of the system and the corresponding target example from the dataset. The average loss over the batch guides the optical parameter update, which is common to all the virtual optical setups. This cycle is repeated until convergence is reached. (c) Identified phase mask solutions for SLM$\#$1 and SLM$\#2$. Identified distances correspond to $z_1=10.14$ cm, $z_2= 5.46$ cm and $z_3=7.54$ cm. Input, detected intensity and expected (ground truth) intensity patterns for (d) a simple geometry, and (e) a complex structure, the \textit{Max Planck Society}'s logo. In both cases, the identified optical design successfully inverts and magnifies $2\times$ the input mask.}
  \label{fig:4f}
\end{figure*} 

\section{Results}\label{sec_results}
In this section, we showcase the virtual optical designs generated by \algo. As benchmarks, we aim to re-discover three foundational experiments, each one covering different areas in optics. By increasing the complexity of the description of light (from scalar to vectorial fields representation), we selected: (1) an optical configuration commonly used to adjust beam and image sizes, (2) beam shaping as employed in STED microscopy \cite{Hell:94}, and (3) the super-resolution technique using optical vortices detailed in Ref. \cite{leuchs}. Finally, we demonstrate the discovery of a new experimental blueprint within a large-scale exploration framework. For the first example, we use a data-driven learning methodology. For the last three, we set-up a discovery scheme where no training data is involved. The showcased solutions in both scenarios are the result from running multiple optimizations.

\subsection{Data-driven rediscovery}
The optical configuration to adjust beam and image sizes comprises two lenses, each one positioned a focal length apart from their respective input and output planes, $f_1$ and $f_2$, respectively, and $f_1 + f_2$ from each other. This arrangement performs optical Fourier transformations of input light with magnifications determined by the ratio $f_2/f_1$. To revisit this design with a magnification of 2$\times$, we encoded the virtual setup depicted in Fig. \ref{fig:4f}a.

The data-driven learning approach is outlined in Fig. \ref{fig:4f}b. This workflow resembles the dynamics of training, in a supervised way, a physical learning system \cite{McMahon_2023, Wright_2022}. The system dynamically adjusts the optical elements to effectively transform the input data into the desired configuration. The cost function is computed as
\begin{equation}
    \mathcal{L}=\text{MSE}(I_\text{Det}, I_\text{GT})\;,
\end{equation}
where MSE is the mean squared error between the detected intensity pattern from the virtual setup, $I_{\text{Det}}$ and the corresponding ground truth from the dataset, $I_{\text{GT}}$. The parameter space comprises $1024\times1024$ pixel phases and $3$ optical distances (a total of $\sim$ 2 million parameters). Further details of the optimization are provided in the Methods section. 

The obtained results, displayed in Fig. \ref{fig:4f}c, depict lens-like quadratic phases. Notably, the reference model traditionally uses two lenses set at specific distances, yet the identified distances don't fulfill such relation. We validate the performance of the identified configuration by imaging the triangle-shaped amplitude mask shown in Fig. \ref{fig:4f}d, not included in the training data. We further demonstrate how the identified configuration generalizes to complex structures by using the \textit{Max Planck Society}'s logo in Fig. \ref{fig:4f}e. In both cases, the detected intensity distribution demonstrates that the optical setup inverts and magnifies the input shape by 2$\times$.

\begin{figure*}[ht!]
  \centering
  \includegraphics[width=0.75\linewidth]{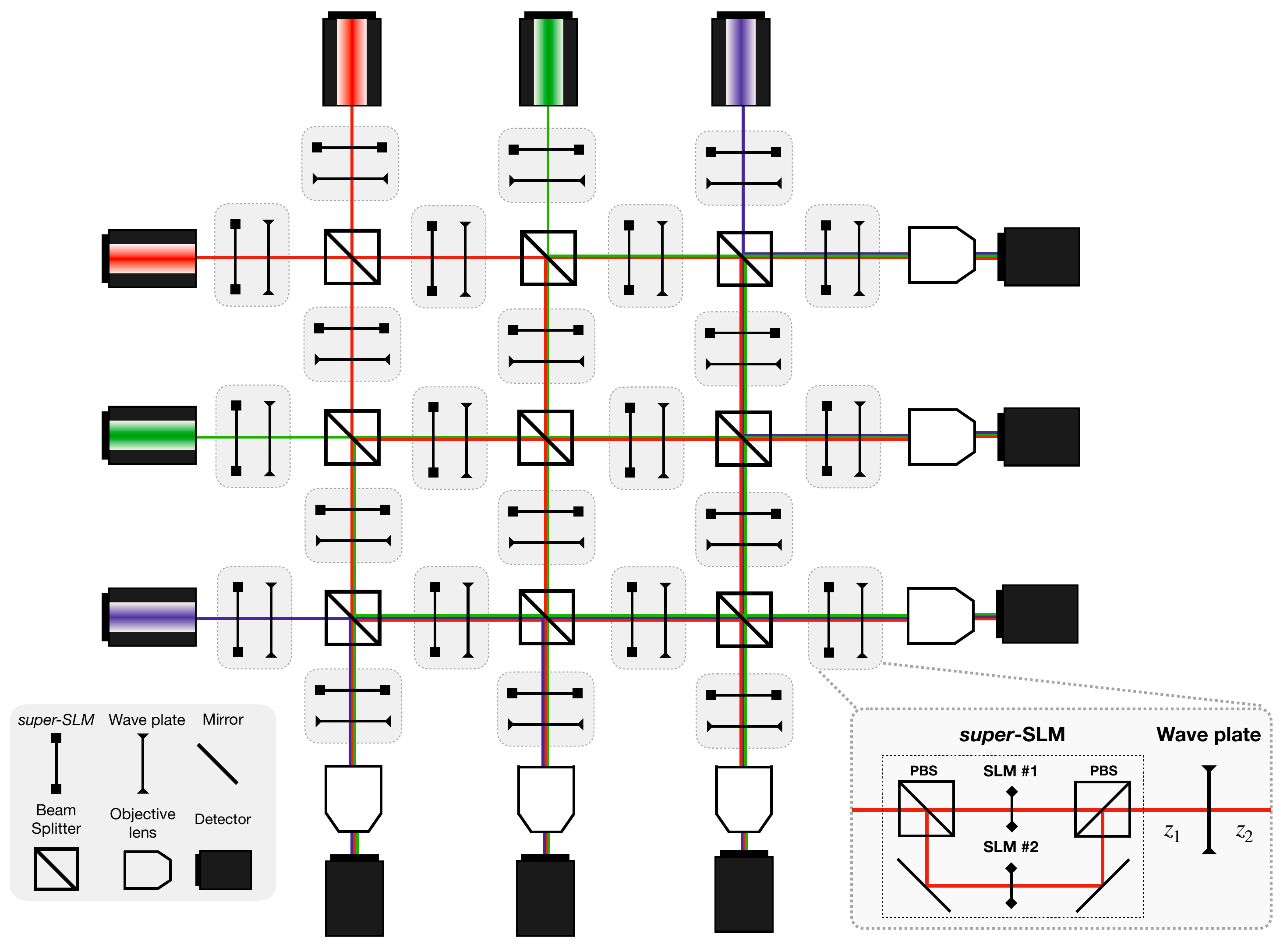}
  \caption{General virtual optical setup for large-scale discovery schemes. Gray boxes represent fundamental building units, each containing a \textit{super-SLM} and a wave plate positioned a distance $z_1$ apart. These units are inter-connected through free propagation distances $z_2$, and beam splitters. The \textit{super}-SLM is a hardware-box-type which consists of two spatial light modulators (SLMs), each one independently imprinting a phase pattern on the horizontal and vertical polarization components of the field. The setup’s complexity and size can be arbitrarily extended by incorporating additional connections, building units, light sources, detectors, etc. The quantization of the large-scale search space generated by this optical setup and the analysis on the computational efficiency of \algo in constructing this optical setup is provided in the Methods section.}
  \label{fig:large-scale}
\end{figure*} 

\subsection{Towards large-scale discovery}

We aim to use \algo to discover new microscopy concepts.  In essence, discovering new experimental configurations entails an hybrid discrete-continuous search problem. The discrete aspect originates from configuring the optical network topology, whereas the continuous part is linked to the settings of optical elements, such as laser power and beam splitter reflectivity. Discrete-continuous optimization is extremely difficult computationally, therefore we invent a way to translate this hybrid discrete-continuous optimization problem into a purely continuous optimization problem which can be solved with efficient gradient-based methods. We design the quasi-universal computational \textit{ansatz} illustrated in Fig. \ref{fig:large-scale}, which is designed in a way that setting different (continuous) parameters leads to different optical setup topologies. Remarkably, for a very discrete approach of available parameters, the number of possible discrete arrangements within this general framework scales up to $\sim 10^{20}$. Details on this derivation can be found in the Methods section.

Now, the task of \algo is to automatically discover new, superior topologies together with their parameter setting, using purely continuous optimization. To achieve this, we initialize the setups with a large and complex optical topology, inspired by other fields that start with highly expressive initial circuits \cite{sim2019expressibility,krenn2021conceptual}. From here, \algo should be able to extract much more complex solutions which humans might not have thought about yet \cite{krenn2022scientific}.

To test our framework, we target \algo to rediscover the concepts of STED microscopy \cite{Hell:94} and the super-resolution (SR) technique using optical vortices \cite{leuchs}. To do so, we conduct an exploration-based optimizing procedure which does not involve training examples. In particular, we first explore the optimization of the optical layout within systems containing pre-defined elements. Later on, we optimize both the optical topology and highly parameterized elements (i.e., SLMs). Early experiments on \algo rediscovering SR-techniques involve the use of fixed initial topologies. The obtained results are provided in the Methods section.

\subsubsection{Loss function}
The loss function, $\mathcal{L}$, is calculated as the inverse of the density of the total detected intensity over a certain threshold, $I_{\varepsilon}$. Thus, minimizing $\mathcal{L}$ aims to maximize the generation of small, high intensity beams. In particular,
\begin{equation}
    \mathcal{L} = \frac{1}{\text{Density}} = \frac{\text{Area}}{I_{\varepsilon}}
    \label{eq:loss}
\end{equation}
where $I_{\varepsilon}$ is the sum of pixel intensity values greater than the threshold value $\varepsilon \cdot i_{\text{max}}$, where $0\leq\varepsilon\leq 1$ and $i_{max}$ corresponds to the maximum detected intensity. The Area corresponds to the total number of camera pixels fulfilling the same condition. The loss function $\mathcal{L}$ is common to all the optical setups henceforth described. Importantly, light gets detected across various devices. Thus, we compute the loss function at each detector and the parameter update is driven by the device demonstrating the minimum loss value. This selection is performed in a fully differentiable manner. Details on the derivation of the loss function are provided in the Methods section. 

\subsubsection{Rediscovery through exploration of optical topologies}
In this section we target \algo to discover optical topologies for STED \cite{Hell:94} and the sharp focus \cite{leuchs} techniques within virtual optical tables that contain randomly positioned phase masks displaying fixed patterns. The goal is to discover the optical topology using the available optimizable optical parameters: beam splitter ratios, distances and wave plate's angles. The detailed description of the optimization processes hereby conducted are provided in the Methods section. 

\begin{figure*}[ht!]
  \centering
  \includegraphics[width=1\linewidth]{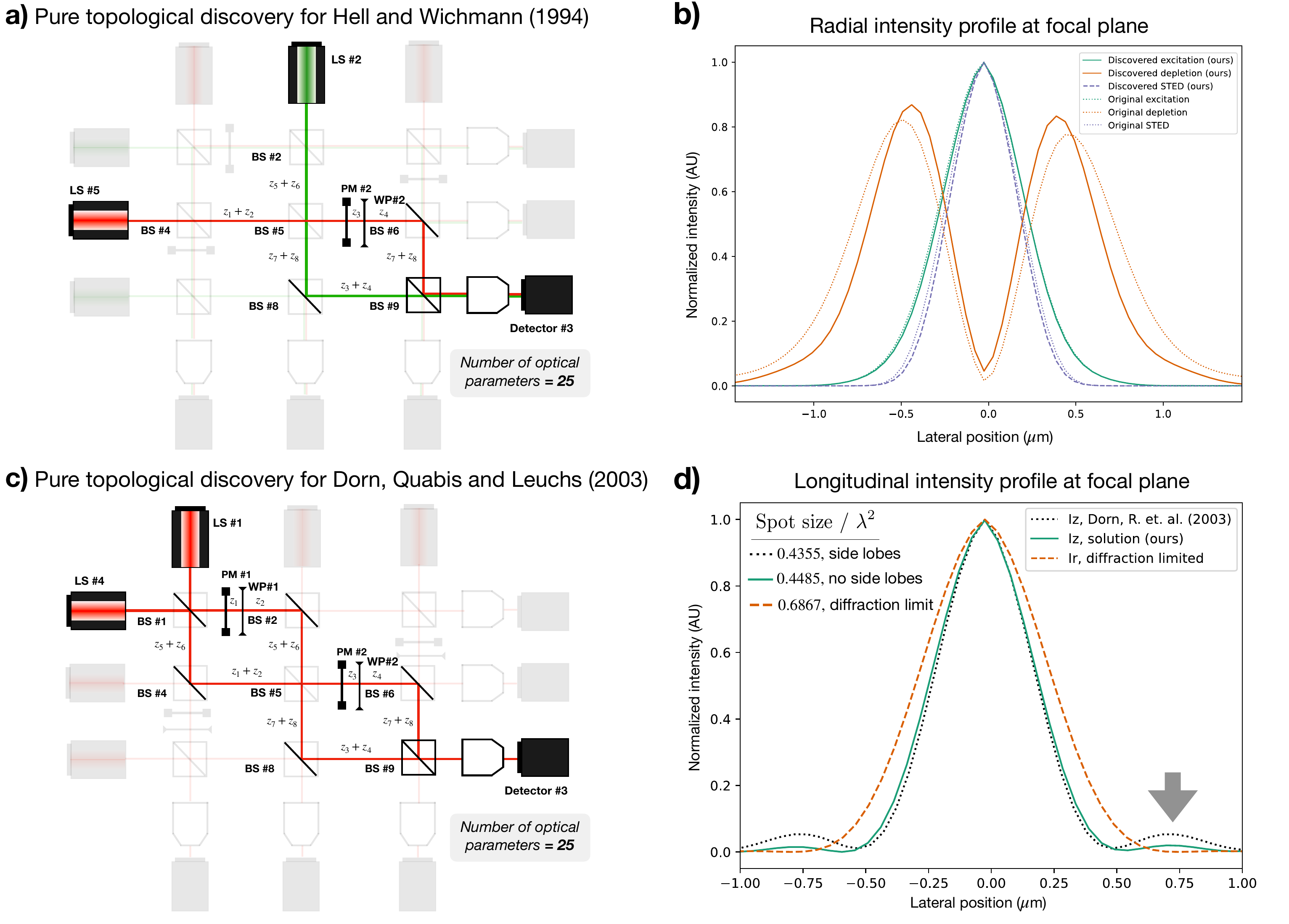}
  \caption{Pure topological discovery within a fully continuous framework. (a) Discovered optical topology for STED microscopy (Hell and Wichmann, 1994). The initial optical setup and phase masks used for this experiment are depicted in Extended Data Figs. \ref{fig:extended_data_init_setup_xl}a and \ref{fig:extended_data_init_setup_xl}c, respectively. The parameter space (25 optical parameters) is defined by 9 beam splitter ratios, 8 distances and 4 wave plates (with variable phase retardance $\eta$ and orientation angle $\theta$). The minimum value of the loss function is demonstrated in detector $\#3$. The setup topology is retrieved from detector $\#3$ following the identified beam splitter ratios across the system. The identified optical parameters correspond to: the beam splitter ratios, in [Transmittance, Reflectance] pairs: BS$\#2$: [0.999, 0.000], BS$\#4$: [0.926, 0.073], BS$\#5$: [0.998, 0.001], BS$\#6$: [0.001, 0.998], BS$\#8$: [0.342, 0.658], and BS$\#9$: [0.664, 0.336]. The wave plate, in radians (2): $\eta = 1.911$, $\theta = -1.697$. The propagation distances (in cm) are $z_1 = 74.70$, $z_2 = 34.31$, $z_3 = 59.07$, $z_4 = 27.13$, $z_5 = 36.13$, $z_6 = 18.69$, $z_7 = 76.03$ and $z_8 = 20.88$. The phase mask $\#2$ corresponds to the radial phase pattern originally used in STED microscopy, which generates a doughnut-shaped beam. (b) Radial intensity profile, $|E_x|^2 + |E_y|^2$, in horizontal beam section: excitation (green), depletion (orange), and super-resolution effective STED beam (dashed blue line). The data corresponding to the original STED spiral phase are indicated with dotted lines. The excitation and depletion beams are diffraction-limited. The effective response breaks the diffraction limit. Lateral position indicates lateral distance from the optical axis. (c) Discovered optical topology for Dorn, Quabis and Leuchs (2003). The initial optical setup and phase masks used for this experiment are depicted in Extended Data Figs. \ref{fig:extended_data_init_setup_xl}b and \ref{fig:extended_data_init_setup_xl}c, respectively. The parameter space (25 optical parameters) is defined by 9 beam splitter ratios, 8 distances and 4 wave plates (with variable phase retardance $\eta$ and orientation angle $\theta$). The minimum value of the loss function is demonstrated in detector $\#3$. The setup topology is retrieved from detector $\#3$ following the identified beam splitter ratios across the system. The identified optical parameters correspond to: the beam splitter ratios, in [Transmittance, Reflectance] pairs: BS$\#1$: [0.000, 0.999], BS$\#2$: [0.000, 0.999], BS$\#4$: [0.037, 0.963], BS$\#5$: [0.999, 0.000], BS$\#6$: [0.030, 0.970], BS$\#8$: [0.001, 0.999], and BS$\#9$: [0.589, 0.411]. The wave plates, in radians (1): $\eta = -0.027$, $\theta = -1.991$; and (2): $\eta = -4.698 $, $\theta = -1.551$. The propagation distances (in cm) are $z_1 = 15.27$, $z_2 = 46.46$, $z_3 = 16.08$, $z_4 = 90.66$, $z_5 = 173.84$, $z_6 = 121.16$, $z_7 = 30.07$ and $z_8 = 70.09$. The phase masks $\#1$ and $\#2$ correspond to the polarization converter demonstrated in Dorn, Quabis, and Leuchs (2003) and a radial phase pattern originally used in STED microscopy, respectively. Both phase patterns generate, independently, a doughnut-shaped beam. (d) Normalized longitudinal intensity profile, $|E_z|^2$, for Dorn, Quabis, and Leuchs (2003) and the identified solution (black dotted, and green lines, respectively) and radial intensity profile, $|E_x|^2+|E_y|^2$, of the diffraction-limited linearly polarized beam (orange dotted line). Lateral position indicates lateral distance from the optical axis. The spot size is computed as $\phi = (\pi/4)\text{FWHM}_x \text{FWHM}_y$, where FWHM denotes for Full Width Half Maximum. The discovered approach breaks the diffraction limit, demonstrating a spot size close to the reference. Remarkably, it does not feature side lobes (indicated with a gray arrow), which can limit practical imaging techniques.}
  \label{fig:XLSetup}
\end{figure*}

STED microscopy \cite{Hell:94, Hell2005} is one of the first discovered techniques that circumvent the classical diffraction limit of light. The key idea of this technique is the use of two diffraction-limited laser beams, one probe to activate (excite) the light emitters of the sample and one, doughnut-shaped beam to deactivate its excitation in a controlled way (depletion). Thus, the ultimately detected light is that of the emitters laying in the central region of the doughnut-shaped beam. This effectively reduces the area of normal fluorescence, which leads to super-resolution imaging. To simulate one of the fundamental concepts of STED without having to rely on time-dependent processes, such as the energy level relaxation times of the excited emitters, we perform a nonlinear modulation of the focused light based on the Beer-Lambert law \cite{Mayerhfer2020TheBL}, commonly used to describe the optical attenuation in light-matter interaction. The details of our model are provided in the Methods section.

We initialize \algo with the virtual optical table in Extended Data Fig. \ref{fig:extended_data_init_setup_xl}a. The loss function corresponds to equation (\ref{eq:loss}) considering the radial component of the effective beam resulting from the STED process. The discovered topology is depicted in Fig. \ref{fig:XLSetup}a. It shapes the depletion beam into a doughnut using the phase mask $\#2$, which corresponds to the spiral phase pattern originally used in STED microscopy. We compute the horizontal cross-section of the focused intensity patterns for both excitation and depletion beams (green and orange lines in Fig. \ref{fig:XLSetup}b, respectively) and the effective beam (dotted blue line in Fig. \ref{fig:XLSetup}b). The behavior across the vertical axis yields similar features. For comparison, we also feature the radial intensity profile of the simulated STED reference. 

The generation of an ultra-sharp focus is a feature that breaks the diffraction limit in the longitudinal direction as demonstrated by Dorn, Quabis and Leuchs in Ref. \cite{leuchs}. This super-resolution is achieved when a radially polarized beam is tightly focused \cite{quabis_2001, Quinteiro2017}. We initialize \algo with the virtual optical table in Extended Data Fig. \ref{fig:extended_data_init_setup_xl}b. The loss function corresponds to equation (\ref{eq:loss}) considering the measured intensity as the field's longitudinal component. The discovered topology is depicted in Fig. \ref{fig:XLSetup}c. It shapes the beam into a doughnut using a polarization converter originally used in Ref. \cite{leuchs} and the spiral phase pattern from STED microscopy (phase masks $\#1$ and $\#2$ in Fig. \ref{fig:XLSetup}c, respectively). The longitudinal intensity profiles for the reference and the discovered solution are depicted in Fig. \ref{fig:XLSetup}d (represented by dotted black and green, respectively). For comparison, we also feature the radial intensity profile of the diffraction-limited beam (dotted orange line in Fig. \ref{fig:XLSetup}d). Remarkably, the identified solution demonstrates a spot size close to the reference and does not feature side lobes, which can limit practical imaging techniques.

In both cases we successfully demonstrate how \algo can explore different topologies in a fully continuous manner: by adjusting the optical parameters, (e.g., beam splitter ratios), the optimizer can ``\textit{turn off}" the optical paths.

Importantly, we are not restricted to the use of $3\times3$ optical grids. We conduct an optimization using an optical grid of $6\times6$. The results are depicted in Extended Data Fig. \ref{fig:6x6} and discussed in the Methods section.

\subsubsection{Rediscovery through exploration in highly parameterized systems}

The results presented thus far predominantly involve optical setups characterized by fixed phase masks. We further demonstrate \algo's potential in highly parameterized, complex optical systems. We target \algo to rediscover the aforementioned SR techniques within the optical table in Extended Data Fig. \ref{fig:extended_data_init_setup_discovery}, where \textit{super-SLM}s are now optimizable. The goal here is to discover both the optical topology and the phase patterns to imprint onto the light beams using the available optimizable optical parameters (i.e., SLMs, distances, beam splitter ratios and wave plate's angles). The optimization processes hereby conducted are detailed in the Methods section. 

\begin{figure*}[ht!]
  \centering
  \includegraphics[width=1\linewidth]{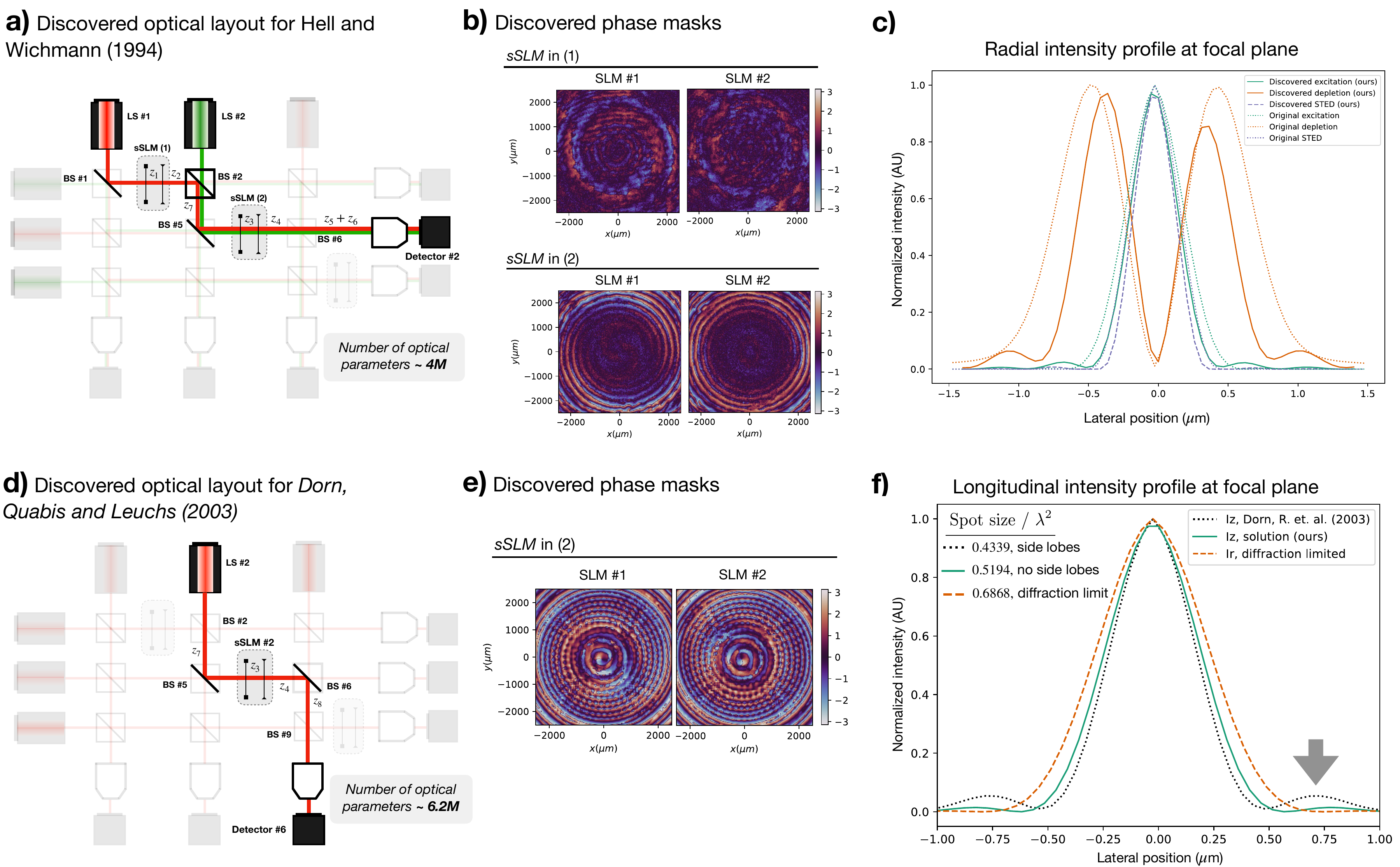}
  \caption{Rediscovery of optical solutions within highly parameterized optical systems. (a) Discovered optical topology for STED microscopy (Hell and Wichmann, 1994). The initial optical setup is depicted in Extended Data Fig. \ref{fig:extended_data_init_setup_discovery}a. The parameter space ($\sim$ 4 million optical parameters) is defined by 3 \textit{super}-SLMs (i.e., 6 SLMs) of $(824\times824)$ pixel resolution and a computational pixel size of $6.06 \mu m$, 9 beam splitter ratios, 8 distances and 3 wave plates (with variable phase retardance $\eta$ and orientation angle $\theta$). The minimum value of the loss function is demonstrated in detector $\#2$. The setup topology is retrieved from detector $\#2$ following the identified beam splitter ratios across the system. The identified optical parameters correspond to: the beam splitter ratios, in [Transmittance, Reflectance] pairs: BS$\#1$: [0.000, 0.999], BS$\#2$: [0.201, 0.799], BS$\#5$: [0.000, 0.999], and BS$\#6$: [0.999, 0.000]. The wave plates, in radians (1): $\eta = -1.39$, $\theta = -1.64$, and (2): $\eta = -1.61$, $\theta = -0.86$. The propagation distances (in cm) are $z_1 = 59.52$, $z_2 = 10.14$, $z_3 = 76.36$, $z_4 = 17.93$, $z_5 = 37.07$, $z_6 = 65.95$, and $z_7 = 38.68$. (b) Discovered phase patterns for sSLM $\#1$ and sSLM $\#2$. (c) Radial intensity profile, $|E_x|^2 + |E_y|^2$, in horizontal beam section: excitation (green), depletion (orange), and super-resolution effective STED beam (dashed blue line). The data corresponding to the original STED experiment - i.e., computed using a spiral phase mask - are indicated with dotted lines. Lateral position indicates lateral distance from the optical axis. (d) Discovered virtual optical setup topology for Dorn, Quabis and Leuchs (2003). The initial optical setup is depicted in Extended Data Fig. \ref{fig:extended_data_init_setup_discovery}b. The parameter space ($\sim$ 6.2 million optical parameters) is defined by 3 \textit{super}-SLMs (i.e., 6 SLMs) of $(1024\times1024)$ pixel resolution and a computational pixel size of $4.8 \mu m$, 9 beam splitter ratios, 8 distances and 3 wave plates (with variable phase retardance $\eta$ and orientation angle $\theta$). The minimum value of the loss function is demonstrated in detector $\#6$. The setup topology is retrieved from detector $\#6$ following the identified beam splitter ratios across the system. The identified optical parameters correspond to: the beam splitter ratios, in [Transmittance, Reflectance] pairs: BS$\#2$: [0.999, 0.000], BS$\#5$: [0.000, 0.999], BS$\#6$: [0.000, 0.999], and BS$\#9$: [0.999, 0.000]. The wave plate's $\eta = 1.51$, $\theta = 3.95$; propagation distances (in cm): $z_3=20.49$, $z_4=63.26$, $z_7=47.92$ and $z_8=31.33$. (c) Discovered phase patterns for sSLM $\#2$. (d) Normalized longitudinal intensity profile, $|E_z|^2$, for Dorn, Quabis, and Leuchs (2003) and the identified solution (black dotted, and green lines, respectively) and radial intensity profile, $|E_x|^2 + |E_y|^2$, of the diffraction-limited linearly polarized beam (orange dotted line). Lateral position indicates lateral distance from the optical axis. The spot size is computed as $\phi = (\pi/4)\text{FWHM}_x \text{FWHM}_y$, where FWHM denotes for Full Width Half Maximum. The discovered approach breaks the diffraction limit, demonstrating a larger spot size as the reference. However, it does not feature side lobes (indicated with a gray arrow), which can limit practical imaging techniques.}
  \label{fig:XLSetup_topologies}
\end{figure*}

The discovered topology and phase patterns for STED microscopy are depicted in Figs. \ref{fig:XLSetup_topologies}a and \ref{fig:XLSetup_topologies}b, respectively. As for STED microscopy, the system imprints a phase singularity onto the depletion beam to produce a doughnut shape. In this case, however, it also modulates the excitation beam. The radial intensity profiles of the discovered solution and the reference experiment are depicted in Fig. \ref{fig:XLSetup_topologies}c. 

The discovered topology phase patterns for Dorn, Quabis and Leuchs are depicted in Figs. \ref{fig:XLSetup_topologies}d and \ref{fig:XLSetup_topologies}e, respectively. These produce a $LG_{2,1}$ Laguerre-Gaussian mode \cite{Rubinsztein-Dunlop_2017}, which demonstrates an intensity pattern of concentric rings with a phase singularity in its center. Surprisingly, \algo found an alternative way to imprint a phase singularity onto the beam and produce pronounced longitudinal components on the focal plane. The longitudinal intensity profiles of the discovered solution and the reference experiment are depicted in Fig. \ref{fig:XLSetup_topologies}f. 

Remarkably, in both cases \algo successfully discovered new optical solutions demonstrating similar performance as the reference experiments.

\subsection{Discovery of a new experimental blueprint}

\begin{figure*}[ht!]
  \centering
  \includegraphics[width=0.7\linewidth]{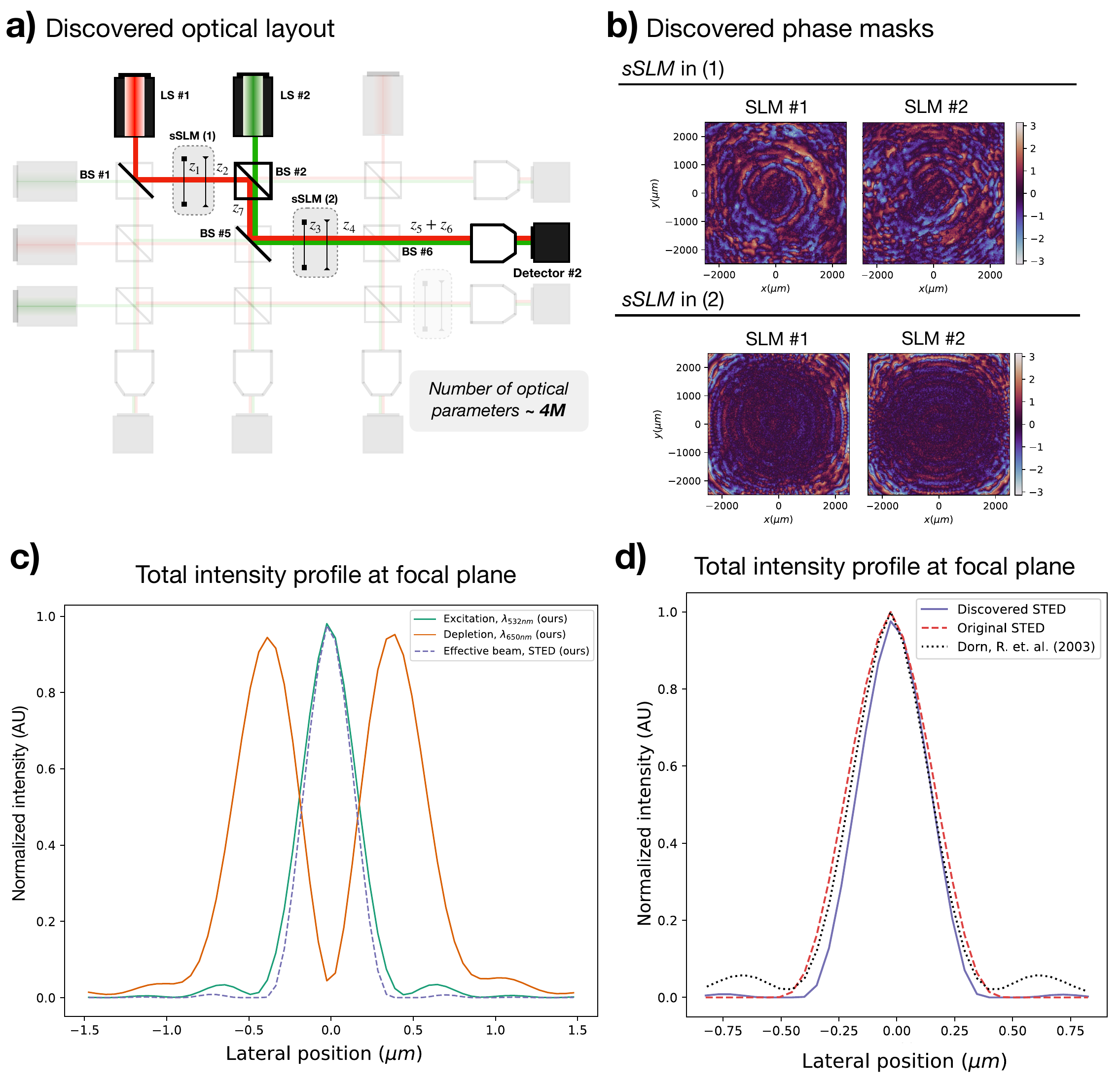}
  \caption{Discovery of a new experimental blueprint within a highly parameterized optical setup. The initial optical setup is depicted in Extended Data Fig. \ref{fig:extended_data_init_setup_discovery}. The parameter space comprises $6\times(824\times824)$ pixel phases, 8 extra optical modulation parameters corresponding to the phase masks and 8 distances (a total of $\sim$ 4 million optical parameters). (a) Discovered optical topology. The minimum value of the loss is demonstrated in detector $\#2$. The setup topology is easily retrieved from detector $\#2$ following the identified beam splitter ratios across the system. The identified optical parameters correspond to: the beam splitter ratios, in [Transmittance, Reflectance] pairs: BS$\#1$: [0.000, 0.999], BS$\#2$: [0.338, 0.662], BS$\#5$: [0.000, 0.999], and BS$\#6$: [0.999, 0.000]. The wave plates, in radians (1): $\eta = 1.09$, $\theta = 0.28$, and (2): $\eta = 0.19$, $\theta = -3.16$. The propagation distances (in cm) are $z_1 = 29.98$, $z_2 = 56.91$, $z_3 = 58.28$, $z_4 = 99.91$, $z_5 = 42.89$, $z_6 = 53.96$, and $z_7 = 50.96$. (b) Discovered phase masks corresponding to the \textit{super}-SLM (\textit{sSLM}) in (1) and (2). (c) Total intensity ($|E_x|^2+|E_y|^2+|E_z|^2$) horizontal cross-section of the detected light beams of $650$ nm (orange), $532$ nm (green), and effective beam emulating stimulated emission (dashed blue). (d) Horizontal cross-section of the normalized total intensity ($|E_x|^2+|E_y|^2+|E_z|^2$) of the effective beam from the discovered solution (blue), the simulated STED reference (dashed red), and the simulated reference (dotted black) using $532$ nm wavelength. The discovered solution outperforms both simulated references for STED microscopy and the sharp focus from Dorn, Quabis and Leuchs (2003).}
  \label{fig:XLSetup_results}
\end{figure*}

Finally, we demonstrate the capability of \algo for genuine discovery. We initialize the system in the virtual setup depicted in Extended Data Fig. \ref{fig:extended_data_init_setup_discovery}. The loss function corresponds to equation (\ref{eq:loss}) considering the total intensity of the effective beam resulting from the emulated STED process. The details of the optimization are provided in the Methods section. The discovered topology and identified phase patterns are depicted in Figs. \ref{fig:XLSetup_results}a and \ref{fig:XLSetup_results}b, respectively. The detected intensity topologies reveal the system generates a doughnut-shaped and a Gaussian-like beams. We compute the vertical cross-section of the focused intensity patterns for both beams and the resulting effective beam (green, orange and dotted blue lines in Fig. \ref{fig:XLSetup_results}c, respectively). The horizontal cross-section exhibits analogous features. We further compare the effective beam intensity with the simulated STED reference \cite{Hell:94} and the the discovered Gaussian-like beam with the simulated sharp focus reference \cite{leuchs}. The obtained results are showcased in Fig. \ref{fig:XLSetup_results}d. Strikingly, the discovered solution exploits the underlying physical concepts of two aforementioned optical systems. On the one hand, it generates a doughnut-shaped ``depletion" beam as demonstrated in Ref. \cite{Hell:94}. On the other hand, it generates a Gaussian-like ``excitation" signal with a sharper focus, achieving smaller effective intensity spots resulting from the STED process. The discovered solution showcases an effective beam profile which is sharper than the simulated STED reference. This occurs due to the enhanced sharpening of the longitudinal component of the excitation beam, which demonstrates similar profile as the simulated sharp focus reference \cite{leuchs}. To the best of our knowledge, this technique has never been discussed in the scientific literature before. Regardless of its physical realizability, this solution demonstrates the ability of \algo to uncover interesting solutions within highly complex systems.

\section{Discussion}

In this work we present \algo, a highly efficient computational framework with seamlessly integrated auto-differentiation capabilities, just-in-time compilation, automatic vectorization and GPU compatibility, for the discovery of novel optical setups in super-resolution microscopy.

We demonstrate the high-performance and efficiency of \algo with a computational speed-up of $\times$ $2.1 \cdot 10^4$ on GPU, and $\times$ $8.4 \cdot 10^2$ on CPU, compared to standard numerical optimization methods. We further prove the accuracy and reliability of our methods by successfully rediscovering three foundational optical experiments. More significantly, within a large-scale discovery framework, \algo identified a novel experimental blueprint that breaks the diffraction limit by integrating the physical principles of two well-known super-resolution techniques.

Having laid the groundwork with \algo for an efficient, versatile optics simulator, many other microscopy and imaging techniques follow as a natural extension. For example, by enabling time information using algorithms iterating over the simulator, light scattering of probe samples can be easily implemented, enabling systems such as iSCAT \cite{iSCAT_vahid2019}, structured illumination microscopy \cite{structured2005}, and localization microscopy \cite{SMLM}. Also, inspired by the work in \cite{deepSTORM_PSF}, we could leverage similar approaches to optimize hardware-software discovery within our extensive framework. Additionally, one could formulate interesting experiments (e.g., in low light conditions) where the use of noise or loss sources (e.g., absorption, vibration, etc.) become relevant. 

Additionally, \algo provides already the basis for an expansion to complex quantum optics microscopy techniques \cite{beyondquantumlimit2013} or other quantum imaging techniques \cite{moreau2019imaging}, as a quantum of light (i.e., a photon) is nothing else than an excitation of the modes of the electromagnetic field. Looking further into the future, one can expect that matter-wave beams (governed by Schrödinger's equation, which is closely related to the paraxial wave equation, a special case of the electromagnetic field) can be simulated in the same framework. This might allow for the AI-based design of microscopy techniques which could harness entirely new ideas combining light and complex matter wave beams such as electron-beams \cite{mihaila2022transverse, kalinin2022machine, kalinin2023human} or coherent beams of high-mass particles \cite{kialka2022roadmap}. Ultimately, bringing so far unexplored concepts from diverse areas of physics to microscopy applications is at the heart of AI-driven discovery in this area, and we hope that this work constitutes a first step in this direction.

\section{Methods}
\subsection{Features and performance of \algo}\label{sec_methods_software}
In this section we provide the detailed description of \algo's simulation features and performance. The simulator enables, among many other features, to define light sources (of any wavelength and power), phase masks (i.e., spatial light modulators, SLMs), polarizers, variable retarders (e.g., wave plates, WPs), diffraction gratings, and high numerical aperture (NA) lenses to replicate strong focusing conditions. Light propagation and diffraction is simulated by two methods, each available for both scalar and vectorial regimes: the fast-Fourier-transform (FFT) based numerical integration of the Rayleigh-Sommerfeld (RS) diffraction integral \cite{RS, vectorial_RS} and the Chirped z-transform (CZT) \cite{hu2020efficient}. The CZT is an accelerated version of the RS algorithm, which allows for arbitrary selection and sampling of the region of interest. These algorithms are based on the FFT and require a reasonable sampling for the calculation to be accurate \cite{FFT_sampling}. In our simulations we consider light sources emitting Gaussian beams of $1.2$ mm beam waist. To avoid possible boundary-generated artifacts during the simulation, we define these beams in larger computational spaces of $4$ mm or $5$ mm. Thus, the pixel resolutions often span $1024\times1024$, or $2048\times2048$.

Some functionalities of \algo's optics simulator (e.g., optical propagation algorithms, planar lens or amplitude masks) are inspired in an open-source NumPy-based Python module for diffraction and interferometry simulation, \textit{Diffractio} \cite{diffractio}, although we have rewritten and modified these approaches to combine them with JAX just-in-time (jit) functionality. In essence, \textit{jit} compilation optimizes sequences of operations together and runs them at once. For this purpose, the first run of a jitted function builds an abstract representation of the sequence of operations specified by the function. This representation encodes the shape and the data-type (dtype) of the arrays - but is agnostic to the values of such arrays. If the input shapes and dtypes are not modified, the abstract structure of the function can be then re-used for subsequent runs, without re-compilation, which allows to execute the subsequent calls faster. However, if the input shape or dtype is modified, the function automatically gets re-compiled. This will cause an extra overhead time due to the extraction of a new abstract structure of the function for the new shapes/dtypes. On top of that, we developed completely new functions (e.g., beam splitters, WPs or propagation through high NA objective lens with CZT methods, to name a few) which significantly expand the software capabilities. The most important hardware addition on the optical simulator are the SLMs, each pixel of which possesses an independent (and variable) phase value. They serve as a universal approximation for phase masks, including lenses, and offer a computational advantage: given a specific pixel resolution, they allow for unrestricted phase design selection. Such flexibility is crucial during the parameter space exploration, as it allows the software to autonomously probe all potential solutions. In addition, we defined under the name of \textit{super-SLM} (\textit{sSLM}) a hardware-box-type which consists of two SLMs, each one independently imprinting a phase mask on the horizontal and vertical polarization components of the field each. 

To evaluate the performance of numerical and auto-differentiation methods we chose to use BFGS (from SciPy's Python library) and Adam (included in the JAX library) as optimizers. Further comparison including SGD (Stochastic-Gradient-Descent), AdaGrad (Adaptive Gradient) and AdamW (Adam with weight decay) is presented in Extended Data Fig. \ref{fig:extended_data_optimizers}. As the optical system, we set-up a Gaussian beam propagating over a distance \textit{z} and interacting with a phase mask. The objective function is the mean squared error between the detected light and the ground truth, characterized by a Gaussian beam with a spiral phase imprinted on its wavefront. We initialize the system with an arbitrary phase mask configuration. We first evaluate the computational time for a single gradient evaluation for numerical and autodiff methods across different computational window sizes (from $10\times 10$ up to $500\times500$ pixels) and devices (CPU and GPU). We keep the default settings for BFGS. For Adam, the step size is set to 0.1. The optimization process is terminated if there is no improvement in the loss value (meaning it has not decreased below the best value recorded), over 50 consecutive iteration steps. For each resolution window, we collect the convergence time of both optimizers and divide it by the total number of gradient evaluations for BFGS and the total number of steps for Adam. The acquired gradient evaluation times correspond to the mean value over 5 runs. Obtained results are depicted in Fig. \ref{fig:performance}c. It is clear how autodiff outperforms numerical methods by up to 4 orders of magnitude on CPU and 5 orders of magnitude when running in the GPU. The advantage over larger sizes is clear given that we run simulations with resolutions of $1024\times1024$ and $2024 \times 2048$ pixels.

We then conduct the evaluation of the convergence time for both methods. We keep the aforementioned settings for the optimizers. We initialize the systems 5 times and compute their mean value. The acquired results are depicted in Fig. \ref{fig:performance}d. On the CPU, numerical methods exhibit exponential scaling in convergence time, reaching about $4.5\cdot 10^4$ seconds (roughly 12 hours) for $250\times250$ pixel resolution. In contrast, autodiff demonstrates superior efficiency, reducing it to roughly 53 seconds. GPU optimization performance is even more pronounced, reaching convergence in 0.24 seconds for $250\times 250$ pixels, and 16 seconds for a resolution of $500\times 500$.

\subsection{Data-driven rediscovery}
In this section we provide the details of the data-driven learning approach outlined in Fig. \ref{fig:4f}b. The training dataset is composed of $18,000$ [input, output] intensity sample pairs. Each sample consists of a Gaussian beam shaped by amplitude masks in various forms (circles, rectangles, squares and rings), with varying sizes and orientations. The corresponding output for each input is an inverted version, magnified by a factor of 2, using homogeneous intensity distribution. The training process involves feeding the input mask into the virtual optical setup. The cost function for each optical setup is computed as the mean squared error between the detected intensity pattern from the virtual setup and the corresponding target mask from the dataset. We select training examples in batches of 10 and evaluate the current setup response and its loss value. The average loss over the batch guides the update of the optical parameters, repeating this cycle until convergence is reached. The parameter space (of $\sim$ 2 million optical parameters) includes the three distances and the phase masks of the two SLMs with a resolution of $1024\times1024$ pixels. We set-up the AdamW optimizer with a step size of 0.01 and a weight decay of $10^{-4}$. The optimization is terminated if there is no improvement in the loss value (i.e., it has not decreased below the best value recorded), over 500 consecutive iteration steps. This condition is checked every 500 steps. We start the optimization with randomly initialized optical parameters with values between 0 and 1. The discovered solution is depicted in Fig. \ref{fig:4f}c. It was identified in about 2.4 hours on the GPU. The loss value evolution over the number of iteration steps is depicted in Extended Data Fig. \ref{fig:Extended_data_loss}a. 

\subsection{Towards large-scale discovery}

In this section we detail the methodology for the optimizations conducted using our quasi-universal computational \textit{ansatz}, a purely continuous framework. We first discuss the enormous search space corresponding to large-scale optical setups and the performance of \algo in building such a general framework. Then, we provide the derivation of the loss function in equation (\ref{eq:loss}) and our simulated emission depletion model. Finally, we provide the details for running the optimizations for STED microscopy and the super-resolution technique exploiting light vortices in three different scenarios, using: (1) fixed optical topologies, (2) optimizable topologies and optical parameters within low parameterized systems, and (3) optimizable topologies and optical parameters within highly parameterized, complex optical setups.

\subsubsection{The large-scale search space}
The large-scale optical setup depicted in Fig. \ref{fig:large-scale} consists 6 light sources that emit linearly polarized Gaussian beams with different wavelengths (e.g., 625 nm, 530 nm and 470 nm). Through 82 vectorial propagation (vectorial Rayleigh-Sommerfeld, VRS), these beams interact with a total of 9 beam splitters, 24 \textit{sSLM}s (i.e., 48 SLMs), 24 wave plates, and get ultimately detected by 6 high NA objective lenses focusing on light detectors. 

We analyze the number of possible discrete arrangements within this general optical setup. For a very discrete approach of beam splitter ratios (either transmit, reflect or have light in both arms) and only allowing the SLMs and wave plates (WP) to be switched ON/OFF (i.e., displaying a constant zero phase or adding zero retardance to the incoming light), the number of possible discrete layouts is of 
\begin{equation}
    N_{\text{Discrete layouts}} = 3^{9_\text{BS}} \cdot 2^{48_\text{SLM}} \cdot 2^{24_\text{WP}} = 2\cdot 10^{20}.
\end{equation}
 
All these are considering that the available beam splitter ratios are restricted to 3 values and the SLMs and wave plates to turn ON/OFF, respectively. In practice, beam splitter ratios and phase values are continuous variables and can take any value (from 0 to 1 and $-\pi$ to $\pi$, respectively) which increases even more the dimension of our search space.

In the following Table \ref{tab:experiment-summary} we present a summary detailing the main properties of the six optimizations conducted within our large-scale ansatz: the number of tunable elements, the dimension of the parameter space and the available number of topologies.

\begin{table}[h!]
\centering
\caption{Outline of the main properties of the six digital experiments conducted within our large-scale ansatz. Displays the total number of tunable elements, the dimension of the parameter space and the available topologies.}
\label{tab:experiment-summary}
\begin{tabular}{cccc}
\toprule
\textbf{\thead{Experiment \\ (Fig. $\#$)}} & \textbf{\thead{$\#$ tunable \\elements}} & \textbf{\thead{Parameter\\ space}} & \textbf{\thead{$\#$ available \\ topologies}} \\ \hline
\hline 
\multirow{2}{*}{\makecell{Fig. 4a\\  Fig. 4c}}  & \multirow{2}{*}{\makecell{21}} & \multirow{2}{*}{25} & \multirow{2}{*}{$10^5$} \\ \\ \hline
\thead{Ext. Data\\ Fig. 7} & \makecell{50} & 52 & $10^{18}$ \\ \hline
\multirow{2}{*}{\makecell{Fig. 5a \\ Fig. 5d}} & \multirow{2}{*}{\makecell{26}}  &  $\sim$ 4 million & \multirow{2}{*}{$10^7$} \\ 
& & $\sim$ 6.3 million \\ \hline 
Fig. 6  & \makecell{26} & $\sim$ 4 million & $10^7$\\
\bottomrule
\end{tabular}
\end{table}

\subsubsection{Performance on large-scale optical setups}
We evaluate the efficiency of \algo by measuring the computational time it takes to construct the large-scale optical setup depicted in Fig. \ref{fig:large-scale} and comparing it with respect to \textit{Diffractio} across different resolutions (from $10\times10$ to $1024\times1024$ pixels) and devices (CPU and GPU). The resulting times, measured across varying computational window sizes, are depicted in Extended Data Fig. \ref{fig:extended_data_performance}. 

\textit{Diffractio} exhibits a notable exponential increase in its computational time past the size of $300\times300$ pixels, showing running times of almost 6 minutes for a resolution of $1024\times1024$ pixels. For smaller sizes, the efficiency of NumPy's dispatch overhead times becomes relevant, making it faster in such regimes. In fact, when considering resolutions up to $200\times200$ pixels, \algo experiences significant overhead times (on both CPU and GPU). These are attributed to JAX's dispatch overhead. Although this time is independent of the size of the arrays, it becomes more pronounced when multiple operations are performed on smaller arrays. However, as array size increases, the dispatch costs diminish, highlighting the benefits of JAX's accelerated linear algebra and just-in-time compilation. When running on CPU, \algo outperforms \textit{Diffractio}, exhibiting superior scalability in both initial and subsequent runs. For example, for $1024\times1024$ pixels, \algo operates in nearly half the time (2.5 minutes) required by \textit{Diffractio}. This advantage becomes even more pronounced when \algo operates on a GPU. The initial run times remain fairly consistent across various computational window sizes, ranging from 14 to 16 seconds. Subsequent runs exhibit similar consistency, with times around 3 to 3.7 seconds up to $500\times500$ pixels and 6 seconds for $1024\times1024$ pixels. 

\subsubsection{Stimulated emission depletion model} 

STED microscopy \cite{Hell:94, Hell2005} is based on excitation and spatially targeted depletion of fluorophores. In order to achieve this, a Gaussian-shaped excitation beam and a doughnut-shaped depletion beam (generated by imprinting a spiral phase into its wavefront) are concentrically overlapped. The depletion beam has zero intensity in the center, where the excitation beam has its maximum. Fluorophores that are not in the center of the beams are forced to emit at the wavelength of the depletion beam. Their emission is spectrally filtered out. Only fluorophores in the center of the beams are allowed to fluoresce normally, and only their emission is ultimately detected. This effectively reduces the area of normal fluorescence, which leads to super-resolution imaging. 

We simulate one of the fundamental concepts of STED microscopy without having to rely on time-dependent processes related to absorption and fluorescence. To do so, we perform a nonlinear modulation of the intensity of the excitation and depletion beams based on the Beer-Lambert law \cite{Mayerhfer2020TheBL}. We define the effective fluorescence that would ultimately be detected as:
\begin{equation}
    I_{\text{eff}}= I_{\text{ex}} \left[1-\beta \left(1-e^{-(I_{\text{dep}}/ I_{\text{ex})}}\right)\right]\;,
    \label{eq:STED}
\end{equation}
where $I_{\text{ex}}$ and $I_{\text{dep}}$ correspond to the excitation and depletion intensities, respectively, and $0\leq \beta \leq 1$ captures the quenching efficiency of the depletion beam. This expression bounds the effect of the depletion beam such that scenarios with negative effective intensity or unrealistically high values are avoided. In particular, assuming a perfect efficiency of the depletion beam in suppressing the excitation (i.e., $\beta=1$), we obtain an expression resembling the Beer-Lambert law:
\begin{equation}
    I_{\text{eff}}= I_{\text{ex}} \cdot e^{-(I_{\text{dep}}/ I_{\text{ex})}}\;.
\end{equation}
Thus, the effective detected light falls off exponentially with the intensity ratio $I_{\text{dep}}/ I_{\text{ex}}$. In the limit case where there is no excitation intensity, $I_{\text{ex}}=0$, the detected light is zero as well, $I_{\text{eff}}=0$. If there is no depletion intensity, $I_\text{dep}=0$, the detected light corresponds to the excitation beam $I_{\text{eff}}=I_{\text{ex}}$. The trivial case of null efficiency in the quenching, $\beta=0$, leads to the same result. 

To evaluate the nonlinear effect we consider $\beta=1$ and $I_{\text{dep}}=\frac{1}{2}I_{\text{ex}}$. From equation (\ref{eq:STED}) we obtain 
\begin{equation}
    I_{\text{eff}}= I_{\text{ex}} e^{-1/2} \approx 0.6 I_{\text{ex}}\;.
\end{equation}
Now, by slightly increasing the depletion energy, e.g., $I_{\text{dep}}=\frac{3}{2}I_{\text{ex}}$, it reads
\begin{equation}
    I_{\text{eff}}= I_{\text{ex}} e^{-3/2} \approx 0.2 I_{\text{ex}}\;.
\end{equation}
Therefore, a small change in the depletion energy causes a large effect in the effective intensity. As a further example, if we set an intermediate efficiency of $\beta=0.5$ and $I_{\text{dep}}=\frac{1}{2}I_{\text{ex}}$ we obtain
\begin{equation}
    I_{\text{eff}}= I_{\text{ex}} \left[\frac{1+ e^{-1/2}}{2} \right] \approx 0.8 I_{\text{ex}}\;.
\end{equation}
which clearly demonstrates the effect of diminishing the efficiency of the suppression. Overall, we successfully imprinted the nonlinear behavior of the quenching for different range of effectiveness, achieving a realistic, bounded physical model for STED.

\subsubsection{Loss function}

The loss function, $\mathcal{L}$, is inversely proportional to the total detected intensity density that is above a specified intensity threshold, $I_{\varepsilon}$. Thus, minimizing $\mathcal{L}$ aims to maximize the generation of small, high intensity beams. In particular, it reads
\begin{equation}
    \mathcal{L} = \frac{1}{\text{Density}} = \frac{\text{Area}}{I_{\varepsilon}}\;.
\end{equation}
The total intensity $I_{\varepsilon}$ above the threshold is computed as
\begin{equation}
    I_{\varepsilon} = \sum_{k,l}^{N} i_{\varepsilon}(k,l) \;,
\end{equation}
where $N$ is the total number of pixels in the camera's sensor and $i_{\varepsilon}(k,l)$ represents the intensity value at each pixel once the threshold condition is applied. This condition is defined as follows:
\begin{equation}
    i_{\varepsilon}(k, l) = 
    \begin{cases} 
    i_{\text{det}}(k, l) & \text{if } i_{\text{det}}(k, l) > \varepsilon i_{\text{max}}\;, \\
    0 & \text{otherwise}\;,
    \end{cases}
\end{equation}
where $i_{\text{det}}(k, l)$ is the intensity value at the \textit{i}-th row and \textit{j}-th column in the detected 2D intensity pattern, $\varepsilon i_{\text{max}}$ (with $0\leq\varepsilon\leq1$) is the threshold value, with $i_{\text{max}}$ being the maximum intensity value in the entire 2D detector array. 

The $\text{Area}$ is determined using a variation of the Heaviside function $\Theta$ applied to $i_{\varepsilon}$, quantifying the area where the intensity is above the threshold:
\begin{equation}
    \text{Area} = \sum_{k,l}^{N} \Theta(i_{\varepsilon}(k, l))\;,
\end{equation}
where $N$ is the total number of pixels in the camera's sensor and $\Theta(i_{\varepsilon}(k, l))$ is defined as:
\begin{equation}
    \Theta(i_{\varepsilon}(k, l)) = 
    \begin{cases} 
    1 & \text{if } i_{\varepsilon}(k, l) > 0\;, \\
    0 & \text{otherwise}\;.
    \end{cases}
    \label{eq:decision}
\end{equation}
Therefore, the loss function can be read as follows:
\begin{equation}
    \mathcal{L} = \frac{1}{\text{Density}} = \frac{\text{Area}}{I_{\varepsilon}} = \frac{\sum_{k,l}^{N} \Theta(i_{\varepsilon}(k, l))}{\sum_{k,l}^{N} i_{\varepsilon}(k,l)}\;.
\end{equation}

It is important to note that the camera pixel selection in equation (\ref{eq:decision}) is a discrete operation. However, JAX offers some interesting capabilities due to its integrated autodiff framework. In particular, control flow operations in JAX are supported and differentiable. Therefore, we compute the loss function in a fully differentiable manner using \texttt{jax.numpy.where()}.

\subsubsection{Rediscovery through exploration within fixed topologies}

To test the capabilities of \algo, we first conduct the rediscovery the beam shaping as employed in STED microscopy and the super-resolution technique employing optical vortices using optical setups with a fixed optical topology.
 
\textbf{STED microscopy:} In this instance, we initialize the system using the optical setup in Extended Data Fig. \ref{fig:extended_data_early_experiment}a. The parameter space corresponds to the $2048\times2048$ pixels of the SLM ($\sim$ 4 million optical parameters), with a pixel size of $1.95$ $\mu m$. The loss function corresponds to equation (\ref{eq:loss}) considering the radial component of the effective beam, $|E_x|^2 + |E_y|^2$ and $\varepsilon=0.7$. We simulate the stimulated emission depletion effect, using equation (\ref{eq:STED}) with the efficiency set to $\beta=1$. We set-up the Adam optimizer with a step size of 0.01 and initialize the system in a random initial phase mask. The optimization is terminated if there is no improvement in the loss value over 500 consecutive iteration steps. This condition is checked every 100 steps. The system converged into a pattern alike to the spiral phase in roughly 7 minutes using a GPU. The phase mask used to compute the reference experiment and the identified solution are depicted in Extended Data Fig. \ref{fig:extended_data_early_experiment}b. To highlight the doughnut shape of the depletion beam, we compute the vertical cross-section of the focused intensity patterns for both excitation and depletion beams (green and orange lines in Extended Data Fig. \ref{fig:extended_data_early_experiment}c, respectively) and the effective beam (dotted blue line in Fig. Extended Data Fig. \ref{fig:extended_data_early_experiment}c). The behavior across the horizontal axis yields similar features. In addition, we systematically change the relative beam intensities as depicted in Extended Data Fig. \ref{fig:extended_data_early_experiment}d. We observe the inverse square root scaling of the effective beam diameter relative to the intensity ratio as demonstrated for STED microscopy \cite{Hell:94}.

While the spiral phase mask features a consistent and gradual phase variation across the spiral, this progression is not as evident in the discovered solution. Furthermore, we would like to emphasize the remarkably low noise contribution on the identified phase pattern. Other solutions presented noisy phase patterns which failed to achieve the essential doughnut-shaped depletion beam. Real-world STED setups demand almost perfect phase patterns and alignment of components; even minor errors can compromise STED. Remarkably, without prior knowledge, our system detected this sensitivity, converging towards a smooth phase pattern. Moreover, despite both the excitation and depletion beams being diffraction-limited, the effective response is sub-diffraction. Such outcomes accentuate the success of \algo in identifying crucial components intrinsic to STED microscopy. 

\textbf{Sharper focus with optical vortices:} In this instance, we initialize the system using the optical setup in Extended Data Fig. \ref{fig:extended_data_early_experiment}e. The parameter space is defined by two SLMs with a resolution of $2048\times2048$ pixels with a pixel size of $2.44$ $\mu m$ each, one WP with with variable phase retardance $\eta$ and orientation angle $\theta$, and three distances ($\sim$ 8.4 million optical parameters). The loss function corresponds to equation (\ref{eq:loss}) considering the intensity of the longitudinal component $|E_z|^2$, and $\varepsilon=0.7$. We set-up the Adam optimizer with a step size of 0.03 and initialize the system with random optical parameters with values between 0 and 1. The stopping condition for the optimizer is the same as for STED microscopy: the optimization is terminated if there is no improvement in the loss value over 500 consecutive iteration steps. This condition is checked every 100 steps. The phase masks to simulate the reference experiment and the identified solution are depicted in Extended Data Figs. \ref{fig:extended_data_early_experiment}f and \ref{fig:extended_data_early_experiment}g, respectively. The system converged after roughly 2 hours using a GPU. The discovered patterns produce an $LG_{2,1}$ Laguerre-Gaussian mode, which demonstrates an intensity pattern of concentric rings with a phase singularity in its center. Surprisingly, \algo found an alternative way to imprint a phase singularity onto the beam and produce pronounced longitudinal components on the focal plane. The longitudinal intensity profiles for the reference and the discovered solution are depicted in Extended Data Fig. \ref{fig:extended_data_early_experiment}h (represented by dotted black and green, respectively). For comparison, we also feature the radial intensity profile of the diffraction-limited beam (dotted orange line in Extended Data Fig. \ref{fig:extended_data_early_experiment}h). The identified solution showcases a radial intensity doughnut shape and surpasses the diffraction limit in the longitudinal component demonstrating a spot size slightly larger than the reference.

\subsubsection{Rediscovery through exploration of optical topologies}

To further prove the capability of \algo to discover new optical topologies of already existing solutions, we conduct two optimizations using our \textit{quasi-universal} optical setup scheme. The goal is to discover the optical topology using the available optimizable optical parameters (i.e., distances, beam splitter ratios and wave plate's angles).

Crucially, light is detected across six different devices. Therefore, we compute the loss function at each detector and the parameter update is driven by the detector that shows the minimum value of the loss. We conduct this selection by using a differentiable, smooth approximation using \texttt{jax.nn.logsumexp()} as:
\begin{verbatim}
def softmin(l_det, beta):
    return - logsumexp(-beta * l_det)/ beta,
\end{verbatim}
where \texttt{l$\_$det} is the array of the loss values corresponding to each detector and \texttt{beta} is the strength of the modulation.

We first target \algo to discover new topologies for STED microscopy \cite{Hell:94} and Dorn, Quabis and Leuchs (2003) \cite{leuchs} using a virtual optical table featuring randomly positioned phase masks displaying fixed (non-optimizable) phase patterns.

In particular, to rediscover the concept of STED microscopy, we initialize the $3\times3$ optical setup in Extended Data Fig. \ref{fig:extended_data_init_setup_xl}a, using light sources emitting linearly polarized light of $632.8$ nm and $530$ nm wavelengths and four phase masks of $824\times824$ pixel resolution (PM$\#1$ to PM$\#4$ in Extended Data Fig. \ref{fig:extended_data_init_setup_xl}a) displaying the fixed phase patterns depicted in \ref{fig:extended_data_init_setup_xl}b. The parameter space (of 25 optical parameters) is defined by 9 symmetric beam splitter ratios, 8 distances and 4 wave plates with variable retardance and orientation (BS$\#1$ to BS$\#9$, $z_1$ to $z_8$ and WP$\#1$ to WP$\#4$ in Extended Data Fig. \ref{fig:extended_data_init_setup_xl}a, respectively). The loss function corresponds to equation (\ref{eq:loss}) considering the radial component of the effective beam resulting from the STED process, $|E_x|^2 + |E_y|^2$ and $\epsilon = 0.5$. We simulate the stimulated emission depletion effect using equation (\ref{eq:STED}) with the efficiency set to $\beta=1$. We set-up the AdamW optimizer with a step size of $10^{-3}$ and a weight decay of $10^{-4}$. The system is initialized with random optical parameters with values between 0 and 1. The optimization is terminated if there is no improvement in the loss value over 500 consecutive iteration steps. This condition is checked every 100 steps. The loss value evolution over the number of iteration steps is depicted in Extended Data Fig. \ref{fig:Extended_data_loss}b. The system converged (in roughly 35 minutes using a GPU) into the topology highlighted in Fig. \ref{fig:XLSetup}a, demonstrating the smallest loss value in the detector $\#$3. To retrieve the discovered topology, one can identify the optical path following the light, in reverse, from the detector $\#3$. First, the beam splitter $BS\#9$ has 66.4$\%$ transmittance, which implies that light is coming from $BS\#8$ and $BS\#6$. In one hand, $BS\#8$ has a reflectance value of 65.8$\%$, meaning that most of the light comes from $BS\#5$. On the other hand, $BS\#6$ has a reflectance value of 99.8$\%$, meaning that light also comes from $BS\#5$ which, in turn, has a transmittance value of 99.8$\%$. Due to $BS\#5$ being fully transmissible, light paths can be easily redirected to $BS\#2$ and $BS\#4$, with transmission values of 99.9$\%$ and 92.6$\%$, respectively. Thus, the Gaussian beam emerging from source $\#2$ goes through the setup without being modulated and light from source $\#5$ interacts with phase mask $\#2$, generating a doughnut-shaped beam in the detector $\#3$.

To rediscover the physical principle exploited in Dorn, Quabis and Leuchs (2003) we initialize the $3\times3$ optical setup in Extended Data Fig. \ref{fig:extended_data_init_setup_xl}b, using light sources emitting linearly polarized light of $632.8$ nm wavelength and four phase masks (of $1024\times1024$ pixel resolution) displaying the fixed phase patterns depicted in \ref{fig:extended_data_init_setup_xl}b. The parameter space (25 optical parameters) is defined by 9 symmetric beam splitter ratios, 8 distances and 4 wave plates (retardance and orientation). The loss function corresponds to equation (\ref{eq:loss}) considering the intensity of the electromagnetic field’s longitudinal component, $|E_z|^2$, and $\epsilon = 0.7$. We set-up the AdamW optimizer with a step size of $10^{-2}$ and a weight decay of $10^{-4}$. The system is initialized with random optical parameters with values between 0 and 1. The optimization is terminated if there is no improvement in the loss value over 500 consecutive iteration steps. This condition is checked every 100 steps. The loss value evolution over the number of iteration steps is depicted in Extended Data Fig. \ref{fig:Extended_data_loss}c. The system converged (in roughly 1 hour using a GPU) into the topology highlighted in Fig. \ref{fig:XLSetup}b, demonstrating the smallest loss value in the detector $\#$3. Alike for the previous example, we retrieve the discovered topology following the light from the detector $\#3$. First, the beam splitter $BS\#9$ has 58.9$\%$ transmittance, which implies that light is coming from $BS\#8$ and $BS\#6$. In one hand, $BS\#8$ has a reflectance value of 99.9$\%$, meaning that light comes from $BS\#5$. On the other hand, $BS\#6$ has a reflectance value of 97.0$\%$, meaning that light also comes from $BS\#5$ which, in turn, has a transmittance value of 99.9$\%$. Due to $BS\#5$ being fully transmissible, light paths can be easily redirected to $BS\#2$ and $BS\#4$, with reflectance values of 99.9$\%$ and 96.3$\%$, respectively. Which leads to $BS\#1$ with a transmittance value of 99.9$\%$. Thus, the light emerging from sources $\#1$ and $\#4$ interacts, independently, with the phase masks $\#1$ and $\#2$, respectively, imprinting phase singularities in the center of the beam generating doughnut-shaped beams. Light gets ultimately combined before detector $\#6$.

Importantly, we are not restricted to the use of $3\times3$ optical grids. Thus, we conduct the same optimization procedure for Dorn, Quabis and Leuchs (2003) this time within a $6\times6$ optical system. The obtained results are depicted in Extended Data Fig. \ref{fig:6x6}. The optical system, depicted in Extended Data Fig. \ref{fig:6x6}a, consists of 12 light sources emitting a $635$ nm wavelength Gaussian beam that are linearly polarized at $45^o$ interacting with 36 beam splitters and four phase masks displaying the fixed phase patterns in Extended Data Fig. \ref{fig:6x6}b. The loss function corresponds to equation (\ref{eq:loss}) considering the intensity of the electromagnetic field’s longitudinal component, $|E_z|^2$, and $\epsilon = 0.7$. We set-up the AdamW optimizer with a step size of $0.05$ and a weight decay of $10^{-4}$. The system is initialized with random optical parameters with values between 0 and 1. The optimization is terminated if there is no improvement in the loss value over 500 consecutive iteration steps. This condition is checked every 100 steps. The system converged (in roughly 1 hour using a GPU) into the topology highlighted in  Extended Data Fig. \ref{fig:6x6}c., demonstrating the smallest loss value in the detector $\#$3. The identified solution displays similar spot size as the reference (see Extended Data Fig. \ref{fig:6x6}d).

\subsubsection{Rediscovery through exploration in highly parameterized systems}

We further task \algo to discover new topologies within larger parameter spaces, this time enabling the system to optimize the SLMs masks instead of having fixed phase patterns. The goal here is to discover both the optical topology using the available optical parameters (i.e., distances, beam splitter ratios and wave plate's angles) and the phase patterns to imprint onto the light beams (i.e., using SLMs).

For this purpose we build the $3\times3$ optical setup depicted in Extended Data Fig. \ref{fig:extended_data_init_setup_discovery}. It consists of six light sources emitting linearly polarized Gaussian beams of wavelengths $650$ nm and $532$ nm. Three building blocks, which contain one \textit{super}-SLM (i.e., two SLMs imprinting independent phase masks to orthogonal polarization states) and a wave plate separated a distance $z$, are placed within the diagonal of the grid (grey boxes in Extended Data Fig. \ref{fig:extended_data_init_setup_discovery}). Light gets ultimately detected across six detectors. As discussed in the previous section, the loss function is computed at each detector, the parameter update is driven by the device demonstrating the minimum value. This selection is conducted in a fully differentiable manner using \texttt{jax.nn.logsumexp()}. 

We first target \algo to rediscover the concept of STED microscopy within the general setup in Extended Data Fig. \ref{fig:extended_data_init_setup_discovery}. The parameter space ($\sim$ 4 million parameters) corresponds to three \textit{super}-SLMs (i.e., 6 SLMs) with a resolution of $824\times824$ pixels with a pixel size of $6.06 \mu m$, three wave plates, eight distances and nine beam splitter ratios. The loss function corresponds to equation (\ref{eq:loss}), in this instance considering the radial intensity of the effective light emerging from the STED process, $|E_x|^2 + |E_y|^2$, and $\varepsilon= 0.5$. We simulate the stimulated emission depletion effect using equation (\ref{eq:STED}) with the efficiency set to $\beta=1$. We set-up the AdamW optimizer with a step size of $10^{-3}$ and a weight decay of $10^{-2}$. The system is initialized with random optical parameters with values between 0 and 1. The optimization is terminated if there is no improvement in the loss value over 500 consecutive iteration steps. This condition is checked every 100 steps. The loss value evolution over the number of iteration steps is depicted in Extended Data Fig. \ref{fig:Extended_data_loss}d. The system converged (in roughly 1.3 hours using a GPU) into the topology highlighted in Fig. \ref{fig:XLSetup_topologies}a, demonstrating the smallest loss value in the detector $\#2$. The discovered topology can be identified from the beam splitter ratios starting from detector $\#2$: $BS\#6$ has 99.9$\%$ transmittance, which implies that light is directly is coming from $BS\#5$. In turn, $BS\#5$ has a reflectance of 99.9$\%$ which redirects light from $BS\#2$, which has a transmittance of 79.9$\%$. This already defines light source $\#2$ as the emitter of the green (excitation) beam. Then, the remaining light comes from $BS\#1$ which shows a 99.9$\%$ reflectance, meaning that the red light comes from source $\#1$.

To rediscover the concept used in Dorn, Quabis and Leuchs \cite{leuchs}, we initialize the system using the general setup in Extended Data Fig. \ref{fig:extended_data_init_setup_discovery}. This time, however, the light sources emitting $532$ nm wavelength are switched to emit $650$ nm. The parameter space ($\sim$ 6.2 million parameters) corresponds to three \textit{super}-SLMs (i.e., 6 SLMs) with a resolution of $1024\times1024$ pixels with a pixel size of $4.8 \mu m$, three wave plates, eight distances and nine beam splitter ratios. The loss function corresponds to equation (\ref{eq:loss}), in this instance considering the intensity from the longitudinal component of the electric field, $|E_z|^2$, and $\varepsilon= 0.7$. We set-up the AdamW optimizer with a step size of $0.05$ and a weight decay of $10^{-5}$. The system is initialized with random optical parameters with values between 0 and 1. The optimization is terminated if there is no improvement in the loss value over 500 consecutive iteration steps. This condition is checked every 100 steps. The loss value evolution over the number of iteration steps is depicted in Extended Data Fig. \ref{fig:Extended_data_loss}e. The system converged (in roughly 35 minutes using a GPU) into the topology highlighted in Fig. \ref{fig:XLSetup_topologies}d, demonstrating the smallest loss value in the detector $\#$6. As for the previous examples, the discovered topology can be identified from the beam splitter ratios: starting from detector $\#6$, $BS\#9$ has 99.9$\%$ transmittance, which implies that light is directly is coming from $BS\#6$. In turn, $BS\#6$ has a reflectance of 99.9$\%$. This means that light has interacted with $BS\#5$ and gone through the sSLM and WP. Then, $BS\#5$ has has a reflectance of 99.9$\%$, which defines the incoming light at $BS\#2$. This, in turn, shows a 99.9$\%$ transmittance, meaning that light comes from the light source $\#2$.

\subsection{Discovery of a new experimental blueprint}

Finally, we demonstrate the capabilities of \algo for genuine discovery. We use the same initial optical setup in Extended Data Fig. \ref{fig:extended_data_init_setup_discovery}. The parameter space ($\sim$ 4 million parameters) corresponds to three \textit{super}-SLMs (i.e., 6 SLMs) with a resolution of $824\times824$ pixels with a pixel size of $6.06 \mu m$, three wave plates, eight distances and nine beam splitter ratios. The loss function corresponds to equation (\ref{eq:loss}), in this instance considering the total intensity of the effective light emerging from the STED process, $|E_x|^2 + |E_y|^2 + |E_z|^2$, and $\varepsilon= 0.5$. We simulate the stimulated emission depletion effect using equation (\ref{eq:STED}) with the efficiency set to $\beta=1$. We set-up the AdamW optimizer with a step size of $10^{-3}$ and a weight decay of $10^{-3}$ and initialize the system with random optical parameters with values between 0 and 1. The optimization is terminated if there is no improvement in the loss value over 500 consecutive iteration steps. This condition is checked every 100 steps. The loss value evolution over the number of iteration steps is depicted in Extended Data Fig. \ref{fig:Extended_data_loss}f. The system converged (in roughly 3.8 hours using a GPU) into the topology highlighted in Fig. \ref{fig:XLSetup_results}a, demonstrating the smallest loss value in the detector $\#2$. The discovered topology can be identified from the beam splitter ratios starting from detector $\#2$: $BS\#6$ has 99.9$\%$ transmittance, which implies that light is directly is coming from $BS\#5$. In turn, $BS\#5$ has a reflectance of 99.9$\%$ which redirects light from $BS\#2$, which has a transmittance of 66.2$\%$. This already defines light source $\#2$ as the emitter of the green (excitation) beam. Then, the remaining light comes from $BS\#1$ which shows a 99.9$\%$ reflectance, meaning that the red light (depletion) comes from source $\#1$.

\section{Data availability}
Data can be readily generated using the Python script provided via GitHub at \href{https://github.com/artificial-scientist-lab/XLuminA}{https://github.com/artificial-scientist-lab/XLuminA}.

\section{Code availability}
The produced code and documentation can be found via GitHub at \href{https://github.com/artificial-scientist-lab/XLuminA}{https://github.com/artificial-scientist-lab/XLuminA}. 

\bibliography{main}

\begin{thebibliography}{76}%
\makeatletter
\providecommand \@ifxundefined [1]{%
 \@ifx{#1\undefined}
}%
\providecommand \@ifnum [1]{%
 \ifnum #1\expandafter \@firstoftwo
 \else \expandafter \@secondoftwo
 \fi
}%
\providecommand \@ifx [1]{%
 \ifx #1\expandafter \@firstoftwo
 \else \expandafter \@secondoftwo
 \fi
}%
\providecommand \natexlab [1]{#1}%
\providecommand \enquote  [1]{``#1''}%
\providecommand \bibnamefont  [1]{#1}%
\providecommand \bibfnamefont [1]{#1}%
\providecommand \citenamefont [1]{#1}%
\providecommand \href@noop [0]{\@secondoftwo}%
\providecommand \href [0]{\begingroup \@sanitize@url \@href}%
\providecommand \@href[1]{\@@startlink{#1}\@@href}%
\providecommand \@@href[1]{\endgroup#1\@@endlink}%
\providecommand \@sanitize@url [0]{\catcode `\\12\catcode `\$12\catcode `\&12\catcode `\#12\catcode `\^12\catcode `\_12\catcode `\%12\relax}%
\providecommand \@@startlink[1]{}%
\providecommand \@@endlink[0]{}%
\providecommand \url  [0]{\begingroup\@sanitize@url \@url }%
\providecommand \@url [1]{\endgroup\@href {#1}{\urlprefix }}%
\providecommand \urlprefix  [0]{URL }%
\providecommand \Eprint [0]{\href }%
\providecommand \doibase [0]{https://doi.org/}%
\providecommand \selectlanguage [0]{\@gobble}%
\providecommand \bibinfo  [0]{\@secondoftwo}%
\providecommand \bibfield  [0]{\@secondoftwo}%
\providecommand \translation [1]{[#1]}%
\providecommand \BibitemOpen [0]{}%
\providecommand \bibitemStop [0]{}%
\providecommand \bibitemNoStop [0]{.\EOS\space}%
\providecommand \EOS [0]{\spacefactor3000\relax}%
\providecommand \BibitemShut  [1]{\csname bibitem#1\endcsname}%
\let\auto@bib@innerbib\@empty
\bibitem [{\citenamefont {Wang}\ \emph {et~al.}(2023)\citenamefont {Wang}, \citenamefont {Fu}, \citenamefont {Du}, \citenamefont {Gao}, \citenamefont {Huang}, \citenamefont {Liu}, \citenamefont {Chandak}, \citenamefont {Liu}, \citenamefont {Van~Katwyk}, \citenamefont {Deac} \emph {et~al.}}]{wang2023scientific}%
  \BibitemOpen
  \bibfield  {author} {\bibinfo {author} {\bibfnamefont {H.}~\bibnamefont {Wang}}, \bibinfo {author} {\bibfnamefont {T.}~\bibnamefont {Fu}}, \bibinfo {author} {\bibfnamefont {Y.}~\bibnamefont {Du}}, \bibinfo {author} {\bibfnamefont {W.}~\bibnamefont {Gao}}, \bibinfo {author} {\bibfnamefont {K.}~\bibnamefont {Huang}}, \bibinfo {author} {\bibfnamefont {Z.}~\bibnamefont {Liu}}, \bibinfo {author} {\bibfnamefont {P.}~\bibnamefont {Chandak}}, \bibinfo {author} {\bibfnamefont {S.}~\bibnamefont {Liu}}, \bibinfo {author} {\bibfnamefont {P.}~\bibnamefont {Van~Katwyk}}, \bibinfo {author} {\bibfnamefont {A.}~\bibnamefont {Deac}}, \emph {et~al.},\ }\bibfield  {title} {\bibinfo {title} {Scientific discovery in the age of artificial intelligence},\ }\href {https://doi.org/10.1038/s41586-023-06221-2} {\bibfield  {journal} {\bibinfo  {journal} {Nature}\ }\textbf {\bibinfo {volume} {620}},\ \bibinfo {pages} {47} (\bibinfo {year} {2023})}\BibitemShut {NoStop}%
\bibitem [{\citenamefont {Krenn}\ \emph {et~al.}(2022)\citenamefont {Krenn}, \citenamefont {Pollice}, \citenamefont {Guo}, \citenamefont {Aldeghi}, \citenamefont {Cervera-Lierta}, \citenamefont {Friederich}, \citenamefont {dos Passos~Gomes}, \citenamefont {H{\"a}se}, \citenamefont {Jinich}, \citenamefont {Nigam} \emph {et~al.}}]{krenn2022scientific}%
  \BibitemOpen
  \bibfield  {author} {\bibinfo {author} {\bibfnamefont {M.}~\bibnamefont {Krenn}}, \bibinfo {author} {\bibfnamefont {R.}~\bibnamefont {Pollice}}, \bibinfo {author} {\bibfnamefont {S.~Y.}\ \bibnamefont {Guo}}, \bibinfo {author} {\bibfnamefont {M.}~\bibnamefont {Aldeghi}}, \bibinfo {author} {\bibfnamefont {A.}~\bibnamefont {Cervera-Lierta}}, \bibinfo {author} {\bibfnamefont {P.}~\bibnamefont {Friederich}}, \bibinfo {author} {\bibfnamefont {G.}~\bibnamefont {dos Passos~Gomes}}, \bibinfo {author} {\bibfnamefont {F.}~\bibnamefont {H{\"a}se}}, \bibinfo {author} {\bibfnamefont {A.}~\bibnamefont {Jinich}}, \bibinfo {author} {\bibfnamefont {A.}~\bibnamefont {Nigam}}, \emph {et~al.},\ }\bibfield  {title} {\bibinfo {title} {On scientific understanding with artificial intelligence},\ }\href {https://doi.org/10.1038/s42254-022-00518-3} {\bibfield  {journal} {\bibinfo  {journal} {Nature Reviews Physics}\ }\textbf {\bibinfo {volume} {4}},\ \bibinfo {pages} {761} (\bibinfo {year} {2022})}\BibitemShut {NoStop}%
\bibitem [{\citenamefont {Wollman}\ \emph {et~al.}(2015)\citenamefont {Wollman}, \citenamefont {Nudd}, \citenamefont {Hedlund},\ and\ \citenamefont {Leake}}]{Antonj_300years}%
  \BibitemOpen
  \bibfield  {author} {\bibinfo {author} {\bibfnamefont {A.~J.~M.}\ \bibnamefont {Wollman}}, \bibinfo {author} {\bibfnamefont {R.}~\bibnamefont {Nudd}}, \bibinfo {author} {\bibfnamefont {E.~G.}\ \bibnamefont {Hedlund}},\ and\ \bibinfo {author} {\bibfnamefont {M.~C.}\ \bibnamefont {Leake}},\ }\bibfield  {title} {\bibinfo {title} {From animaculum to single molecules: 300 years of the light microscope},\ }\href {https://doi.org/10.1098/rsob.150019} {\bibfield  {journal} {\bibinfo  {journal} {Open Biology}\ }\textbf {\bibinfo {volume} {5}} (\bibinfo {year} {2015})}\BibitemShut {NoStop}%
\bibitem [{\citenamefont {Reigoto}\ \emph {et~al.}(2021)\citenamefont {Reigoto}, \citenamefont {Andrade}, \citenamefont {Seixas}, \citenamefont {Costa},\ and\ \citenamefont {Mermelstein}}]{AR_microscopy}%
  \BibitemOpen
  \bibfield  {author} {\bibinfo {author} {\bibfnamefont {A.~M.}\ \bibnamefont {Reigoto}}, \bibinfo {author} {\bibfnamefont {S.~A.}\ \bibnamefont {Andrade}}, \bibinfo {author} {\bibfnamefont {M.~C. R.~R.}\ \bibnamefont {Seixas}}, \bibinfo {author} {\bibfnamefont {M.~L.}\ \bibnamefont {Costa}},\ and\ \bibinfo {author} {\bibfnamefont {C.}~\bibnamefont {Mermelstein}},\ }\bibfield  {title} {\bibinfo {title} {A comparative study on the use of microscopy in pharmacology and cell biology research},\ }\href {https://doi.org/10.1371/journal.pone.0245795} {\bibfield  {journal} {\bibinfo  {journal} {PLOS ONE}\ }\textbf {\bibinfo {volume} {16}},\ \bibinfo {pages} {1} (\bibinfo {year} {2021})}\BibitemShut {NoStop}%
\bibitem [{\citenamefont {Weisenburger}\ and\ \citenamefont {Sandoghdar}(2015)}]{Sandoghdar_light_microscopy}%
  \BibitemOpen
  \bibfield  {author} {\bibinfo {author} {\bibfnamefont {S.}~\bibnamefont {Weisenburger}}\ and\ \bibinfo {author} {\bibfnamefont {V.}~\bibnamefont {Sandoghdar}},\ }\bibfield  {title} {\bibinfo {title} {Light microscopy: an ongoing contemporary revolution},\ }\href {https://doi.org/10.1080/00107514.2015.1026557} {\bibfield  {journal} {\bibinfo  {journal} {Contemporary Physics}\ }\textbf {\bibinfo {volume} {56}},\ \bibinfo {pages} {123} (\bibinfo {year} {2015})}\BibitemShut {NoStop}%
\bibitem [{\citenamefont {Bullen}(2008)}]{Bullen_microdrug}%
  \BibitemOpen
  \bibfield  {author} {\bibinfo {author} {\bibfnamefont {A.}~\bibnamefont {Bullen}},\ }\bibfield  {title} {\bibinfo {title} {Microscopic imaging techniques for drug discovery},\ }\href {https://doi.org/10.1038/nrd2446} {\bibfield  {journal} {\bibinfo  {journal} {Nature Reviews Drug Discovery}\ }\textbf {\bibinfo {volume} {7}},\ \bibinfo {pages} {54} (\bibinfo {year} {2008})}\BibitemShut {NoStop}%
\bibitem [{\citenamefont {Antony}\ \emph {et~al.}(2013)\citenamefont {Antony}, \citenamefont {Trefois}, \citenamefont {Stojanovic}, \citenamefont {Baumuratov},\ and\ \citenamefont {Kozak}}]{PMA_lightmicro}%
  \BibitemOpen
  \bibfield  {author} {\bibinfo {author} {\bibfnamefont {P.}~\bibnamefont {Antony}}, \bibinfo {author} {\bibfnamefont {C.}~\bibnamefont {Trefois}}, \bibinfo {author} {\bibfnamefont {A.}~\bibnamefont {Stojanovic}}, \bibinfo {author} {\bibfnamefont {A.}~\bibnamefont {Baumuratov}},\ and\ \bibinfo {author} {\bibfnamefont {K.}~\bibnamefont {Kozak}},\ }\bibfield  {title} {\bibinfo {title} {Light microscopy applications in systems biology: opportunities and challenges},\ }\href {https://doi.org/10.1186/1478-811X-11-24} {\bibfield  {journal} {\bibinfo  {journal} {Cell Communication and Signaling}\ }\textbf {\bibinfo {volume} {11}} (\bibinfo {year} {2013})}\BibitemShut {NoStop}%
\bibitem [{\citenamefont {Grimm}\ and\ \citenamefont {Lavis}(2022)}]{Grimm_review}%
  \BibitemOpen
  \bibfield  {author} {\bibinfo {author} {\bibfnamefont {J.~B.}\ \bibnamefont {Grimm}}\ and\ \bibinfo {author} {\bibfnamefont {L.~D.}\ \bibnamefont {Lavis}},\ }\bibfield  {title} {\bibinfo {title} {Caveat fluorophore: an insiders' guide to small-molecule fluorescent labels},\ }\href {https://doi.org/10.1038/s41592-021-01338-6} {\bibfield  {journal} {\bibinfo  {journal} {Nature Methods}\ }\textbf {\bibinfo {volume} {19}} (\bibinfo {year} {2022})}\BibitemShut {NoStop}%
\bibitem [{\citenamefont {M.}\ and\ \citenamefont {Palmer}(2014)}]{Palmer_fluorophores}%
  \BibitemOpen
  \bibfield  {author} {\bibinfo {author} {\bibfnamefont {D.~K.}\ \bibnamefont {M.}}\ and\ \bibinfo {author} {\bibfnamefont {A.~E.}\ \bibnamefont {Palmer}},\ }\bibfield  {title} {\bibinfo {title} {Advances in fluorescence labeling strategies for dynamic cellular imaging},\ }\href {https://doi.org/10.1038/nchembio.1556} {\bibfield  {journal} {\bibinfo  {journal} {Nature Chemical Biology}\ }\textbf {\bibinfo {volume} {10}} (\bibinfo {year} {2014})}\BibitemShut {NoStop}%
\bibitem [{\citenamefont {Hell}\ and\ \citenamefont {Wichmann}(1994)}]{Hell:94}%
  \BibitemOpen
  \bibfield  {author} {\bibinfo {author} {\bibfnamefont {S.~W.}\ \bibnamefont {Hell}}\ and\ \bibinfo {author} {\bibfnamefont {J.}~\bibnamefont {Wichmann}},\ }\bibfield  {title} {\bibinfo {title} {Breaking the diffraction resolution limit by stimulated emission: stimulated-emission-depletion fluorescence microscopy},\ }\href {https://doi.org/10.1364/OL.19.000780} {\bibfield  {journal} {\bibinfo  {journal} {Optics Letters}\ }\textbf {\bibinfo {volume} {19}},\ \bibinfo {pages} {780} (\bibinfo {year} {1994})}\BibitemShut {NoStop}%
\bibitem [{\citenamefont {Betzig}\ \emph {et~al.}(2006)\citenamefont {Betzig}, \citenamefont {Patterson}, \citenamefont {Sougrat}, \citenamefont {Lindwasser}, \citenamefont {Olenych}, \citenamefont {Bonifacino}, \citenamefont {Davidson}, \citenamefont {Lippincott-Schwartz},\ and\ \citenamefont {Hess}}]{PALM}%
  \BibitemOpen
  \bibfield  {author} {\bibinfo {author} {\bibfnamefont {E.}~\bibnamefont {Betzig}}, \bibinfo {author} {\bibfnamefont {G.~H.}\ \bibnamefont {Patterson}}, \bibinfo {author} {\bibfnamefont {R.}~\bibnamefont {Sougrat}}, \bibinfo {author} {\bibfnamefont {O.~W.}\ \bibnamefont {Lindwasser}}, \bibinfo {author} {\bibfnamefont {S.}~\bibnamefont {Olenych}}, \bibinfo {author} {\bibfnamefont {J.~S.}\ \bibnamefont {Bonifacino}}, \bibinfo {author} {\bibfnamefont {M.~W.}\ \bibnamefont {Davidson}}, \bibinfo {author} {\bibfnamefont {J.}~\bibnamefont {Lippincott-Schwartz}},\ and\ \bibinfo {author} {\bibfnamefont {H.~F.}\ \bibnamefont {Hess}},\ }\bibfield  {title} {\bibinfo {title} {Imaging intracellular fluorescent proteins at nanometer resolution},\ }\href {https://doi.org/10.1126/science.1127344} {\bibfield  {journal} {\bibinfo  {journal} {Science}\ }\textbf {\bibinfo {volume} {313}},\ \bibinfo {pages} {1642} (\bibinfo {year} {2006})}\BibitemShut {NoStop}%
\bibitem [{\citenamefont {Hess}\ \emph {et~al.}(2006)\citenamefont {Hess}, \citenamefont {Girirajan},\ and\ \citenamefont {Mason}}]{f-PALM}%
  \BibitemOpen
  \bibfield  {author} {\bibinfo {author} {\bibfnamefont {S.~T.}\ \bibnamefont {Hess}}, \bibinfo {author} {\bibfnamefont {T.~P.}\ \bibnamefont {Girirajan}},\ and\ \bibinfo {author} {\bibfnamefont {M.~D.}\ \bibnamefont {Mason}},\ }\bibfield  {title} {\bibinfo {title} {Ultra-high resolution imaging by fluorescence photoactivation localization microscopy},\ }\href {https://doi.org/https://doi.org/10.1529/biophysj.106.091116} {\bibfield  {journal} {\bibinfo  {journal} {Biophysical Journal}\ }\textbf {\bibinfo {volume} {91}},\ \bibinfo {pages} {4258} (\bibinfo {year} {2006})}\BibitemShut {NoStop}%
\bibitem [{\citenamefont {Rust}\ \emph {et~al.}(2006)\citenamefont {Rust}, \citenamefont {Bates},\ and\ \citenamefont {Zhuang}}]{STORM}%
  \BibitemOpen
  \bibfield  {author} {\bibinfo {author} {\bibfnamefont {M.}~\bibnamefont {Rust}}, \bibinfo {author} {\bibfnamefont {M.}~\bibnamefont {Bates}},\ and\ \bibinfo {author} {\bibfnamefont {X.}~\bibnamefont {Zhuang}},\ }\bibfield  {title} {\bibinfo {title} {Sub-diffraction-limit imaging by stochastic optical reconstruction microscopy (storm)},\ }\href {https://doi.org/10.1038/nmeth929} {\bibfield  {journal} {\bibinfo  {journal} {Nature Methods}\ }\textbf {\bibinfo {volume} {3}},\ \bibinfo {pages} {793–796} (\bibinfo {year} {2006})}\BibitemShut {NoStop}%
\bibitem [{\citenamefont {van~de Linde}\ \emph {et~al.}(2011)\citenamefont {van~de Linde}, \citenamefont {Löschberger}, \citenamefont {Klein}, \citenamefont {Heidbreder}, \citenamefont {Wolter}, \citenamefont {Heilemann},\ and\ \citenamefont {Sauer}}]{dSTORM}%
  \BibitemOpen
  \bibfield  {author} {\bibinfo {author} {\bibfnamefont {S.}~\bibnamefont {van~de Linde}}, \bibinfo {author} {\bibfnamefont {A.}~\bibnamefont {Löschberger}}, \bibinfo {author} {\bibfnamefont {T.}~\bibnamefont {Klein}}, \bibinfo {author} {\bibfnamefont {M.}~\bibnamefont {Heidbreder}}, \bibinfo {author} {\bibfnamefont {S.}~\bibnamefont {Wolter}}, \bibinfo {author} {\bibfnamefont {M.}~\bibnamefont {Heilemann}},\ and\ \bibinfo {author} {\bibfnamefont {M.}~\bibnamefont {Sauer}},\ }\bibfield  {title} {\bibinfo {title} {Direct stochastic optical reconstruction microscopy with standard fluorescent probes},\ }\href {https://doi.org/10.1038/nprot.2011.336} {\bibfield  {journal} {\bibinfo  {journal} {Nature Protocols}\ }\textbf {\bibinfo {volume} {6}},\ \bibinfo {pages} {991–1009} (\bibinfo {year} {2011})}\BibitemShut {NoStop}%
\bibitem [{\citenamefont {Gustafsson}(2005{\natexlab{a}})}]{SIM}%
  \BibitemOpen
  \bibfield  {author} {\bibinfo {author} {\bibfnamefont {M.~G.~L.}\ \bibnamefont {Gustafsson}},\ }\bibfield  {title} {\bibinfo {title} {Nonlinear structured-illumination microscopy: Wide-field fluorescence imaging with theoretically unlimited resolution},\ }\href {https://doi.org/10.1073/pnas.0406877102} {\bibfield  {journal} {\bibinfo  {journal} {Proceedings of the National Academy of Sciences}\ }\textbf {\bibinfo {volume} {102}},\ \bibinfo {pages} {13081} (\bibinfo {year} {2005}{\natexlab{a}})}\BibitemShut {NoStop}%
\bibitem [{\citenamefont {Balzarotti}\ \emph {et~al.}(2017)\citenamefont {Balzarotti}, \citenamefont {Eilers}, \citenamefont {Gwosch}, \citenamefont {Gynnå}, \citenamefont {Westphal}, \citenamefont {Stefani}, \citenamefont {Elf},\ and\ \citenamefont {Hell}}]{MINFLUX}%
  \BibitemOpen
  \bibfield  {author} {\bibinfo {author} {\bibfnamefont {F.}~\bibnamefont {Balzarotti}}, \bibinfo {author} {\bibfnamefont {Y.}~\bibnamefont {Eilers}}, \bibinfo {author} {\bibfnamefont {K.~C.}\ \bibnamefont {Gwosch}}, \bibinfo {author} {\bibfnamefont {A.~H.}\ \bibnamefont {Gynnå}}, \bibinfo {author} {\bibfnamefont {V.}~\bibnamefont {Westphal}}, \bibinfo {author} {\bibfnamefont {F.~D.}\ \bibnamefont {Stefani}}, \bibinfo {author} {\bibfnamefont {J.}~\bibnamefont {Elf}},\ and\ \bibinfo {author} {\bibfnamefont {S.~W.}\ \bibnamefont {Hell}},\ }\bibfield  {title} {\bibinfo {title} {Nanometer resolution imaging and tracking of fluorescent molecules with minimal photon fluxes},\ }\href {https://doi.org/10.1126/science.aak9913} {\bibfield  {journal} {\bibinfo  {journal} {Science}\ }\textbf {\bibinfo {volume} {355}},\ \bibinfo {pages} {606} (\bibinfo {year} {2017})}\BibitemShut {NoStop}%
\bibitem [{\citenamefont {Möckl}\ \emph {et~al.}(2019)\citenamefont {Möckl}, \citenamefont {Pedram}, \citenamefont {Roy}, \citenamefont {Krishnan}, \citenamefont {Gustavsson}, \citenamefont {Dorigo}, \citenamefont {Bertozzi},\ and\ \citenamefont {Moerner}}]{MOCKL201957}%
  \BibitemOpen
  \bibfield  {author} {\bibinfo {author} {\bibfnamefont {L.}~\bibnamefont {Möckl}}, \bibinfo {author} {\bibfnamefont {K.}~\bibnamefont {Pedram}}, \bibinfo {author} {\bibfnamefont {A.~R.}\ \bibnamefont {Roy}}, \bibinfo {author} {\bibfnamefont {V.}~\bibnamefont {Krishnan}}, \bibinfo {author} {\bibfnamefont {A.-K.}\ \bibnamefont {Gustavsson}}, \bibinfo {author} {\bibfnamefont {O.}~\bibnamefont {Dorigo}}, \bibinfo {author} {\bibfnamefont {C.~R.}\ \bibnamefont {Bertozzi}},\ and\ \bibinfo {author} {\bibfnamefont {W.}~\bibnamefont {Moerner}},\ }\bibfield  {title} {\bibinfo {title} {Quantitative super-resolution microscopy of the mammalian glycocalyx},\ }\href {https://doi.org/https://doi.org/10.1016/j.devcel.2019.04.035} {\bibfield  {journal} {\bibinfo  {journal} {Developmental Cell}\ }\textbf {\bibinfo {volume} {50}},\ \bibinfo {pages} {57} (\bibinfo {year} {2019})}\BibitemShut {NoStop}%
\bibitem [{\citenamefont {Xu}\ \emph {et~al.}(2013)\citenamefont {Xu}, \citenamefont {Zhong},\ and\ \citenamefont {Zhuang}}]{KeXu2013}%
  \BibitemOpen
  \bibfield  {author} {\bibinfo {author} {\bibfnamefont {K.}~\bibnamefont {Xu}}, \bibinfo {author} {\bibfnamefont {G.}~\bibnamefont {Zhong}},\ and\ \bibinfo {author} {\bibfnamefont {X.}~\bibnamefont {Zhuang}},\ }\bibfield  {title} {\bibinfo {title} {Actin, spectrin, and associated proteins form a periodic cytoskeletal structure in axons},\ }\href {https://doi.org/10.1126/science.1232251} {\bibfield  {journal} {\bibinfo  {journal} {Science}\ }\textbf {\bibinfo {volume} {339}},\ \bibinfo {pages} {452} (\bibinfo {year} {2013})}\BibitemShut {NoStop}%
\bibitem [{\citenamefont {Yildiz}\ \emph {et~al.}(2003)\citenamefont {Yildiz}, \citenamefont {Forkey}, \citenamefont {McKinney}, \citenamefont {Ha}, \citenamefont {Goldman},\ and\ \citenamefont {Selvin}}]{Yildiz2003}%
  \BibitemOpen
  \bibfield  {author} {\bibinfo {author} {\bibfnamefont {A.}~\bibnamefont {Yildiz}}, \bibinfo {author} {\bibfnamefont {J.~N.}\ \bibnamefont {Forkey}}, \bibinfo {author} {\bibfnamefont {S.~A.}\ \bibnamefont {McKinney}}, \bibinfo {author} {\bibfnamefont {T.}~\bibnamefont {Ha}}, \bibinfo {author} {\bibfnamefont {Y.~E.}\ \bibnamefont {Goldman}},\ and\ \bibinfo {author} {\bibfnamefont {P.~R.}\ \bibnamefont {Selvin}},\ }\bibfield  {title} {\bibinfo {title} {Myosin v walks hand-over-hand: Single fluorophore imaging with 1.5-nm localization},\ }\href {https://doi.org/10.1126/science.1084398} {\bibfield  {journal} {\bibinfo  {journal} {Science}\ }\textbf {\bibinfo {volume} {300}},\ \bibinfo {pages} {2061} (\bibinfo {year} {2003})}\BibitemShut {NoStop}%
\bibitem [{\citenamefont {Zhang}\ \emph {et~al.}(2015)\citenamefont {Zhang}, \citenamefont {Lucas}, \citenamefont {Song}, \citenamefont {Beberwyck}, \citenamefont {Fu}, \citenamefont {Xu},\ and\ \citenamefont {Alivisatos}}]{Zhang2015}%
  \BibitemOpen
  \bibfield  {author} {\bibinfo {author} {\bibfnamefont {Y.}~\bibnamefont {Zhang}}, \bibinfo {author} {\bibfnamefont {J.~M.}\ \bibnamefont {Lucas}}, \bibinfo {author} {\bibfnamefont {P.}~\bibnamefont {Song}}, \bibinfo {author} {\bibfnamefont {B.}~\bibnamefont {Beberwyck}}, \bibinfo {author} {\bibfnamefont {Q.}~\bibnamefont {Fu}}, \bibinfo {author} {\bibfnamefont {W.}~\bibnamefont {Xu}},\ and\ \bibinfo {author} {\bibfnamefont {A.~P.}\ \bibnamefont {Alivisatos}},\ }\bibfield  {title} {\bibinfo {title} {Superresolution fluorescence mapping of single-nanoparticle catalysts reveals spatiotemporal variations in surface reactivity},\ }\href {https://doi.org/10.1073/pnas.1502005112} {\bibfield  {journal} {\bibinfo  {journal} {Proceedings of the National Academy of Sciences}\ }\textbf {\bibinfo {volume} {112}},\ \bibinfo {pages} {8959} (\bibinfo {year} {2015})}\BibitemShut {NoStop}%
\bibitem [{\citenamefont {Müller}\ \emph {et~al.}(2019)\citenamefont {Müller}, \citenamefont {Müller}, \citenamefont {Hammer}, \citenamefont {Barner-Kowollik}, \citenamefont {Wegener},\ and\ \citenamefont {Blasco}}]{Mueller2019}%
  \BibitemOpen
  \bibfield  {author} {\bibinfo {author} {\bibfnamefont {P.}~\bibnamefont {Müller}}, \bibinfo {author} {\bibfnamefont {R.}~\bibnamefont {Müller}}, \bibinfo {author} {\bibfnamefont {L.}~\bibnamefont {Hammer}}, \bibinfo {author} {\bibfnamefont {C.}~\bibnamefont {Barner-Kowollik}}, \bibinfo {author} {\bibfnamefont {M.}~\bibnamefont {Wegener}},\ and\ \bibinfo {author} {\bibfnamefont {E.}~\bibnamefont {Blasco}},\ }\bibfield  {title} {\bibinfo {title} {Sted-inspired laser lithography based on photoswitchable spirothiopyran moieties},\ }\href {https://doi.org/10.1021/acs.chemmater.8b04696} {\bibfield  {journal} {\bibinfo  {journal} {Chemistry of Materials}\ }\textbf {\bibinfo {volume} {31}},\ \bibinfo {pages} {1966} (\bibinfo {year} {2019})}\BibitemShut {NoStop}%
\bibitem [{\citenamefont {Bradbury}\ \emph {et~al.}(2018)\citenamefont {Bradbury}, \citenamefont {Frostig}, \citenamefont {Hawkins}, \citenamefont {Johnson}, \citenamefont {Leary}, \citenamefont {Maclaurin}, \citenamefont {Necula}, \citenamefont {Paszke}, \citenamefont {Vander{P}las}, \citenamefont {Wanderman-{M}ilne},\ and\ \citenamefont {Zhang}}]{jax2018github}%
  \BibitemOpen
  \bibfield  {author} {\bibinfo {author} {\bibfnamefont {J.}~\bibnamefont {Bradbury}}, \bibinfo {author} {\bibfnamefont {R.}~\bibnamefont {Frostig}}, \bibinfo {author} {\bibfnamefont {P.}~\bibnamefont {Hawkins}}, \bibinfo {author} {\bibfnamefont {M.~J.}\ \bibnamefont {Johnson}}, \bibinfo {author} {\bibfnamefont {C.}~\bibnamefont {Leary}}, \bibinfo {author} {\bibfnamefont {D.}~\bibnamefont {Maclaurin}}, \bibinfo {author} {\bibfnamefont {G.}~\bibnamefont {Necula}}, \bibinfo {author} {\bibfnamefont {A.}~\bibnamefont {Paszke}}, \bibinfo {author} {\bibfnamefont {J.}~\bibnamefont {Vander{P}las}}, \bibinfo {author} {\bibfnamefont {S.}~\bibnamefont {Wanderman-{M}ilne}},\ and\ \bibinfo {author} {\bibfnamefont {Q.}~\bibnamefont {Zhang}},\ }\href {http://github.com/google/jax} {\bibinfo {title} {{JAX}: composable transformations of {P}ython+{N}um{P}y programs}} (\bibinfo {year} {2018})\BibitemShut {NoStop}%
\bibitem [{\citenamefont {Baydin}\ \emph {et~al.}(2018)\citenamefont {Baydin}, \citenamefont {Pearlmutter}, \citenamefont {Radul},\ and\ \citenamefont {Siskind}}]{autodiffsurvey}%
  \BibitemOpen
  \bibfield  {author} {\bibinfo {author} {\bibfnamefont {A.~G.}\ \bibnamefont {Baydin}}, \bibinfo {author} {\bibfnamefont {B.~A.}\ \bibnamefont {Pearlmutter}}, \bibinfo {author} {\bibfnamefont {A.~A.}\ \bibnamefont {Radul}},\ and\ \bibinfo {author} {\bibfnamefont {J.~M.}\ \bibnamefont {Siskind}},\ }\bibfield  {title} {\bibinfo {title} {Automatic differentiation in machine learning: a survey},\ }\href {http://jmlr.org/papers/v18/17-468.html} {\bibfield  {journal} {\bibinfo  {journal} {Journal of Machine Learning Research}\ }\textbf {\bibinfo {volume} {18}},\ \bibinfo {pages} {1} (\bibinfo {year} {2018})}\BibitemShut {NoStop}%
\bibitem [{\citenamefont {Möckl}\ \emph {et~al.}(2014)\citenamefont {Möckl}, \citenamefont {Lamb},\ and\ \citenamefont {Bräuchle}}]{sted_moeckl}%
  \BibitemOpen
  \bibfield  {author} {\bibinfo {author} {\bibfnamefont {L.}~\bibnamefont {Möckl}}, \bibinfo {author} {\bibfnamefont {D.~C.}\ \bibnamefont {Lamb}},\ and\ \bibinfo {author} {\bibfnamefont {C.}~\bibnamefont {Bräuchle}},\ }\bibfield  {title} {\bibinfo {title} {Super-resolved fluorescence microscopy: Nobel prize in chemistry 2014 for eric betzig, stefan hell, and william e. moerner},\ }\href {https://doi.org/10.1002/anie.201410265} {\bibfield  {journal} {\bibinfo  {journal} {Angewandte Chemie International Edition}\ }\textbf {\bibinfo {volume} {53}},\ \bibinfo {pages} {13972} (\bibinfo {year} {2014})}\BibitemShut {NoStop}%
\bibitem [{\citenamefont {Dorn}\ \emph {et~al.}(2003)\citenamefont {Dorn}, \citenamefont {Quabis},\ and\ \citenamefont {Leuchs}}]{leuchs}%
  \BibitemOpen
  \bibfield  {author} {\bibinfo {author} {\bibfnamefont {R.}~\bibnamefont {Dorn}}, \bibinfo {author} {\bibfnamefont {S.}~\bibnamefont {Quabis}},\ and\ \bibinfo {author} {\bibfnamefont {G.}~\bibnamefont {Leuchs}},\ }\bibfield  {title} {\bibinfo {title} {Sharper focus for a radially polarized light beam},\ }\href {https://doi.org/10.1103/PhysRevLett.91.233901} {\bibfield  {journal} {\bibinfo  {journal} {Physical Review Letters}\ }\textbf {\bibinfo {volume} {91}},\ \bibinfo {pages} {233901} (\bibinfo {year} {2003})}\BibitemShut {NoStop}%
\bibitem [{\citenamefont {Herath}\ \emph {et~al.}(2023)\citenamefont {Herath}, \citenamefont {Haputhanthri}, \citenamefont {Hettiarachchi}, \citenamefont {Kariyawasam}, \citenamefont {Ahmad}, \citenamefont {Ahmad}, \citenamefont {Ahluwalia}, \citenamefont {Edussooriya},\ and\ \citenamefont {Wadduwage}}]{differentiable_microscope}%
  \BibitemOpen
  \bibfield  {author} {\bibinfo {author} {\bibfnamefont {K.}~\bibnamefont {Herath}}, \bibinfo {author} {\bibfnamefont {U.}~\bibnamefont {Haputhanthri}}, \bibinfo {author} {\bibfnamefont {R.}~\bibnamefont {Hettiarachchi}}, \bibinfo {author} {\bibfnamefont {H.}~\bibnamefont {Kariyawasam}}, \bibinfo {author} {\bibfnamefont {R.~N.}\ \bibnamefont {Ahmad}}, \bibinfo {author} {\bibfnamefont {A.}~\bibnamefont {Ahmad}}, \bibinfo {author} {\bibfnamefont {B.~S.}\ \bibnamefont {Ahluwalia}}, \bibinfo {author} {\bibfnamefont {C.~U.~S.}\ \bibnamefont {Edussooriya}},\ and\ \bibinfo {author} {\bibfnamefont {D.~N.}\ \bibnamefont {Wadduwage}},\ }\href@noop {} {\bibinfo {title} {Differentiable microscopy designs an all optical phase retrieval microscope}} (\bibinfo {year} {2023}),\ \Eprint {https://arxiv.org/abs/2203.14944} {arXiv:2203.14944 [physics.optics]} \BibitemShut {NoStop}%
\bibitem [{\citenamefont {Yanny}\ \emph {et~al.}(2020)\citenamefont {Yanny}, \citenamefont {Antipa}, \citenamefont {Liberti}, \citenamefont {Dehaeck}, \citenamefont {Monakhova}, \citenamefont {Liu}, \citenamefont {Shen}, \citenamefont {Ng},\ and\ \citenamefont {Waller}}]{miniscope3D}%
  \BibitemOpen
  \bibfield  {author} {\bibinfo {author} {\bibfnamefont {K.}~\bibnamefont {Yanny}}, \bibinfo {author} {\bibfnamefont {N.}~\bibnamefont {Antipa}}, \bibinfo {author} {\bibfnamefont {W.}~\bibnamefont {Liberti}}, \bibinfo {author} {\bibfnamefont {S.}~\bibnamefont {Dehaeck}}, \bibinfo {author} {\bibfnamefont {K.}~\bibnamefont {Monakhova}}, \bibinfo {author} {\bibfnamefont {F.~L.}\ \bibnamefont {Liu}}, \bibinfo {author} {\bibfnamefont {K.}~\bibnamefont {Shen}}, \bibinfo {author} {\bibfnamefont {R.}~\bibnamefont {Ng}},\ and\ \bibinfo {author} {\bibfnamefont {L.}~\bibnamefont {Waller}},\ }\bibfield  {title} {\bibinfo {title} {Miniscope3d: optimized single-shot miniature 3d fluorescence microscopy},\ }\href {https://doi.org/10.1038/s41377-020-00403-7} {\bibfield  {journal} {\bibinfo  {journal} {Light: Science \& Applications}\ }\textbf {\bibinfo {volume} {9}} (\bibinfo {year} {2020})}\BibitemShut {NoStop}%
\bibitem [{\citenamefont {Wang}\ \emph {et~al.}(2019)\citenamefont {Wang}, \citenamefont {Ye}, \citenamefont {Shen}, \citenamefont {Moringo}, \citenamefont {Dutta}, \citenamefont {Robinson},\ and\ \citenamefont {Landes}}]{Wang:19}%
  \BibitemOpen
  \bibfield  {author} {\bibinfo {author} {\bibfnamefont {W.}~\bibnamefont {Wang}}, \bibinfo {author} {\bibfnamefont {F.}~\bibnamefont {Ye}}, \bibinfo {author} {\bibfnamefont {H.}~\bibnamefont {Shen}}, \bibinfo {author} {\bibfnamefont {N.~A.}\ \bibnamefont {Moringo}}, \bibinfo {author} {\bibfnamefont {C.}~\bibnamefont {Dutta}}, \bibinfo {author} {\bibfnamefont {J.~T.}\ \bibnamefont {Robinson}},\ and\ \bibinfo {author} {\bibfnamefont {C.~F.}\ \bibnamefont {Landes}},\ }\bibfield  {title} {\bibinfo {title} {Generalized method to design phase masks for 3d super-resolution microscopy},\ }\href {https://doi.org/10.1364/OE.27.003799} {\bibfield  {journal} {\bibinfo  {journal} {Opt. Express}\ }\textbf {\bibinfo {volume} {27}},\ \bibinfo {pages} {3799} (\bibinfo {year} {2019})}\BibitemShut {NoStop}%
\bibitem [{\citenamefont {Nehme}\ \emph {et~al.}(2020)\citenamefont {Nehme}, \citenamefont {Freedman}, \citenamefont {Gordon}, \citenamefont {Ferdman}, \citenamefont {Weiss}, \citenamefont {Alalouf}, \citenamefont {Naor}, \citenamefont {Orange}, \citenamefont {Michaeli},\ and\ \citenamefont {Shechtman}}]{deepSTORM_PSF}%
  \BibitemOpen
  \bibfield  {author} {\bibinfo {author} {\bibfnamefont {E.}~\bibnamefont {Nehme}}, \bibinfo {author} {\bibfnamefont {D.}~\bibnamefont {Freedman}}, \bibinfo {author} {\bibfnamefont {R.}~\bibnamefont {Gordon}}, \bibinfo {author} {\bibfnamefont {B.}~\bibnamefont {Ferdman}}, \bibinfo {author} {\bibfnamefont {L.~E.}\ \bibnamefont {Weiss}}, \bibinfo {author} {\bibfnamefont {O.}~\bibnamefont {Alalouf}}, \bibinfo {author} {\bibfnamefont {T.}~\bibnamefont {Naor}}, \bibinfo {author} {\bibfnamefont {R.}~\bibnamefont {Orange}}, \bibinfo {author} {\bibfnamefont {T.}~\bibnamefont {Michaeli}},\ and\ \bibinfo {author} {\bibfnamefont {Y.}~\bibnamefont {Shechtman}},\ }\bibfield  {title} {\bibinfo {title} {Deepstorm3d: dense 3d localization microscopy and psf design by deep learning},\ }\href {https://doi.org/10.1038/s41592-020-0853-5} {\bibfield  {journal} {\bibinfo  {journal} {Nature Methods}\ }\textbf {\bibinfo {volume} {17}},\ \bibinfo {pages} {734–740} (\bibinfo {year} {2020})}\BibitemShut {NoStop}%
\bibitem [{\citenamefont {Fu}\ \emph {et~al.}(2022)\citenamefont {Fu}, \citenamefont {Li}, \citenamefont {Zhou}, \citenamefont {He}, \citenamefont {Liu}, \citenamefont {Hao},\ and\ \citenamefont {Li}}]{Fu:22}%
  \BibitemOpen
  \bibfield  {author} {\bibinfo {author} {\bibfnamefont {S.}~\bibnamefont {Fu}}, \bibinfo {author} {\bibfnamefont {M.}~\bibnamefont {Li}}, \bibinfo {author} {\bibfnamefont {L.}~\bibnamefont {Zhou}}, \bibinfo {author} {\bibfnamefont {Y.}~\bibnamefont {He}}, \bibinfo {author} {\bibfnamefont {X.}~\bibnamefont {Liu}}, \bibinfo {author} {\bibfnamefont {X.}~\bibnamefont {Hao}},\ and\ \bibinfo {author} {\bibfnamefont {Y.}~\bibnamefont {Li}},\ }\bibfield  {title} {\bibinfo {title} {Deformable mirror based optimal psf engineering for 3d super-resolution imaging},\ }\href {https://doi.org/10.1364/OL.460949} {\bibfield  {journal} {\bibinfo  {journal} {Opt. Lett.}\ }\textbf {\bibinfo {volume} {47}},\ \bibinfo {pages} {3031} (\bibinfo {year} {2022})}\BibitemShut {NoStop}%
\bibitem [{\citenamefont {Jia}\ \emph {et~al.}(2014)\citenamefont {Jia}, \citenamefont {Vaughan},\ and\ \citenamefont {Zhuang}}]{jia2014}%
  \BibitemOpen
  \bibfield  {author} {\bibinfo {author} {\bibfnamefont {S.}~\bibnamefont {Jia}}, \bibinfo {author} {\bibfnamefont {J.~C.}\ \bibnamefont {Vaughan}},\ and\ \bibinfo {author} {\bibfnamefont {X.}~\bibnamefont {Zhuang}},\ }\bibfield  {title} {\bibinfo {title} {Isotropic three-dimensional super-resolution imaging with a self-bending point spread function},\ }\href {https://doi.org/https://doi.org/10.1038/nphoton.2014.13} {\bibfield  {journal} {\bibinfo  {journal} {Nature Photonics}\ }\textbf {\bibinfo {volume} {8}},\ \bibinfo {pages} {302–306} (\bibinfo {year} {2014})}\BibitemShut {NoStop}%
\bibitem [{\citenamefont {Izeddin}\ \emph {et~al.}(2012)\citenamefont {Izeddin}, \citenamefont {Beheiry}, \citenamefont {Andilla}, \citenamefont {Ciepielewski}, \citenamefont {Darzacq},\ and\ \citenamefont {Dahan}}]{Izeddin:12}%
  \BibitemOpen
  \bibfield  {author} {\bibinfo {author} {\bibfnamefont {I.}~\bibnamefont {Izeddin}}, \bibinfo {author} {\bibfnamefont {M.~E.}\ \bibnamefont {Beheiry}}, \bibinfo {author} {\bibfnamefont {J.}~\bibnamefont {Andilla}}, \bibinfo {author} {\bibfnamefont {D.}~\bibnamefont {Ciepielewski}}, \bibinfo {author} {\bibfnamefont {X.}~\bibnamefont {Darzacq}},\ and\ \bibinfo {author} {\bibfnamefont {M.}~\bibnamefont {Dahan}},\ }\bibfield  {title} {\bibinfo {title} {Psf shaping using adaptive optics for three-dimensional single-molecule super-resolution imaging and tracking},\ }\href {https://doi.org/10.1364/OE.20.004957} {\bibfield  {journal} {\bibinfo  {journal} {Opt. Express}\ }\textbf {\bibinfo {volume} {20}},\ \bibinfo {pages} {4957} (\bibinfo {year} {2012})}\BibitemShut {NoStop}%
\bibitem [{\citenamefont {Krenn}\ \emph {et~al.}(2016)\citenamefont {Krenn}, \citenamefont {Malik}, \citenamefont {Fickler}, \citenamefont {Lapkiewicz},\ and\ \citenamefont {Zeilinger}}]{krenn2016automated}%
  \BibitemOpen
  \bibfield  {author} {\bibinfo {author} {\bibfnamefont {M.}~\bibnamefont {Krenn}}, \bibinfo {author} {\bibfnamefont {M.}~\bibnamefont {Malik}}, \bibinfo {author} {\bibfnamefont {R.}~\bibnamefont {Fickler}}, \bibinfo {author} {\bibfnamefont {R.}~\bibnamefont {Lapkiewicz}},\ and\ \bibinfo {author} {\bibfnamefont {A.}~\bibnamefont {Zeilinger}},\ }\bibfield  {title} {\bibinfo {title} {Automated search for new quantum experiments},\ }\href {https://doi.org/10.1103/PhysRevLett.116.090405} {\bibfield  {journal} {\bibinfo  {journal} {Phys. Rev. Lett.}\ }\textbf {\bibinfo {volume} {116}},\ \bibinfo {pages} {090405} (\bibinfo {year} {2016})}\BibitemShut {NoStop}%
\bibitem [{\citenamefont {Knott}(2016)}]{knott2016search}%
  \BibitemOpen
  \bibfield  {author} {\bibinfo {author} {\bibfnamefont {P.}~\bibnamefont {Knott}},\ }\bibfield  {title} {\bibinfo {title} {A search algorithm for quantum state engineering and metrology},\ }\href {https://doi.org/10.1088/1367-2630/18/7/073033} {\bibfield  {journal} {\bibinfo  {journal} {New Journal of Physics}\ }\textbf {\bibinfo {volume} {18}},\ \bibinfo {pages} {073033} (\bibinfo {year} {2016})}\BibitemShut {NoStop}%
\bibitem [{\citenamefont {Ruiz-Gonzalez}\ \emph {et~al.}(2022)\citenamefont {Ruiz-Gonzalez}, \citenamefont {Arlt}, \citenamefont {Petermann}, \citenamefont {Sayyad}, \citenamefont {Jaouni}, \citenamefont {Karimi}, \citenamefont {Tischler}, \citenamefont {Gu},\ and\ \citenamefont {Krenn}}]{ruiz2022digital}%
  \BibitemOpen
  \bibfield  {author} {\bibinfo {author} {\bibfnamefont {C.}~\bibnamefont {Ruiz-Gonzalez}}, \bibinfo {author} {\bibfnamefont {S.}~\bibnamefont {Arlt}}, \bibinfo {author} {\bibfnamefont {J.}~\bibnamefont {Petermann}}, \bibinfo {author} {\bibfnamefont {S.}~\bibnamefont {Sayyad}}, \bibinfo {author} {\bibfnamefont {T.}~\bibnamefont {Jaouni}}, \bibinfo {author} {\bibfnamefont {E.}~\bibnamefont {Karimi}}, \bibinfo {author} {\bibfnamefont {N.}~\bibnamefont {Tischler}}, \bibinfo {author} {\bibfnamefont {X.}~\bibnamefont {Gu}},\ and\ \bibinfo {author} {\bibfnamefont {M.}~\bibnamefont {Krenn}},\ }\href@noop {} {\bibinfo {title} {Digital discovery of 100 diverse quantum experiments with pytheus}} (\bibinfo {year} {2022}),\ \Eprint {https://arxiv.org/abs/2210.09980} {arXiv:2210.09980 [quant-ph]} \BibitemShut {NoStop}%
\bibitem [{\citenamefont {Valcarce}\ \emph {et~al.}(2023)\citenamefont {Valcarce}, \citenamefont {Sekatski}, \citenamefont {Gouzien}, \citenamefont {Melnikov},\ and\ \citenamefont {Sangouard}}]{valcarce2023automated}%
  \BibitemOpen
  \bibfield  {author} {\bibinfo {author} {\bibfnamefont {X.}~\bibnamefont {Valcarce}}, \bibinfo {author} {\bibfnamefont {P.}~\bibnamefont {Sekatski}}, \bibinfo {author} {\bibfnamefont {E.}~\bibnamefont {Gouzien}}, \bibinfo {author} {\bibfnamefont {A.}~\bibnamefont {Melnikov}},\ and\ \bibinfo {author} {\bibfnamefont {N.}~\bibnamefont {Sangouard}},\ }\bibfield  {title} {\bibinfo {title} {Automated design of quantum-optical experiments for device-independent quantum key distribution},\ }\href {https://doi.org/10.1103/PhysRevA.107.062607} {\bibfield  {journal} {\bibinfo  {journal} {Phys. Rev. A}\ }\textbf {\bibinfo {volume} {107}},\ \bibinfo {pages} {062607} (\bibinfo {year} {2023})}\BibitemShut {NoStop}%
\bibitem [{\citenamefont {Krenn}\ \emph {et~al.}(2020)\citenamefont {Krenn}, \citenamefont {Erhard},\ and\ \citenamefont {Zeilinger}}]{QO_space}%
  \BibitemOpen
  \bibfield  {author} {\bibinfo {author} {\bibfnamefont {M.}~\bibnamefont {Krenn}}, \bibinfo {author} {\bibfnamefont {M.}~\bibnamefont {Erhard}},\ and\ \bibinfo {author} {\bibfnamefont {A.}~\bibnamefont {Zeilinger}},\ }\bibfield  {title} {\bibinfo {title} {Computer-inspired quantum experiments},\ }\href {https://doi.org/10.1038/s42254-020-0230-4} {\bibfield  {journal} {\bibinfo  {journal} {Nature Review Physics}\ }\textbf {\bibinfo {volume} {2}},\ \bibinfo {pages} {649} (\bibinfo {year} {2020})}\BibitemShut {NoStop}%
\bibitem [{\citenamefont {Killoran}\ \emph {et~al.}(2019)\citenamefont {Killoran}, \citenamefont {Izaac}, \citenamefont {Quesada}, \citenamefont {Bergholm}, \citenamefont {Amy},\ and\ \citenamefont {Weedbrook}}]{killoran2019strawberry}%
  \BibitemOpen
  \bibfield  {author} {\bibinfo {author} {\bibfnamefont {N.}~\bibnamefont {Killoran}}, \bibinfo {author} {\bibfnamefont {J.}~\bibnamefont {Izaac}}, \bibinfo {author} {\bibfnamefont {N.}~\bibnamefont {Quesada}}, \bibinfo {author} {\bibfnamefont {V.}~\bibnamefont {Bergholm}}, \bibinfo {author} {\bibfnamefont {M.}~\bibnamefont {Amy}},\ and\ \bibinfo {author} {\bibfnamefont {C.}~\bibnamefont {Weedbrook}},\ }\bibfield  {title} {\bibinfo {title} {Strawberry {F}ields: {A} {S}oftware {P}latform for {P}hotonic {Q}uantum {C}omputing},\ }\href {https://doi.org/10.22331/q-2019-03-11-129} {\bibfield  {journal} {\bibinfo  {journal} {{Quantum}}\ }\textbf {\bibinfo {volume} {3}},\ \bibinfo {pages} {129} (\bibinfo {year} {2019})}\BibitemShut {NoStop}%
\bibitem [{\citenamefont {Molesky}\ \emph {et~al.}(2018)\citenamefont {Molesky}, \citenamefont {Lin}, \citenamefont {Piggott}, \citenamefont {Jin}, \citenamefont {Vuckovi{\'c}},\ and\ \citenamefont {Rodriguez}}]{molesky2018inverse}%
  \BibitemOpen
  \bibfield  {author} {\bibinfo {author} {\bibfnamefont {S.}~\bibnamefont {Molesky}}, \bibinfo {author} {\bibfnamefont {Z.}~\bibnamefont {Lin}}, \bibinfo {author} {\bibfnamefont {A.~Y.}\ \bibnamefont {Piggott}}, \bibinfo {author} {\bibfnamefont {W.}~\bibnamefont {Jin}}, \bibinfo {author} {\bibfnamefont {J.}~\bibnamefont {Vuckovi{\'c}}},\ and\ \bibinfo {author} {\bibfnamefont {A.~W.}\ \bibnamefont {Rodriguez}},\ }\bibfield  {title} {\bibinfo {title} {Inverse design in nanophotonics},\ }\href {https://doi.org/10.1038/s41566-018-0246-9} {\bibfield  {journal} {\bibinfo  {journal} {Nature Photonics}\ }\textbf {\bibinfo {volume} {12}},\ \bibinfo {pages} {659} (\bibinfo {year} {2018})}\BibitemShut {NoStop}%
\bibitem [{\citenamefont {So}\ \emph {et~al.}(2020)\citenamefont {So}, \citenamefont {Badloe}, \citenamefont {Noh}, \citenamefont {Bravo-Abad},\ and\ \citenamefont {Rho}}]{so2020deep}%
  \BibitemOpen
  \bibfield  {author} {\bibinfo {author} {\bibfnamefont {S.}~\bibnamefont {So}}, \bibinfo {author} {\bibfnamefont {T.}~\bibnamefont {Badloe}}, \bibinfo {author} {\bibfnamefont {J.}~\bibnamefont {Noh}}, \bibinfo {author} {\bibfnamefont {J.}~\bibnamefont {Bravo-Abad}},\ and\ \bibinfo {author} {\bibfnamefont {J.}~\bibnamefont {Rho}},\ }\bibfield  {title} {\bibinfo {title} {Deep learning enabled inverse design in nanophotonics},\ }\href {https://doi.org/10.1515/nanoph-2019-0474} {\bibfield  {journal} {\bibinfo  {journal} {Nanophotonics}\ }\textbf {\bibinfo {volume} {9}},\ \bibinfo {pages} {1041} (\bibinfo {year} {2020})}\BibitemShut {NoStop}%
\bibitem [{\citenamefont {Sapra}\ \emph {et~al.}(2020)\citenamefont {Sapra}, \citenamefont {Yang}, \citenamefont {Vercruysse}, \citenamefont {Leedle}, \citenamefont {Black}, \citenamefont {England}, \citenamefont {Su}, \citenamefont {Trivedi}, \citenamefont {Miao}, \citenamefont {Solgaard} \emph {et~al.}}]{sapra2020chip}%
  \BibitemOpen
  \bibfield  {author} {\bibinfo {author} {\bibfnamefont {N.~V.}\ \bibnamefont {Sapra}}, \bibinfo {author} {\bibfnamefont {K.~Y.}\ \bibnamefont {Yang}}, \bibinfo {author} {\bibfnamefont {D.}~\bibnamefont {Vercruysse}}, \bibinfo {author} {\bibfnamefont {K.~J.}\ \bibnamefont {Leedle}}, \bibinfo {author} {\bibfnamefont {D.~S.}\ \bibnamefont {Black}}, \bibinfo {author} {\bibfnamefont {R.~J.}\ \bibnamefont {England}}, \bibinfo {author} {\bibfnamefont {L.}~\bibnamefont {Su}}, \bibinfo {author} {\bibfnamefont {R.}~\bibnamefont {Trivedi}}, \bibinfo {author} {\bibfnamefont {Y.}~\bibnamefont {Miao}}, \bibinfo {author} {\bibfnamefont {O.}~\bibnamefont {Solgaard}}, \emph {et~al.},\ }\bibfield  {title} {\bibinfo {title} {On-chip integrated laser-driven particle accelerator},\ }\href {https://doi.org/10.1126/science.aay5734} {\bibfield  {journal} {\bibinfo  {journal} {Science}\ }\textbf {\bibinfo {volume} {367}},\ \bibinfo {pages} {79} (\bibinfo {year} {2020})}\BibitemShut {NoStop}%
\bibitem [{\citenamefont {Su}\ \emph {et~al.}(2018)\citenamefont {Su}, \citenamefont {Piggott}, \citenamefont {Sapra}, \citenamefont {Petykiewicz},\ and\ \citenamefont {Vuckovic}}]{su2018inverse}%
  \BibitemOpen
  \bibfield  {author} {\bibinfo {author} {\bibfnamefont {L.}~\bibnamefont {Su}}, \bibinfo {author} {\bibfnamefont {A.~Y.}\ \bibnamefont {Piggott}}, \bibinfo {author} {\bibfnamefont {N.~V.}\ \bibnamefont {Sapra}}, \bibinfo {author} {\bibfnamefont {J.}~\bibnamefont {Petykiewicz}},\ and\ \bibinfo {author} {\bibfnamefont {J.}~\bibnamefont {Vuckovic}},\ }\bibfield  {title} {\bibinfo {title} {Inverse design and demonstration of a compact on-chip narrowband three-channel wavelength demultiplexer},\ }\href {https://doi.org/10.1021/acsphotonics.7b00987} {\bibfield  {journal} {\bibinfo  {journal} {ACS Photonics}\ }\textbf {\bibinfo {volume} {5}},\ \bibinfo {pages} {301} (\bibinfo {year} {2018})}\BibitemShut {NoStop}%
\bibitem [{\citenamefont {Hughes}\ \emph {et~al.}(2018)\citenamefont {Hughes}, \citenamefont {Minkov}, \citenamefont {Williamson},\ and\ \citenamefont {Fan}}]{SFan_ID2018}%
  \BibitemOpen
  \bibfield  {author} {\bibinfo {author} {\bibfnamefont {T.~W.}\ \bibnamefont {Hughes}}, \bibinfo {author} {\bibfnamefont {M.}~\bibnamefont {Minkov}}, \bibinfo {author} {\bibfnamefont {I.~A.~D.}\ \bibnamefont {Williamson}},\ and\ \bibinfo {author} {\bibfnamefont {S.}~\bibnamefont {Fan}},\ }\bibfield  {title} {\bibinfo {title} {Adjoint method and inverse design for nonlinear nanophotonic devices},\ }\href {https://doi.org/10.1021/acsphotonics.8b01522} {\bibfield  {journal} {\bibinfo  {journal} {ACS Photonics}\ }\textbf {\bibinfo {volume} {5}},\ \bibinfo {pages} {4781} (\bibinfo {year} {2018})}\BibitemShut {NoStop}%
\bibitem [{\citenamefont {Minkov}\ \emph {et~al.}(2020)\citenamefont {Minkov}, \citenamefont {Williamson}, \citenamefont {Andreani}, \citenamefont {Gerace}, \citenamefont {Lou}, \citenamefont {Song}, \citenamefont {Hughes},\ and\ \citenamefont {Fan}}]{SFan_AD2020}%
  \BibitemOpen
  \bibfield  {author} {\bibinfo {author} {\bibfnamefont {M.}~\bibnamefont {Minkov}}, \bibinfo {author} {\bibfnamefont {I.~A.~D.}\ \bibnamefont {Williamson}}, \bibinfo {author} {\bibfnamefont {L.~C.}\ \bibnamefont {Andreani}}, \bibinfo {author} {\bibfnamefont {D.}~\bibnamefont {Gerace}}, \bibinfo {author} {\bibfnamefont {B.}~\bibnamefont {Lou}}, \bibinfo {author} {\bibfnamefont {A.~Y.}\ \bibnamefont {Song}}, \bibinfo {author} {\bibfnamefont {T.~W.}\ \bibnamefont {Hughes}},\ and\ \bibinfo {author} {\bibfnamefont {S.}~\bibnamefont {Fan}},\ }\bibfield  {title} {\bibinfo {title} {Inverse design of photonic crystals through automatic differentiation},\ }\href {https://doi.org/10.1021/acsphotonics.0c00327} {\bibfield  {journal} {\bibinfo  {journal} {ACS Photonics}\ }\textbf {\bibinfo {volume} {7}},\ \bibinfo {pages} {1729} (\bibinfo {year} {2020})}\BibitemShut {NoStop}%
\bibitem [{\citenamefont {Lesina}\ \emph {et~al.}(2015)\citenamefont {Lesina}, \citenamefont {Vaccari}, \citenamefont {Berini},\ and\ \citenamefont {Ramunno}}]{lesina2015convergence}%
  \BibitemOpen
  \bibfield  {author} {\bibinfo {author} {\bibfnamefont {A.~C.}\ \bibnamefont {Lesina}}, \bibinfo {author} {\bibfnamefont {A.}~\bibnamefont {Vaccari}}, \bibinfo {author} {\bibfnamefont {P.}~\bibnamefont {Berini}},\ and\ \bibinfo {author} {\bibfnamefont {L.}~\bibnamefont {Ramunno}},\ }\bibfield  {title} {\bibinfo {title} {On the convergence and accuracy of the fdtd method for nanoplasmonics},\ }\href@noop {} {\bibfield  {journal} {\bibinfo  {journal} {Optics Express}\ }\textbf {\bibinfo {volume} {23}},\ \bibinfo {pages} {10481} (\bibinfo {year} {2015})}\BibitemShut {NoStop}%
\bibitem [{\citenamefont {Brea}(2019)}]{diffractio}%
  \BibitemOpen
  \bibfield  {author} {\bibinfo {author} {\bibfnamefont {L.~M.~S.}\ \bibnamefont {Brea}},\ }\href {https://pypi.org/project/diffractio/} {\bibinfo {title} {Diffractio, python module for diffraction and interference optics}} (\bibinfo {year} {2019})\BibitemShut {NoStop}%
\bibitem [{\citenamefont {Freise}\ \emph {et~al.}(2013)\citenamefont {Freise}, \citenamefont {Brown},\ and\ \citenamefont {Bond}}]{Finesse}%
  \BibitemOpen
  \bibfield  {author} {\bibinfo {author} {\bibfnamefont {A.}~\bibnamefont {Freise}}, \bibinfo {author} {\bibfnamefont {D.}~\bibnamefont {Brown}},\ and\ \bibinfo {author} {\bibfnamefont {C.}~\bibnamefont {Bond}},\ }\href@noop {} {\bibinfo {title} {Finesse, frequency domain interferometer simulation software}} (\bibinfo {year} {2013}),\ \Eprint {https://arxiv.org/abs/1306.2973} {arXiv:1306.2973 [physics.comp-ph]} \BibitemShut {NoStop}%
\bibitem [{\citenamefont {{Perrin}}\ \emph {et~al.}(2012)\citenamefont {{Perrin}}, \citenamefont {{Soummer}}, \citenamefont {{Elliott}}, \citenamefont {{Lallo}},\ and\ \citenamefont {{Sivaramakrishnan}}}]{POPPY2012}%
  \BibitemOpen
  \bibfield  {author} {\bibinfo {author} {\bibfnamefont {M.~D.}\ \bibnamefont {{Perrin}}}, \bibinfo {author} {\bibfnamefont {R.}~\bibnamefont {{Soummer}}}, \bibinfo {author} {\bibfnamefont {E.~M.}\ \bibnamefont {{Elliott}}}, \bibinfo {author} {\bibfnamefont {M.~D.}\ \bibnamefont {{Lallo}}},\ and\ \bibinfo {author} {\bibfnamefont {A.}~\bibnamefont {{Sivaramakrishnan}}},\ }\bibfield  {title} {\bibinfo {title} {Simulating point spread functions for the james webb space telescope with webbpsf},\ }\href {https://doi.org/10.1117/12.925230} {\bibfield  {journal} {\bibinfo  {journal} {Space Telescopes and Instrumentation 2012: Optical, Infrared, and Millimeter Wave}\ }\textbf {\bibinfo {volume} {8442}} (\bibinfo {year} {2012})}\BibitemShut {NoStop}%
\bibitem [{\citenamefont {Fontaine}\ \emph {et~al.}(2019)\citenamefont {Fontaine}, \citenamefont {Ryf}, \citenamefont {Chen}, \citenamefont {Neilson}, \citenamefont {Kim},\ and\ \citenamefont {Carpenter}}]{LGsorter}%
  \BibitemOpen
  \bibfield  {author} {\bibinfo {author} {\bibfnamefont {N.~K.}\ \bibnamefont {Fontaine}}, \bibinfo {author} {\bibfnamefont {R.}~\bibnamefont {Ryf}}, \bibinfo {author} {\bibfnamefont {H.}~\bibnamefont {Chen}}, \bibinfo {author} {\bibfnamefont {D.~T.}\ \bibnamefont {Neilson}}, \bibinfo {author} {\bibfnamefont {K.}~\bibnamefont {Kim}},\ and\ \bibinfo {author} {\bibfnamefont {J.}~\bibnamefont {Carpenter}},\ }\bibfield  {title} {\bibinfo {title} {Laguerre-gaussian mode sorter},\ }\href {https://doi.org/10.1038%2Fs41467-019-09840-4} {\bibfield  {journal} {\bibinfo  {journal} {Nature Communications}\ }\textbf {\bibinfo {volume} {10}} (\bibinfo {year} {2019})}\BibitemShut {NoStop}%
\bibitem [{\citenamefont {Labroille}\ \emph {et~al.}(2014)\citenamefont {Labroille}, \citenamefont {Denolle}, \citenamefont {Jian}, \citenamefont {Genevaux}, \citenamefont {Treps},\ and\ \citenamefont {Morizur}}]{Labroille:14}%
  \BibitemOpen
  \bibfield  {author} {\bibinfo {author} {\bibfnamefont {G.}~\bibnamefont {Labroille}}, \bibinfo {author} {\bibfnamefont {B.}~\bibnamefont {Denolle}}, \bibinfo {author} {\bibfnamefont {P.}~\bibnamefont {Jian}}, \bibinfo {author} {\bibfnamefont {P.}~\bibnamefont {Genevaux}}, \bibinfo {author} {\bibfnamefont {N.}~\bibnamefont {Treps}},\ and\ \bibinfo {author} {\bibfnamefont {J.-F.}\ \bibnamefont {Morizur}},\ }\bibfield  {title} {\bibinfo {title} {Efficient and mode selective spatial mode multiplexer based on multi-plane light conversion},\ }\href {https://doi.org/10.1364/OE.22.015599} {\bibfield  {journal} {\bibinfo  {journal} {Opt. Express}\ }\textbf {\bibinfo {volume} {22}},\ \bibinfo {pages} {15599} (\bibinfo {year} {2014})}\BibitemShut {NoStop}%
\bibitem [{\citenamefont {Flam-Shepherd}\ \emph {et~al.}(2022)\citenamefont {Flam-Shepherd}, \citenamefont {Wu}, \citenamefont {Gu}, \citenamefont {Cervera-Lierta}, \citenamefont {Krenn},\ and\ \citenamefont {Aspuru-Guzik}}]{VAE}%
  \BibitemOpen
  \bibfield  {author} {\bibinfo {author} {\bibfnamefont {D.}~\bibnamefont {Flam-Shepherd}}, \bibinfo {author} {\bibfnamefont {T.~C.}\ \bibnamefont {Wu}}, \bibinfo {author} {\bibfnamefont {X.}~\bibnamefont {Gu}}, \bibinfo {author} {\bibfnamefont {A.}~\bibnamefont {Cervera-Lierta}}, \bibinfo {author} {\bibfnamefont {M.}~\bibnamefont {Krenn}},\ and\ \bibinfo {author} {\bibfnamefont {A.}~\bibnamefont {Aspuru-Guzik}},\ }\bibfield  {title} {\bibinfo {title} {Learning interpretable representations of entanglement in quantum optics experiments using deep generative models},\ }\href {https://doi.org/10.1038%2Fs42256-022-00493-5} {\bibfield  {journal} {\bibinfo  {journal} {Nature Machine Intelligence}\ }\textbf {\bibinfo {volume} {4}},\ \bibinfo {pages} {544} (\bibinfo {year} {2022})}\BibitemShut {NoStop}%
\bibitem [{\citenamefont {Nocedal}\ and\ \citenamefont {Wright}()}]{nocedal1999numerical}%
  \BibitemOpen
  \bibfield  {author} {\bibinfo {author} {\bibfnamefont {J.}~\bibnamefont {Nocedal}}\ and\ \bibinfo {author} {\bibfnamefont {S.~J.}\ \bibnamefont {Wright}},\ }\href@noop {} {\emph {\bibinfo {title} {Numerical optimization}}}\ (\bibinfo  {publisher} {Springer})\BibitemShut {NoStop}%
\bibitem [{\citenamefont {Kingma}\ and\ \citenamefont {Ba}(2017)}]{kingma2017adam}%
  \BibitemOpen
  \bibfield  {author} {\bibinfo {author} {\bibfnamefont {D.~P.}\ \bibnamefont {Kingma}}\ and\ \bibinfo {author} {\bibfnamefont {J.}~\bibnamefont {Ba}},\ }\href@noop {} {\bibinfo {title} {Adam: A method for stochastic optimization}} (\bibinfo {year} {2017}),\ \Eprint {https://arxiv.org/abs/1412.6980} {arXiv:1412.6980 [cs.LG]} \BibitemShut {NoStop}%
\bibitem [{\citenamefont {Virtanen}\ \emph {et~al.}(2020)\citenamefont {Virtanen}, \citenamefont {Gommers}, \citenamefont {Oliphant}, \citenamefont {Haberland}, \citenamefont {Reddy}, \citenamefont {Cournapeau}, \citenamefont {Burovski}, \citenamefont {Peterson}, \citenamefont {Weckesser}, \citenamefont {Bright}, \citenamefont {{van der Walt}}, \citenamefont {Brett}, \citenamefont {Wilson}, \citenamefont {Millman}, \citenamefont {Mayorov}, \citenamefont {Nelson}, \citenamefont {Jones}, \citenamefont {Kern}, \citenamefont {Larson}, \citenamefont {Carey}, \citenamefont {Polat}, \citenamefont {Feng}, \citenamefont {Moore}, \citenamefont {{VanderPlas}}, \citenamefont {Laxalde}, \citenamefont {Perktold}, \citenamefont {Cimrman}, \citenamefont {Henriksen}, \citenamefont {Quintero}, \citenamefont {Harris}, \citenamefont {Archibald}, \citenamefont {Ribeiro}, \citenamefont {Pedregosa}, \citenamefont {{van Mulbregt}},\ and\ \citenamefont {{SciPy 1.0 Contributors}}}]{2020SciPy-NMeth}%
  \BibitemOpen
  \bibfield  {author} {\bibinfo {author} {\bibfnamefont {P.}~\bibnamefont {Virtanen}}, \bibinfo {author} {\bibfnamefont {R.}~\bibnamefont {Gommers}}, \bibinfo {author} {\bibfnamefont {T.~E.}\ \bibnamefont {Oliphant}}, \bibinfo {author} {\bibfnamefont {M.}~\bibnamefont {Haberland}}, \bibinfo {author} {\bibfnamefont {T.}~\bibnamefont {Reddy}}, \bibinfo {author} {\bibfnamefont {D.}~\bibnamefont {Cournapeau}}, \bibinfo {author} {\bibfnamefont {E.}~\bibnamefont {Burovski}}, \bibinfo {author} {\bibfnamefont {P.}~\bibnamefont {Peterson}}, \bibinfo {author} {\bibfnamefont {W.}~\bibnamefont {Weckesser}}, \bibinfo {author} {\bibfnamefont {J.}~\bibnamefont {Bright}}, \bibinfo {author} {\bibfnamefont {S.~J.}\ \bibnamefont {{van der Walt}}}, \bibinfo {author} {\bibfnamefont {M.}~\bibnamefont {Brett}}, \bibinfo {author} {\bibfnamefont {J.}~\bibnamefont {Wilson}}, \bibinfo {author} {\bibfnamefont {K.~J.}\ \bibnamefont {Millman}}, \bibinfo {author} {\bibfnamefont {N.}~\bibnamefont {Mayorov}}, \bibinfo {author} {\bibfnamefont
  {A.~R.~J.}\ \bibnamefont {Nelson}}, \bibinfo {author} {\bibfnamefont {E.}~\bibnamefont {Jones}}, \bibinfo {author} {\bibfnamefont {R.}~\bibnamefont {Kern}}, \bibinfo {author} {\bibfnamefont {E.}~\bibnamefont {Larson}}, \bibinfo {author} {\bibfnamefont {C.~J.}\ \bibnamefont {Carey}}, \bibinfo {author} {\bibfnamefont {{\.I}.}~\bibnamefont {Polat}}, \bibinfo {author} {\bibfnamefont {Y.}~\bibnamefont {Feng}}, \bibinfo {author} {\bibfnamefont {E.~W.}\ \bibnamefont {Moore}}, \bibinfo {author} {\bibfnamefont {J.}~\bibnamefont {{VanderPlas}}}, \bibinfo {author} {\bibfnamefont {D.}~\bibnamefont {Laxalde}}, \bibinfo {author} {\bibfnamefont {J.}~\bibnamefont {Perktold}}, \bibinfo {author} {\bibfnamefont {R.}~\bibnamefont {Cimrman}}, \bibinfo {author} {\bibfnamefont {I.}~\bibnamefont {Henriksen}}, \bibinfo {author} {\bibfnamefont {E.~A.}\ \bibnamefont {Quintero}}, \bibinfo {author} {\bibfnamefont {C.~R.}\ \bibnamefont {Harris}}, \bibinfo {author} {\bibfnamefont {A.~M.}\ \bibnamefont {Archibald}}, \bibinfo {author}
  {\bibfnamefont {A.~H.}\ \bibnamefont {Ribeiro}}, \bibinfo {author} {\bibfnamefont {F.}~\bibnamefont {Pedregosa}}, \bibinfo {author} {\bibfnamefont {P.}~\bibnamefont {{van Mulbregt}}},\ and\ \bibinfo {author} {\bibnamefont {{SciPy 1.0 Contributors}}},\ }\bibfield  {title} {\bibinfo {title} {{{SciPy} 1.0: Fundamental Algorithms for Scientific Computing in Python}},\ }\href {https://doi.org/10.1038/s41592-019-0686-2} {\bibfield  {journal} {\bibinfo  {journal} {Nature Methods}\ }\textbf {\bibinfo {volume} {17}},\ \bibinfo {pages} {261} (\bibinfo {year} {2020})}\BibitemShut {NoStop}%
\bibitem [{\citenamefont {McMahon}(2023)}]{McMahon_2023}%
  \BibitemOpen
  \bibfield  {author} {\bibinfo {author} {\bibfnamefont {P.~L.}\ \bibnamefont {McMahon}},\ }\bibfield  {title} {\bibinfo {title} {The physics of optical computing},\ }\bibfield  {journal} {\bibinfo  {journal} {Nature Reviews Physics}\ }\textbf {\bibinfo {volume} {5}},\ \href {https://doi.org/10.1038/s42254-023-00645-5} {10.1038/s42254-023-00645-5} (\bibinfo {year} {2023})\BibitemShut {NoStop}%
\bibitem [{\citenamefont {Wright}\ \emph {et~al.}(2022)\citenamefont {Wright}, \citenamefont {Onodera}, \citenamefont {Stein}, \citenamefont {Wang}, \citenamefont {Schachter}, \citenamefont {Hu},\ and\ \citenamefont {McMahon}}]{Wright_2022}%
  \BibitemOpen
  \bibfield  {author} {\bibinfo {author} {\bibfnamefont {L.~G.}\ \bibnamefont {Wright}}, \bibinfo {author} {\bibfnamefont {T.}~\bibnamefont {Onodera}}, \bibinfo {author} {\bibfnamefont {M.~M.}\ \bibnamefont {Stein}}, \bibinfo {author} {\bibfnamefont {T.}~\bibnamefont {Wang}}, \bibinfo {author} {\bibfnamefont {D.~T.}\ \bibnamefont {Schachter}}, \bibinfo {author} {\bibfnamefont {Z.}~\bibnamefont {Hu}},\ and\ \bibinfo {author} {\bibfnamefont {P.~L.}\ \bibnamefont {McMahon}},\ }\bibfield  {title} {\bibinfo {title} {Deep physical neural networks trained with backpropagation},\ }\href {https://doi.org/10.1038/s41586-021-04223-6} {\bibfield  {journal} {\bibinfo  {journal} {Nature}\ }\textbf {\bibinfo {volume} {601}},\ \bibinfo {pages} {549–555} (\bibinfo {year} {2022})}\BibitemShut {NoStop}%
\bibitem [{\citenamefont {Sim}\ \emph {et~al.}(2019)\citenamefont {Sim}, \citenamefont {Johnson},\ and\ \citenamefont {Aspuru-Guzik}}]{sim2019expressibility}%
  \BibitemOpen
  \bibfield  {author} {\bibinfo {author} {\bibfnamefont {S.}~\bibnamefont {Sim}}, \bibinfo {author} {\bibfnamefont {P.~D.}\ \bibnamefont {Johnson}},\ and\ \bibinfo {author} {\bibfnamefont {A.}~\bibnamefont {Aspuru-Guzik}},\ }\bibfield  {title} {\bibinfo {title} {Expressibility and entangling capability of parameterized quantum circuits for hybrid quantum-classical algorithms},\ }\href {https://doi.org/10.1002/qute.201900070} {\bibfield  {journal} {\bibinfo  {journal} {Advanced Quantum Technologies}\ }\textbf {\bibinfo {volume} {2}},\ \bibinfo {pages} {1900070} (\bibinfo {year} {2019})}\BibitemShut {NoStop}%
\bibitem [{\citenamefont {Krenn}\ \emph {et~al.}(2021)\citenamefont {Krenn}, \citenamefont {Kottmann}, \citenamefont {Tischler},\ and\ \citenamefont {Aspuru-Guzik}}]{krenn2021conceptual}%
  \BibitemOpen
  \bibfield  {author} {\bibinfo {author} {\bibfnamefont {M.}~\bibnamefont {Krenn}}, \bibinfo {author} {\bibfnamefont {J.~S.}\ \bibnamefont {Kottmann}}, \bibinfo {author} {\bibfnamefont {N.}~\bibnamefont {Tischler}},\ and\ \bibinfo {author} {\bibfnamefont {A.}~\bibnamefont {Aspuru-Guzik}},\ }\bibfield  {title} {\bibinfo {title} {Conceptual understanding through efficient automated design of quantum optical experiments},\ }\href {https://doi.org/10.1103/PhysRevX.11.031044} {\bibfield  {journal} {\bibinfo  {journal} {Phys. Rev. X}\ }\textbf {\bibinfo {volume} {11}},\ \bibinfo {pages} {031044} (\bibinfo {year} {2021})}\BibitemShut {NoStop}%
\bibitem [{\citenamefont {Hofmann}\ \emph {et~al.}(2005)\citenamefont {Hofmann}, \citenamefont {Eggeling}, \citenamefont {Jakobs},\ and\ \citenamefont {Hell}}]{Hell2005}%
  \BibitemOpen
  \bibfield  {author} {\bibinfo {author} {\bibfnamefont {M.}~\bibnamefont {Hofmann}}, \bibinfo {author} {\bibfnamefont {C.}~\bibnamefont {Eggeling}}, \bibinfo {author} {\bibfnamefont {S.}~\bibnamefont {Jakobs}},\ and\ \bibinfo {author} {\bibfnamefont {S.~W.}\ \bibnamefont {Hell}},\ }\bibfield  {title} {\bibinfo {title} {Breaking the diffraction barrier in fluorescence microscopy at low light intensities by using reversibly photoswitchable proteins},\ }\href {https://doi.org/10.1073/pnas.0506010102} {\bibfield  {journal} {\bibinfo  {journal} {Proceedings of the National Academy of Sciences}\ }\textbf {\bibinfo {volume} {102}},\ \bibinfo {pages} {17565} (\bibinfo {year} {2005})}\BibitemShut {NoStop}%
\bibitem [{\citenamefont {Mayerh{\"o}fer}\ \emph {et~al.}(2020)\citenamefont {Mayerh{\"o}fer}, \citenamefont {Pahlow},\ and\ \citenamefont {Popp}}]{Mayerhfer2020TheBL}%
  \BibitemOpen
  \bibfield  {author} {\bibinfo {author} {\bibfnamefont {T.~G.}\ \bibnamefont {Mayerh{\"o}fer}}, \bibinfo {author} {\bibfnamefont {S.}~\bibnamefont {Pahlow}},\ and\ \bibinfo {author} {\bibfnamefont {J.}~\bibnamefont {Popp}},\ }\bibfield  {title} {\bibinfo {title} {The bouguer‐beer‐lambert law: Shining light on the obscure},\ }\href {https://api.semanticscholar.org/CorpusID:220520649} {\bibfield  {journal} {\bibinfo  {journal} {Chemphyschem}\ }\textbf {\bibinfo {volume} {21}},\ \bibinfo {pages} {2029 } (\bibinfo {year} {2020})}\BibitemShut {NoStop}%
\bibitem [{\citenamefont {Quabis}\ \emph {et~al.}(2000)\citenamefont {Quabis}, \citenamefont {Dorn}, \citenamefont {Eberler}, \citenamefont {Glöckl},\ and\ \citenamefont {Leuchs}}]{quabis_2001}%
  \BibitemOpen
  \bibfield  {author} {\bibinfo {author} {\bibfnamefont {S.}~\bibnamefont {Quabis}}, \bibinfo {author} {\bibfnamefont {R.}~\bibnamefont {Dorn}}, \bibinfo {author} {\bibfnamefont {M.}~\bibnamefont {Eberler}}, \bibinfo {author} {\bibfnamefont {O.}~\bibnamefont {Glöckl}},\ and\ \bibinfo {author} {\bibfnamefont {G.}~\bibnamefont {Leuchs}},\ }\bibfield  {title} {\bibinfo {title} {Focusing light to a tighter spot},\ }\href {https://doi.org/https://doi.org/10.1016/S0030-4018(99)00729-4} {\bibfield  {journal} {\bibinfo  {journal} {Optics Communications}\ }\textbf {\bibinfo {volume} {179}},\ \bibinfo {pages} {1} (\bibinfo {year} {2000})}\BibitemShut {NoStop}%
\bibitem [{\citenamefont {Quinteiro}\ \emph {et~al.}(2017)\citenamefont {Quinteiro}, \citenamefont {Schmidt-Kaler},\ and\ \citenamefont {Schmiegelow}}]{Quinteiro2017}%
  \BibitemOpen
  \bibfield  {author} {\bibinfo {author} {\bibfnamefont {G.~F.}\ \bibnamefont {Quinteiro}}, \bibinfo {author} {\bibfnamefont {F.}~\bibnamefont {Schmidt-Kaler}},\ and\ \bibinfo {author} {\bibfnamefont {C.~T.}\ \bibnamefont {Schmiegelow}},\ }\bibfield  {title} {\bibinfo {title} {Twisted-light--ion interaction: The role of longitudinal fields},\ }\href {https://doi.org/10.1103/PhysRevLett.119.253203} {\bibfield  {journal} {\bibinfo  {journal} {Phys. Rev. Lett.}\ }\textbf {\bibinfo {volume} {119}},\ \bibinfo {pages} {253203} (\bibinfo {year} {2017})}\BibitemShut {NoStop}%
\bibitem [{\citenamefont {Rubinsztein-Dunlop}\ \emph {et~al.}(2016)\citenamefont {Rubinsztein-Dunlop}, \citenamefont {Forbes}, \citenamefont {Berry}, \citenamefont {Dennis}, \citenamefont {Andrews}, \citenamefont {Mansuripur}, \citenamefont {Denz}, \citenamefont {Alpmann}, \citenamefont {Banzer}, \citenamefont {Bauer}, \citenamefont {Karimi}, \citenamefont {Marrucci}, \citenamefont {Padgett}, \citenamefont {Ritsch-Marte}, \citenamefont {Litchinitser}, \citenamefont {Bigelow}, \citenamefont {Rosales-Guzmán}, \citenamefont {Belmonte}, \citenamefont {Torres}, \citenamefont {Neely}, \citenamefont {Baker}, \citenamefont {Gordon}, \citenamefont {Stilgoe}, \citenamefont {Romero}, \citenamefont {White}, \citenamefont {Fickler}, \citenamefont {Willner}, \citenamefont {Xie}, \citenamefont {McMorran},\ and\ \citenamefont {Weiner}}]{Rubinsztein-Dunlop_2017}%
  \BibitemOpen
  \bibfield  {author} {\bibinfo {author} {\bibfnamefont {H.}~\bibnamefont {Rubinsztein-Dunlop}}, \bibinfo {author} {\bibfnamefont {A.}~\bibnamefont {Forbes}}, \bibinfo {author} {\bibfnamefont {M.~V.}\ \bibnamefont {Berry}}, \bibinfo {author} {\bibfnamefont {M.~R.}\ \bibnamefont {Dennis}}, \bibinfo {author} {\bibfnamefont {D.~L.}\ \bibnamefont {Andrews}}, \bibinfo {author} {\bibfnamefont {M.}~\bibnamefont {Mansuripur}}, \bibinfo {author} {\bibfnamefont {C.}~\bibnamefont {Denz}}, \bibinfo {author} {\bibfnamefont {C.}~\bibnamefont {Alpmann}}, \bibinfo {author} {\bibfnamefont {P.}~\bibnamefont {Banzer}}, \bibinfo {author} {\bibfnamefont {T.}~\bibnamefont {Bauer}}, \bibinfo {author} {\bibfnamefont {E.}~\bibnamefont {Karimi}}, \bibinfo {author} {\bibfnamefont {L.}~\bibnamefont {Marrucci}}, \bibinfo {author} {\bibfnamefont {M.}~\bibnamefont {Padgett}}, \bibinfo {author} {\bibfnamefont {M.}~\bibnamefont {Ritsch-Marte}}, \bibinfo {author} {\bibfnamefont {N.~M.}\ \bibnamefont {Litchinitser}}, \bibinfo {author}
  {\bibfnamefont {N.~P.}\ \bibnamefont {Bigelow}}, \bibinfo {author} {\bibfnamefont {C.}~\bibnamefont {Rosales-Guzmán}}, \bibinfo {author} {\bibfnamefont {A.}~\bibnamefont {Belmonte}}, \bibinfo {author} {\bibfnamefont {J.~P.}\ \bibnamefont {Torres}}, \bibinfo {author} {\bibfnamefont {T.~W.}\ \bibnamefont {Neely}}, \bibinfo {author} {\bibfnamefont {M.}~\bibnamefont {Baker}}, \bibinfo {author} {\bibfnamefont {R.}~\bibnamefont {Gordon}}, \bibinfo {author} {\bibfnamefont {A.~B.}\ \bibnamefont {Stilgoe}}, \bibinfo {author} {\bibfnamefont {J.}~\bibnamefont {Romero}}, \bibinfo {author} {\bibfnamefont {A.~G.}\ \bibnamefont {White}}, \bibinfo {author} {\bibfnamefont {R.}~\bibnamefont {Fickler}}, \bibinfo {author} {\bibfnamefont {A.~E.}\ \bibnamefont {Willner}}, \bibinfo {author} {\bibfnamefont {G.}~\bibnamefont {Xie}}, \bibinfo {author} {\bibfnamefont {B.}~\bibnamefont {McMorran}},\ and\ \bibinfo {author} {\bibfnamefont {A.~M.}\ \bibnamefont {Weiner}},\ }\bibfield  {title} {\bibinfo {title} {Roadmap on structured
  light},\ }\href {https://doi.org/10.1088/2040-8978/19/1/013001} {\bibfield  {journal} {\bibinfo  {journal} {Journal of Optics}\ }\textbf {\bibinfo {volume} {19}},\ \bibinfo {pages} {013001} (\bibinfo {year} {2016})}\BibitemShut {NoStop}%
\bibitem [{\citenamefont {Taylor}\ and\ \citenamefont {Sandoghdar}(2019)}]{iSCAT_vahid2019}%
  \BibitemOpen
  \bibfield  {author} {\bibinfo {author} {\bibfnamefont {R.~W.}\ \bibnamefont {Taylor}}\ and\ \bibinfo {author} {\bibfnamefont {V.}~\bibnamefont {Sandoghdar}},\ }\bibfield  {title} {\bibinfo {title} {Interferometric scattering microscopy: Seeing single nanoparticles and molecules via rayleigh scattering},\ }\href {https://doi.org/10.1021/acs.nanolett.9b01822} {\bibfield  {journal} {\bibinfo  {journal} {Nano Letters}\ }\textbf {\bibinfo {volume} {19}},\ \bibinfo {pages} {4827} (\bibinfo {year} {2019})}\BibitemShut {NoStop}%
\bibitem [{\citenamefont {Gustafsson}(2005{\natexlab{b}})}]{structured2005}%
  \BibitemOpen
  \bibfield  {author} {\bibinfo {author} {\bibfnamefont {M.~G.~L.}\ \bibnamefont {Gustafsson}},\ }\bibfield  {title} {\bibinfo {title} {Nonlinear structured-illumination microscopy: Wide-field fluorescence imaging with theoretically unlimited resolution},\ }\href {https://doi.org/10.1073/pnas.0406877102} {\bibfield  {journal} {\bibinfo  {journal} {Proceedings of the National Academy of Sciences}\ }\textbf {\bibinfo {volume} {102}},\ \bibinfo {pages} {13081} (\bibinfo {year} {2005}{\natexlab{b}})}\BibitemShut {NoStop}%
\bibitem [{\citenamefont {Lelek}\ \emph {et~al.}(2021)\citenamefont {Lelek}, \citenamefont {Gyparaki}, \citenamefont {Beliu}, \citenamefont {Schueder}, \citenamefont {Griffié}, \citenamefont {Manley}, \citenamefont {Jungmann}, \citenamefont {Sauer}, \citenamefont {Lakadamyali},\ and\ \citenamefont {Zimmer}}]{SMLM}%
  \BibitemOpen
  \bibfield  {author} {\bibinfo {author} {\bibfnamefont {M.}~\bibnamefont {Lelek}}, \bibinfo {author} {\bibfnamefont {M.~T.}\ \bibnamefont {Gyparaki}}, \bibinfo {author} {\bibfnamefont {G.}~\bibnamefont {Beliu}}, \bibinfo {author} {\bibfnamefont {F.}~\bibnamefont {Schueder}}, \bibinfo {author} {\bibfnamefont {J.}~\bibnamefont {Griffié}}, \bibinfo {author} {\bibfnamefont {S.}~\bibnamefont {Manley}}, \bibinfo {author} {\bibfnamefont {R.}~\bibnamefont {Jungmann}}, \bibinfo {author} {\bibfnamefont {M.}~\bibnamefont {Sauer}}, \bibinfo {author} {\bibfnamefont {M.}~\bibnamefont {Lakadamyali}},\ and\ \bibinfo {author} {\bibfnamefont {C.}~\bibnamefont {Zimmer}},\ }\bibfield  {title} {\bibinfo {title} {Single-molecule localization microscopy},\ }\href {https://doi.org/10.1038/s43586-021-00038-x} {\bibfield  {journal} {\bibinfo  {journal} {Nature Reviews Methods Primers}\ }\textbf {\bibinfo {volume} {1}} (\bibinfo {year} {2021})}\BibitemShut {NoStop}%
\bibitem [{\citenamefont {Taylor}\ \emph {et~al.}(2013)\citenamefont {Taylor}, \citenamefont {Janousek}, \citenamefont {Daria}, \citenamefont {Knittel}, \citenamefont {Hage}, \citenamefont {Bachor},\ and\ \citenamefont {Bowen}}]{beyondquantumlimit2013}%
  \BibitemOpen
  \bibfield  {author} {\bibinfo {author} {\bibfnamefont {M.~A.}\ \bibnamefont {Taylor}}, \bibinfo {author} {\bibfnamefont {J.}~\bibnamefont {Janousek}}, \bibinfo {author} {\bibfnamefont {V.}~\bibnamefont {Daria}}, \bibinfo {author} {\bibfnamefont {J.}~\bibnamefont {Knittel}}, \bibinfo {author} {\bibfnamefont {B.}~\bibnamefont {Hage}}, \bibinfo {author} {\bibfnamefont {H.-A.}\ \bibnamefont {Bachor}},\ and\ \bibinfo {author} {\bibfnamefont {W.~P.}\ \bibnamefont {Bowen}},\ }\bibfield  {title} {\bibinfo {title} {Biological measurement beyond the quantum limit},\ }\href {https://doi.org/10.1038/nphoton.2012.346} {\bibfield  {journal} {\bibinfo  {journal} {Nature Photonics}\ }\textbf {\bibinfo {volume} {7}},\ \bibinfo {pages} {229} (\bibinfo {year} {2013})}\BibitemShut {NoStop}%
\bibitem [{\citenamefont {Moreau}\ \emph {et~al.}(2019)\citenamefont {Moreau}, \citenamefont {Toninelli}, \citenamefont {Gregory},\ and\ \citenamefont {Padgett}}]{moreau2019imaging}%
  \BibitemOpen
  \bibfield  {author} {\bibinfo {author} {\bibfnamefont {P.-A.}\ \bibnamefont {Moreau}}, \bibinfo {author} {\bibfnamefont {E.}~\bibnamefont {Toninelli}}, \bibinfo {author} {\bibfnamefont {T.}~\bibnamefont {Gregory}},\ and\ \bibinfo {author} {\bibfnamefont {M.~J.}\ \bibnamefont {Padgett}},\ }\bibfield  {title} {\bibinfo {title} {Imaging with quantum states of light},\ }\href {https://doi.org/10.1038/s42254-019-0056-0} {\bibfield  {journal} {\bibinfo  {journal} {Nature Reviews Physics}\ }\textbf {\bibinfo {volume} {1}},\ \bibinfo {pages} {367} (\bibinfo {year} {2019})}\BibitemShut {NoStop}%
\bibitem [{\citenamefont {Chirita~Mihaila}\ \emph {et~al.}(2022)\citenamefont {Chirita~Mihaila}, \citenamefont {Weber}, \citenamefont {Schneller}, \citenamefont {Grandits}, \citenamefont {Nimmrichter},\ and\ \citenamefont {Juffmann}}]{mihaila2022transverse}%
  \BibitemOpen
  \bibfield  {author} {\bibinfo {author} {\bibfnamefont {M.~C.}\ \bibnamefont {Chirita~Mihaila}}, \bibinfo {author} {\bibfnamefont {P.}~\bibnamefont {Weber}}, \bibinfo {author} {\bibfnamefont {M.}~\bibnamefont {Schneller}}, \bibinfo {author} {\bibfnamefont {L.}~\bibnamefont {Grandits}}, \bibinfo {author} {\bibfnamefont {S.}~\bibnamefont {Nimmrichter}},\ and\ \bibinfo {author} {\bibfnamefont {T.}~\bibnamefont {Juffmann}},\ }\bibfield  {title} {\bibinfo {title} {Transverse electron-beam shaping with light},\ }\href {https://doi.org/10.1103/PhysRevX.12.031043} {\bibfield  {journal} {\bibinfo  {journal} {Phys. Rev. X}\ }\textbf {\bibinfo {volume} {12}},\ \bibinfo {pages} {031043} (\bibinfo {year} {2022})}\BibitemShut {NoStop}%
\bibitem [{\citenamefont {Kalinin}\ \emph {et~al.}(2022)\citenamefont {Kalinin}, \citenamefont {Ophus}, \citenamefont {Voyles}, \citenamefont {Erni}, \citenamefont {Kepaptsoglou}, \citenamefont {Grillo}, \citenamefont {Lupini}, \citenamefont {Oxley}, \citenamefont {Schwenker}, \citenamefont {Chan} \emph {et~al.}}]{kalinin2022machine}%
  \BibitemOpen
  \bibfield  {author} {\bibinfo {author} {\bibfnamefont {S.~V.}\ \bibnamefont {Kalinin}}, \bibinfo {author} {\bibfnamefont {C.}~\bibnamefont {Ophus}}, \bibinfo {author} {\bibfnamefont {P.~M.}\ \bibnamefont {Voyles}}, \bibinfo {author} {\bibfnamefont {R.}~\bibnamefont {Erni}}, \bibinfo {author} {\bibfnamefont {D.}~\bibnamefont {Kepaptsoglou}}, \bibinfo {author} {\bibfnamefont {V.}~\bibnamefont {Grillo}}, \bibinfo {author} {\bibfnamefont {A.~R.}\ \bibnamefont {Lupini}}, \bibinfo {author} {\bibfnamefont {M.~P.}\ \bibnamefont {Oxley}}, \bibinfo {author} {\bibfnamefont {E.}~\bibnamefont {Schwenker}}, \bibinfo {author} {\bibfnamefont {M.~K.}\ \bibnamefont {Chan}}, \emph {et~al.},\ }\bibfield  {title} {\bibinfo {title} {Machine learning in scanning transmission electron microscopy},\ }\href {https://doi.org/10.1038/s43586-022-00095-w} {\bibfield  {journal} {\bibinfo  {journal} {Nature Reviews Methods Primers}\ }\textbf {\bibinfo {volume} {2}},\ \bibinfo {pages} {11} (\bibinfo {year} {2022})}\BibitemShut {NoStop}%
\bibitem [{\citenamefont {Kalinin}\ \emph {et~al.}(2023)\citenamefont {Kalinin}, \citenamefont {Liu}, \citenamefont {Biswas}, \citenamefont {Duscher}, \citenamefont {Pratiush}, \citenamefont {Roccapriore}, \citenamefont {Ziatdinov},\ and\ \citenamefont {Vasudevan}}]{kalinin2023human}%
  \BibitemOpen
  \bibfield  {author} {\bibinfo {author} {\bibfnamefont {S.~V.}\ \bibnamefont {Kalinin}}, \bibinfo {author} {\bibfnamefont {Y.}~\bibnamefont {Liu}}, \bibinfo {author} {\bibfnamefont {A.}~\bibnamefont {Biswas}}, \bibinfo {author} {\bibfnamefont {G.}~\bibnamefont {Duscher}}, \bibinfo {author} {\bibfnamefont {U.}~\bibnamefont {Pratiush}}, \bibinfo {author} {\bibfnamefont {K.}~\bibnamefont {Roccapriore}}, \bibinfo {author} {\bibfnamefont {M.}~\bibnamefont {Ziatdinov}},\ and\ \bibinfo {author} {\bibfnamefont {R.}~\bibnamefont {Vasudevan}},\ }\bibfield  {title} {\bibinfo {title} {Human-in-the-loop: The future of machine learning in automated electron microscopy},\ }\href {https://arxiv.org/abs/2310.05018} {\bibfield  {journal} {\bibinfo  {journal} {arXiv:2310.05018}\ } (\bibinfo {year} {2023})}\BibitemShut {NoStop}%
\bibitem [{\citenamefont {Kia{\l}ka}\ \emph {et~al.}(2022)\citenamefont {Kia{\l}ka}, \citenamefont {Fein}, \citenamefont {Pedalino}, \citenamefont {Gerlich},\ and\ \citenamefont {Arndt}}]{kialka2022roadmap}%
  \BibitemOpen
  \bibfield  {author} {\bibinfo {author} {\bibfnamefont {F.}~\bibnamefont {Kia{\l}ka}}, \bibinfo {author} {\bibfnamefont {Y.~Y.}\ \bibnamefont {Fein}}, \bibinfo {author} {\bibfnamefont {S.}~\bibnamefont {Pedalino}}, \bibinfo {author} {\bibfnamefont {S.}~\bibnamefont {Gerlich}},\ and\ \bibinfo {author} {\bibfnamefont {M.}~\bibnamefont {Arndt}},\ }\bibfield  {title} {\bibinfo {title} {A roadmap for universal high-mass matter-wave interferometry},\ }\href {https://doi.org/10.1116/5.0080940} {\bibfield  {journal} {\bibinfo  {journal} {AVS Quantum Science}\ }\textbf {\bibinfo {volume} {4}} (\bibinfo {year} {2022})}\BibitemShut {NoStop}%
\bibitem [{\citenamefont {Shen}\ and\ \citenamefont {Wang}(2006)}]{RS}%
  \BibitemOpen
  \bibfield  {author} {\bibinfo {author} {\bibfnamefont {F.}~\bibnamefont {Shen}}\ and\ \bibinfo {author} {\bibfnamefont {A.}~\bibnamefont {Wang}},\ }\bibfield  {title} {\bibinfo {title} {Fast-fourier-transform based numerical integration method for the rayleigh-sommerfeld diffraction formula},\ }\href {https://doi.org/10.1364/AO.45.001102} {\bibfield  {journal} {\bibinfo  {journal} {Applied Optics}\ }\textbf {\bibinfo {volume} {45}},\ \bibinfo {pages} {1102} (\bibinfo {year} {2006})}\BibitemShut {NoStop}%
\bibitem [{\citenamefont {Ye}\ \emph {et~al.}(2013)\citenamefont {Ye}, \citenamefont {Qiu}, \citenamefont {Huang}, \citenamefont {Teng}, \citenamefont {Luk'Yanchuk},\ and\ \citenamefont {Yeo}}]{vectorial_RS}%
  \BibitemOpen
  \bibfield  {author} {\bibinfo {author} {\bibfnamefont {H.}~\bibnamefont {Ye}}, \bibinfo {author} {\bibfnamefont {C.-W.}\ \bibnamefont {Qiu}}, \bibinfo {author} {\bibfnamefont {K.}~\bibnamefont {Huang}}, \bibinfo {author} {\bibfnamefont {J.}~\bibnamefont {Teng}}, \bibinfo {author} {\bibfnamefont {B.}~\bibnamefont {Luk'Yanchuk}},\ and\ \bibinfo {author} {\bibfnamefont {S.}~\bibnamefont {Yeo}},\ }\bibfield  {title} {\bibinfo {title} {Creation of a longitudinally polarized subwavelength hotspot with an ultra-thin planar lens: Vectorial rayleigh-sommerfeld method},\ }\href {https://doi.org/10.1088/1612-2011/10/6/065004} {\bibfield  {journal} {\bibinfo  {journal} {Laser Physics Letters}\ }\textbf {\bibinfo {volume} {10}} (\bibinfo {year} {2013})}\BibitemShut {NoStop}%
\bibitem [{\citenamefont {Hu}\ \emph {et~al.}(2020)\citenamefont {Hu}, \citenamefont {Wang}, \citenamefont {Wang} \emph {et~al.}}]{hu2020efficient}%
  \BibitemOpen
  \bibfield  {author} {\bibinfo {author} {\bibfnamefont {Y.}~\bibnamefont {Hu}}, \bibinfo {author} {\bibfnamefont {Z.}~\bibnamefont {Wang}}, \bibinfo {author} {\bibfnamefont {X.}~\bibnamefont {Wang}}, \emph {et~al.},\ }\bibfield  {title} {\bibinfo {title} {Efficient full-path optical calculation of scalar and vector diffraction using the bluestein method},\ }\href {https://doi.org/10.1038/s41377-020-00362-z} {\bibfield  {journal} {\bibinfo  {journal} {Light: Science \& Applications}\ }\textbf {\bibinfo {volume} {9}},\ \bibinfo {pages} {119} (\bibinfo {year} {2020})}\BibitemShut {NoStop}%
\bibitem [{\citenamefont {Li}\ \emph {et~al.}(2002)\citenamefont {Li}, \citenamefont {Fan},\ and\ \citenamefont {Fu}}]{FFT_sampling}%
  \BibitemOpen
  \bibfield  {author} {\bibinfo {author} {\bibfnamefont {J.}~\bibnamefont {Li}}, \bibinfo {author} {\bibfnamefont {Z.}~\bibnamefont {Fan}},\ and\ \bibinfo {author} {\bibfnamefont {Y.}~\bibnamefont {Fu}},\ }\bibfield  {title} {\bibinfo {title} {{FFT calculation for Fresnel diffraction and energy conservation criterion of sampling quality}},\ }in\ \href {https://doi.org/10.1117/12.482883} {\emph {\bibinfo {booktitle} {Lasers in Material Processing and Manufacturing}}},\ Vol.\ \bibinfo {volume} {4915}\ (\bibinfo  {publisher} {SPIE},\ \bibinfo {year} {2002})\ pp.\ \bibinfo {pages} {180 -- 186}\BibitemShut {NoStop}%
\end{thebibliography}%

\newpage
\onecolumngrid
\newpage
\section*{Extended data}
{\renewcommand{\figurename}{Extended Data Fig.}
\setcounter{figure}{0}

\begin{figure*}[hb!]
  \centering
  \includegraphics[width=1\linewidth]{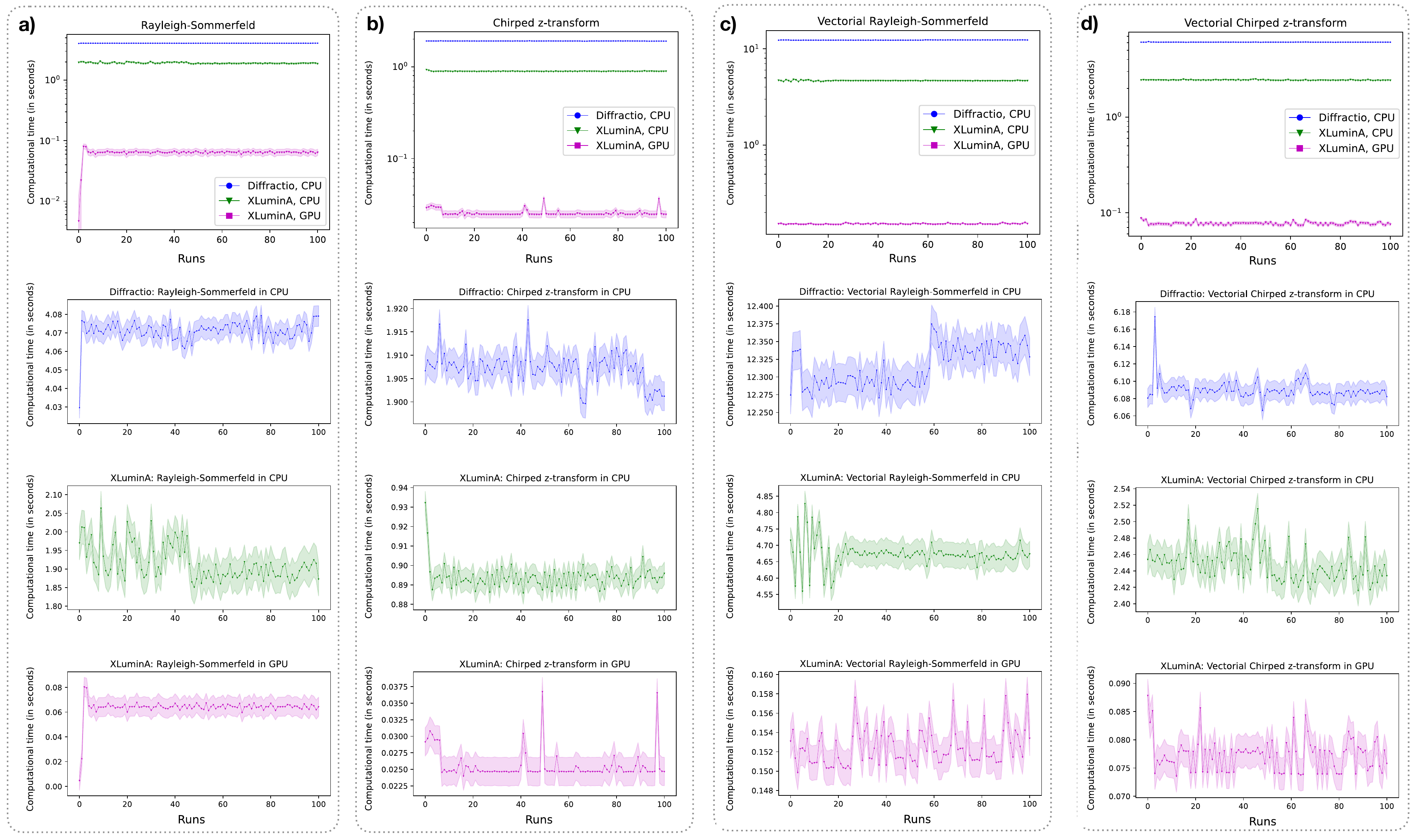}
  \caption{Execution time over 100 runs, within a resolution of $2048\times2048$ pixels for the propagation methods of, in columns, (a) Rayleigh-Sommerfeld, (b) Chirped z-transform, (c) Vectorial Rayleigh-Sommerfeld and (d) Vectorial Chirped z-transform. Times for \textit{Diffractio} are depicted in blue dots. First row corresponds to the log-scale representation. Times for \algo on CPU and GPU correspond to green triangles and magenta squares, respectively. Shades denote for the standard deviations values. Among the minor oscillations of the running times, the Rayleigh-Sommerfeld algorithm behavior stands out. When running in \textit{Diffractio}, it shows an increase of the execution time of 0.005 seconds from the first run to subsequent runs. Similar behavior can be observed for the Rayleigh-Sommerfeld (RS) algorithm in \algo when executing in the GPU. This time, however, the increase occurs during the first 5 runs, from an first execution time of 0.0047 seconds to stabilize in 0.06 seconds for subsequent runs. This behavior is not present when executing \algo's RS in the CPU. The origin of this factor of 10 is still unknown. We believe a further optimization on the RS propagation algorithm will improve the execution time. }
  \label{fig:extended_data_propagation_runs}
\end{figure*}

\begin{figure*}[hb!]
  \centering
  \includegraphics[width=1\linewidth]{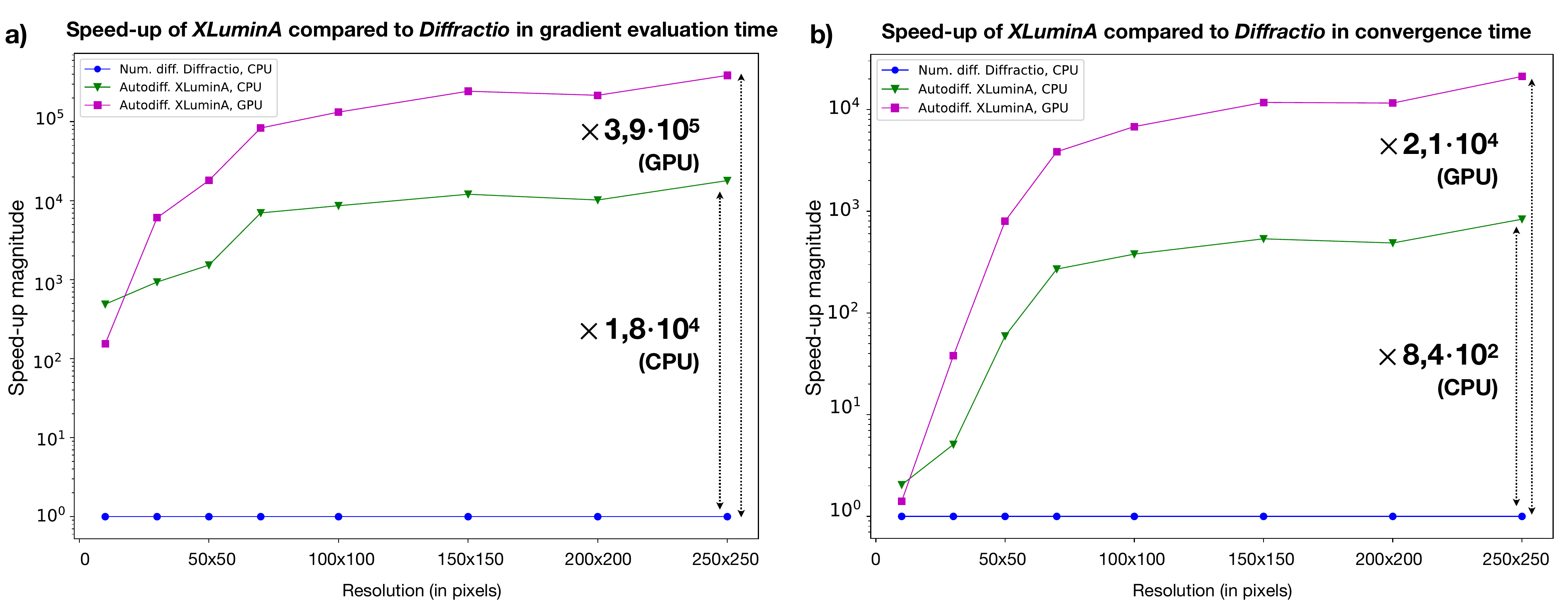}
  \caption{Speed-up magnitudes of \algo (auto-differentiation) compared to \textit{Diffractio} (numerical methods) across different resolutions in (a) single gradient evaluation and (b) convergence time. Numerical differentiation methods are computed using \textit{Diffractio}’s optical simulator (blue dots) and auto-differentiation (green triangles for CPU and magenta squares for GPU) on \algo. Autodiff. outperforms numerical methods in gradient evaluation by up to 4 orders of magnitude when running on CPU and 5 orders of magnitude when running on the GPU for a resolution of $250\times250$ pixels. In convergence time, autodiff outperforms numerical methods by almost 3 orders of magnitude in the CPU and 4 orders of magnitude on the GPU.}
  \label{fig:extended_data_speedup}
\end{figure*}

\begin{figure*}[hb!]
  \centering
  \includegraphics[width=1\linewidth]{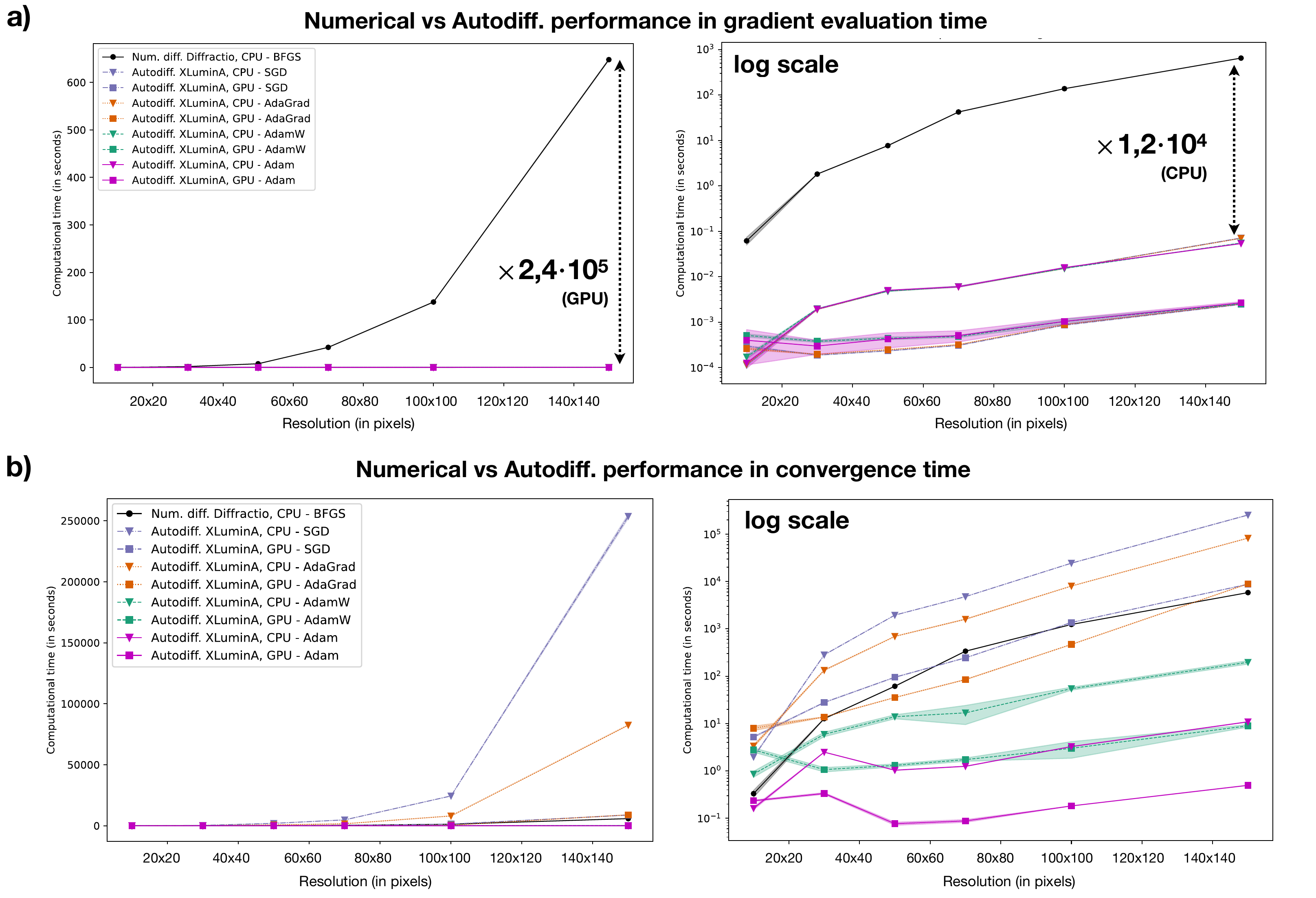}
  \caption{Performance of \algo (auto-differentiation) compared to \textit{Diffractio} (numerical methods) across different resolutions and optimizers in (a) single gradient evaluation and (b) convergence time. Data corresponds to the average time over 5 runs. Numerical differentiation is computed using \textit{Diffractio}’s optical simulator and the Broyden-Fletcher-Goldfarb-Shanno (BFGS) optimizer (black dots) and auto-differentiation (triangles for CPU and squares for GPU) on \algo. The Stochastic-Gradient-Descent (SGD), Adaptive Gradient (AdaGrad), Adaptive moment estimation with weight decay (AdamW) and Adaptive moment estimation (Adam) correspond to blue (dash-dot line), orange (dotted line), green (dash line) and magenta (continuous line), respectively. Shaded regions correspond to standard deviation values. The step size is set to $0.1$ and is common to all the optimizers. For AdamW, the weight decay is set to $10^{-4}$. The stopping condition is common to all the frameworks: the optimization is terminated if there is no improvement in the loss value (i.e., it has not decreased below the best value recorded), over 500 consecutive iteration steps. This condition is checked every 100 steps. The use of \algo with autodiff methods improves the gradient evaluation time by a factor of $\times 2.4 \cdot 10^5$ in the GPU and a factor of $\times 1.2 \cdot 10^4$ on the CPU for resolutions of $150\times150$ pixels. This behavior is common to all the tested optimizers (Adam, AdamW, SGD and AdaGrad). When evaluating the convergence time, the use of Autodiff methods on \algo using the Adam and AdamW optimizers improve the performance with respect to numerical methods by a factor of $\times 1.1 \cdot 10^4$ and $\times 6.5 \cdot 10^2$ in the GPU, respectively, for a resolution of $150\times150$ pixels. The performance of Adam and AdamW in the CPU demonstrates factors of $\times 5.8 \cdot 10^2$ and $\times 2.9 \cdot 10^1$, respectively, for the same resolution. Remarkably, the use of \textit{Diffractio} with numerical methods (BFGS) outperforms both AdaGrad and SGD in convergence time. In particular, numerical methods outperform AdaGrad by a factor of $\times 1.53$ in the GPU and $\times 14.14$ in the CPU, for a resolution of $150\times150$ pixels. Similar behavior is demonstrated for SGD: numerical methods outperform it by a factor of $\times 1.50$ in the GPU and $\times 43.50$ in the CPU, for a resolution of $150\times150$ pixels. Overall, the use of Autodiff methods (in particular, using Adam or AdamW) within GPU-accelerated frameworks is a more appropriate choice to conduct efficient optimization.}

  \label{fig:extended_data_optimizers}
\end{figure*}

\begin{figure*}[ht!]
  \centering
  \includegraphics[width=1\linewidth]{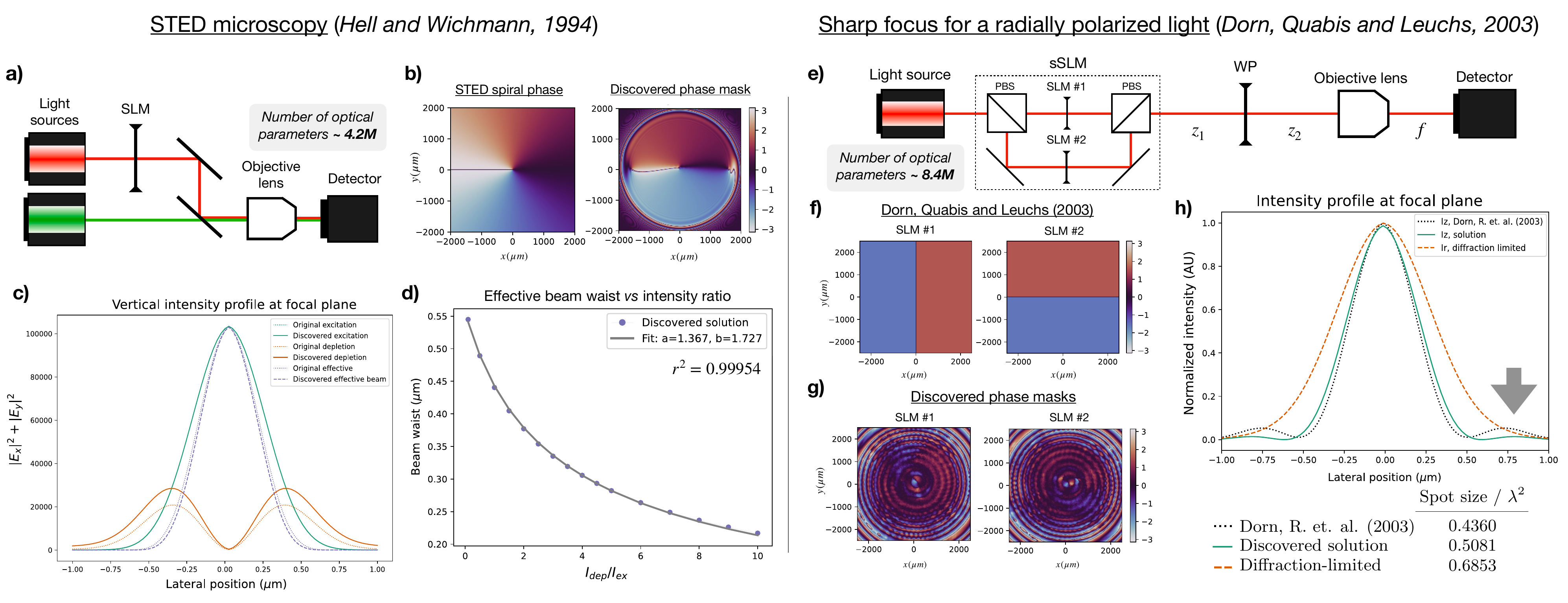}
  \caption{Rediscovery of beam-shaping technique as employed in STED microscopy and the super-resolution technique employing optical vortices to generate a longitudinal sharp focus. (a) Virtual optical setup for Hell and Wichmann (1994). It consists of two light sources generating Gaussian beams corresponding to the depletion and excitation beams, linearly polarized in orthogonal directions, with wavelengths of $650$ nm and $532$ nm, respectively. Within the depletion beam's optical path, we place an SLM with a resolution of $2048\times2048$ and a computational pixel size of $1.95$ $\mu m$. After propagating some set distance, a 0.9 NA objective lens focuses both light beams into the detector screen of $0.05$ $\mu m$ pixel size. The parameter space for the optimization is defined by the SLM ($\sim$ 4.2 million optical parameters). (b) STED spiral phase (Hell and Wichmann, 1994) and discovered phase mask. This solution was found in roughly 7 minutes using a GPU. (c) Radial intensity profile in vertical beam section: excitation (green), depletion (orange), and super-resolution effective STED beam (discontinuous blue line). The data corresponding to the original STED spiral phase are indicated with dotted lines. The excitation and depletion beams are diffraction-limited. The effective response breaks the diffraction limit. Lateral position indicates lateral distance from the optical axis. (d) Effective beam waist (in $\mu$m) as a function of depletion and excitation intensity ratio ($I_{dep}/I_{ex}$). The fit function corresponds to $f(x) = (a\sqrt{b+x})^{-1}$, with $r^2 = 0.99954$ with $a = 1.367$ and $b = 1.727$. We observe the expected inverse square root scaling as originally demonstrated in STED microscopy: $D = \lambda (2\text{NA}\sqrt{1+I/I_{Sat}})^{-1}$. (e) Virtual optical setup for Dorn, Quabis and Leuchs (2004). It consists of a light source emitting a $635$ nm wavelength Gaussian beam that is linearly polarized. The original optical elements are replaced by an \textit{super-SLM} (sSLM), which consists of two SLMs, each one independently imprinting a phase mask on the horizontal and vertical polarization components of the field. PBS denotes for polarization beam splitter. Each component of the \textit{sSLM} has a resolution of $2048\times2048$ pixels and a computational pixel size of $2.44$ $\mu m$. Additionally, we place a wave plate (WP) with variable phase retardance $\eta$ and orientation angle $\theta$. The beam then passes through a 0.9 NA objective lens before reaching the detector screen of $0.05$ $\mu m$ pixel size; $z_1$ and $z_2$ denote for distances. The parameter space ($\sim$ 8.4 million optical parameters) is defined by the two SLMs, the WP, and the distances. (f) Phase masks for simulating the reference experiment. The optical parameters for WP are set to 0 and the distances $z_1$ and $z_2$ to $40$ mm and $3000$ mm, respectively. (g) Discovered phase patterns. Discovered optical parameters for WP's retardance, orientation, and propagation distances correspond to $-1.23$ rad, $2.32$ rad, $800$ mm and $710$ mm, respectively. This solution was identified in roughly 2 hours using a GPU. (h) Normalized longitudinal intensity profile, $|E_z|^2$, for Dorn, Quabis, and Leuchs (2003) and the identified solution (black dotted, and green lines, respectively) and radial intensity profile, $I_r$, of the diffraction-limited linearly polarized beam (orange dotted line). Lateral position indicates lateral distance from the optical axis. The spot size is computed as $\phi = (\pi/4)\text{FWHM}_x \text{FWHM}_y$, where FWHM denotes for Full Width Half Maximum. The discovered approach breaks the diffraction limit, demonstrating a similar spot size as the reference. Remarkably, it does not feature side lobes (indicated with a gray arrow), which can limit practical imaging techniques.}
  \label{fig:extended_data_early_experiment}
\end{figure*} 

\begin{figure*}[hb!]
  \centering
  \includegraphics[width=0.8\linewidth]{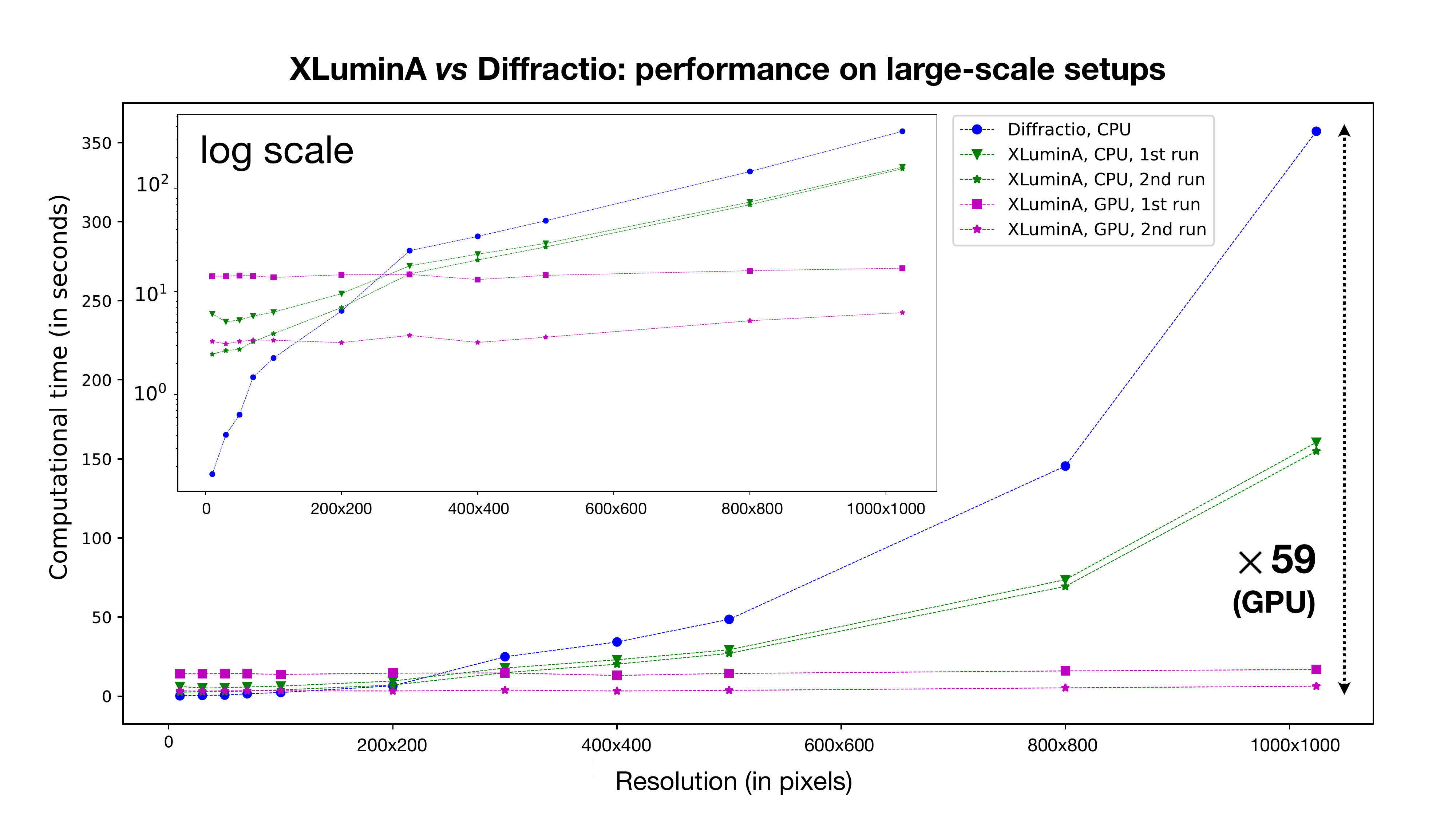}
  \caption{Computational time (in seconds) for generating the large-scale setup for \textit{Diffractio} (blue dots) and \algo across different resolutions on CPU (green triangles) and GPU (magenta squares). For \algo, first run curves are decorated with triangles and second runs (pre-compiled jitted functions) with stars. The first run of a jitted function encodes, in an abstract representation, the shape of the arrays (i.e., pixel resolution). If the input shape remains constant, such abstract structure is re-used and there is no need of re-compiling, which allows to execute the subsequent runs faster. \textit{Diffractio} exhibits a notable exponential increase in its computational time, showing running times of almost 6 minutes for a resolution of $1024\times1024$. When running on CPU, \algo outperforms \textit{Diffractio}, exhibiting superior scalability in both initial and subsequent runs. For example, for $1024\times1024$ pixels, \algo operates in nearly half the time (2.5 minutes) required by \textit{Diffractio}. This advantage becomes even more pronounced when \algo operates on the GPU, demonstrating a speedup factor of $\times$59 for a resolution of $1024\times1024$ pixels.}
  \label{fig:extended_data_performance}
\end{figure*}

\begin{figure*}[hb!]
  \centering
  \includegraphics[width=1\linewidth]{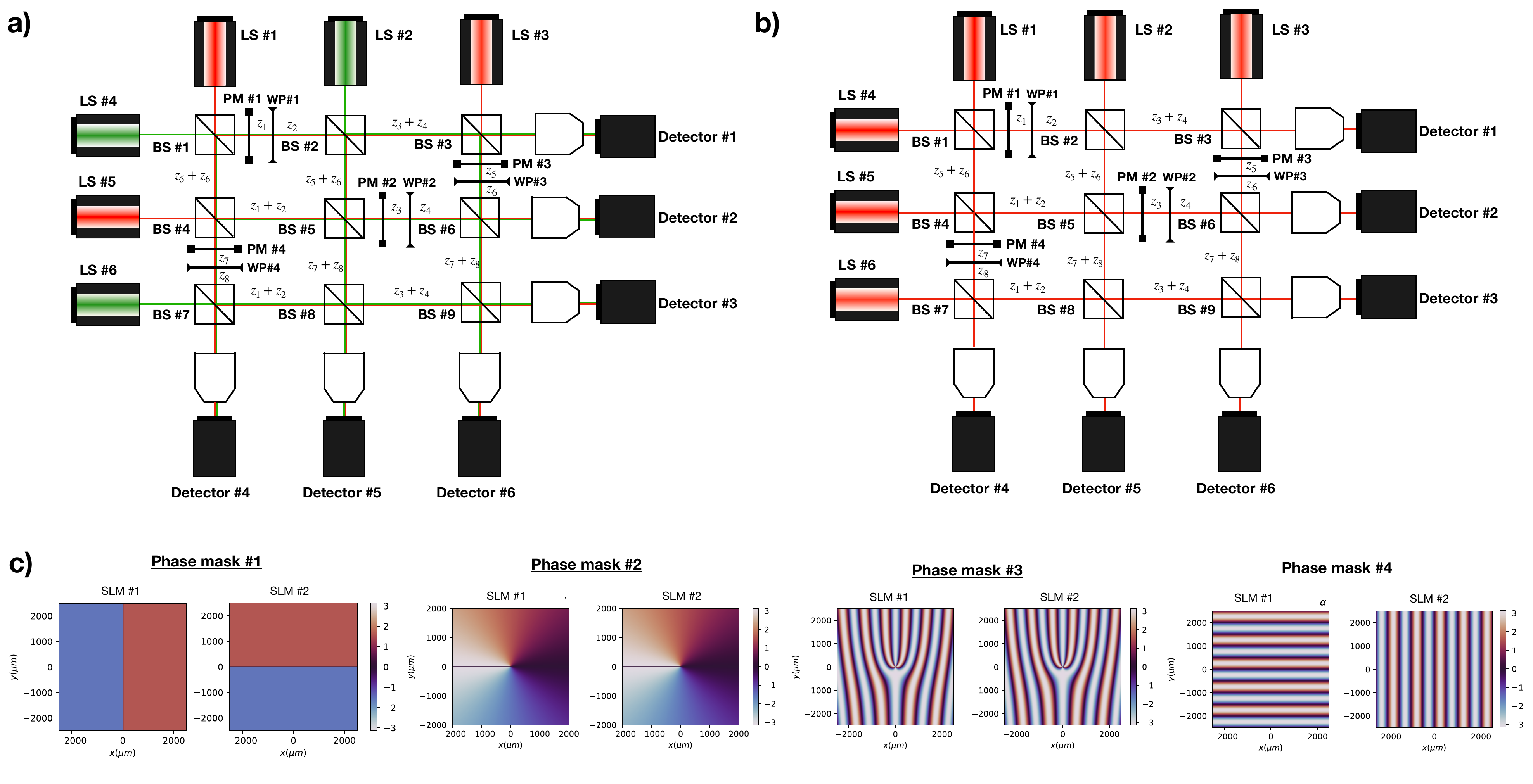}
  \caption{Topological discovery within a large-scale optical setup. The optical elements corresponding to phase masks (PM$\#1$ - PM$\#4$) remain fixed during the optimization. The parameter space (25 optical parameters) is defined by nine beam splitter ratios (BS$\#1$ - BS$\#9$), eight distances ($z_1$ - $z_8$) and four wave plates with variable applied retardance and orientation). After interacting with a high NA objective lens, light gets detected across six detectors ($\#1$ - $\#6$) with $0.05\mu m$ pixel size screen. The parameter update is driven by the camera demonstrating the minimum loss value. (a) Initial virtual setup for STED microscopy (Hell and Wichmann, 1994). The light sources emit $632.8$ nm and $530$ nm wavelength Gaussian beams that are linearly polarized at $45^o$. (b) Initial virtual optical setup for Dorn, Quabis and Leuchs (2003). It consists of six light sources emitting a $632.8$ nm wavelength Gaussian beam that are linearly polarized at $45^o$. (c) Fixed phase patterns corresponding to each phase mask placed on \textit{super}-SLMs (from $\#1$ to $\#4$). The phase mask $\#1$ corresponds to the radial phase pattern originally used in STED microscopy. Phase mask $\#2$ is the polarization converter demonstrated in Dorn, Quabis, and Leuchs (2003). Phase mask $\#3$ corresponds to a forked grating of $p=2$. Phase mask $\#4$ correspond to horizontal and vertical gratings.}
  \label{fig:extended_data_init_setup_xl}
\end{figure*}

\begin{figure*}[hb!]
  \centering
  \includegraphics[width=1\linewidth]{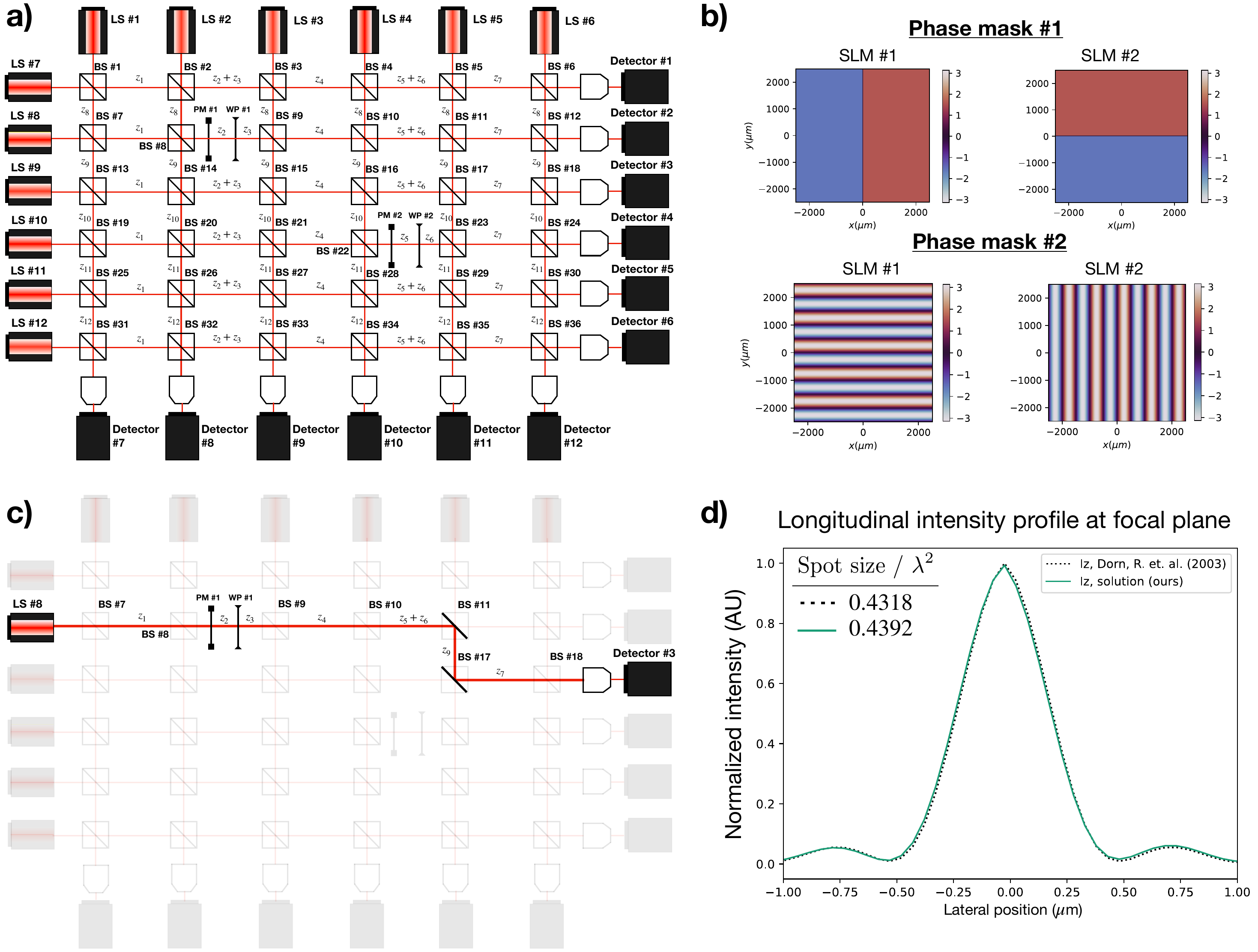}
  \caption{Pure topological discovery within a $6\times6$ large-scale optical setup. The optical elements corresponding to phase masks (PM$\#1$ and PM$\#2$) remain fixed during the optimization. The parameter space is defined by 36 beam splitter ratios (BS$\#1$ - BS$\#36$), 12 distances ($z_1$ - $z_{12}$) and 2 wave plates with variable applied retardance and orientation). After interacting with a high NA objective lens, light gets detected across 12 detectors ($\#1$ - $\#6$) with $0.05\mu m$ pixel size screen. The parameter update is driven by the camera demonstrating the minimum loss value. (a) Initial virtual setup for Dorn, Quabis and Leuchs (2003). It consists of 12 light sources emitting a $635$ nm wavelength Gaussian beam that are linearly polarized at $45^o$. (b) Fixed phase patterns corresponding to each phase mask placed on \textit{super}-SLMs. The phase mask $\#1$ corresponds to the polarization converter demonstrated in Dorn, Quabis, and Leuchs (2003). Phase mask $\#2$ correspond to horizontal and vertical gratings. (c) Discovered topology. The minimum value of the loss is demonstrated in detector $\#3$. The beam splitter ratios, in [Transmittance, Reflectance] pairs correspond to: BS$\#7$ [0.999, 0.001], BS$\#8$ [0.996, 0.004], BS$\#9$ [0.841, 0.159], BS$\#10$ [0.791, 0.209], BS$\#11$ [0.015, 0.984], BS$\#17$ [0.280, 0.720], and BS$\#18$ [0.962, 0.038]. Wave plate's (in radians) $\eta=3.13$ and $\theta=3.13$. Distances (in cm) correspond to $z_1=167.15$, $z_2=133.78$, $z_3=102.45$, $z_4=65.06$, $z_5=81.97$, $z_6=71.99$, $z_7=104.96$ and $z_9=169.96$. (d) Normalized longitudinal intensity profile, $|E_z|^2$, for Dorn, Quabis, and Leuchs (2003) and the identified solution (black dotted, and green lines, respectively). Lateral position indicates lateral distance from the optical axis. The spot size is computed as $\phi = (\pi/4)\text{FWHM}_x \text{FWHM}_y$, where FWHM denotes for Full Width Half Maximum.}
  \label{fig:6x6}
\end{figure*}

\begin{figure*}[hb!]
  \centering
  \includegraphics[width=0.8\linewidth]{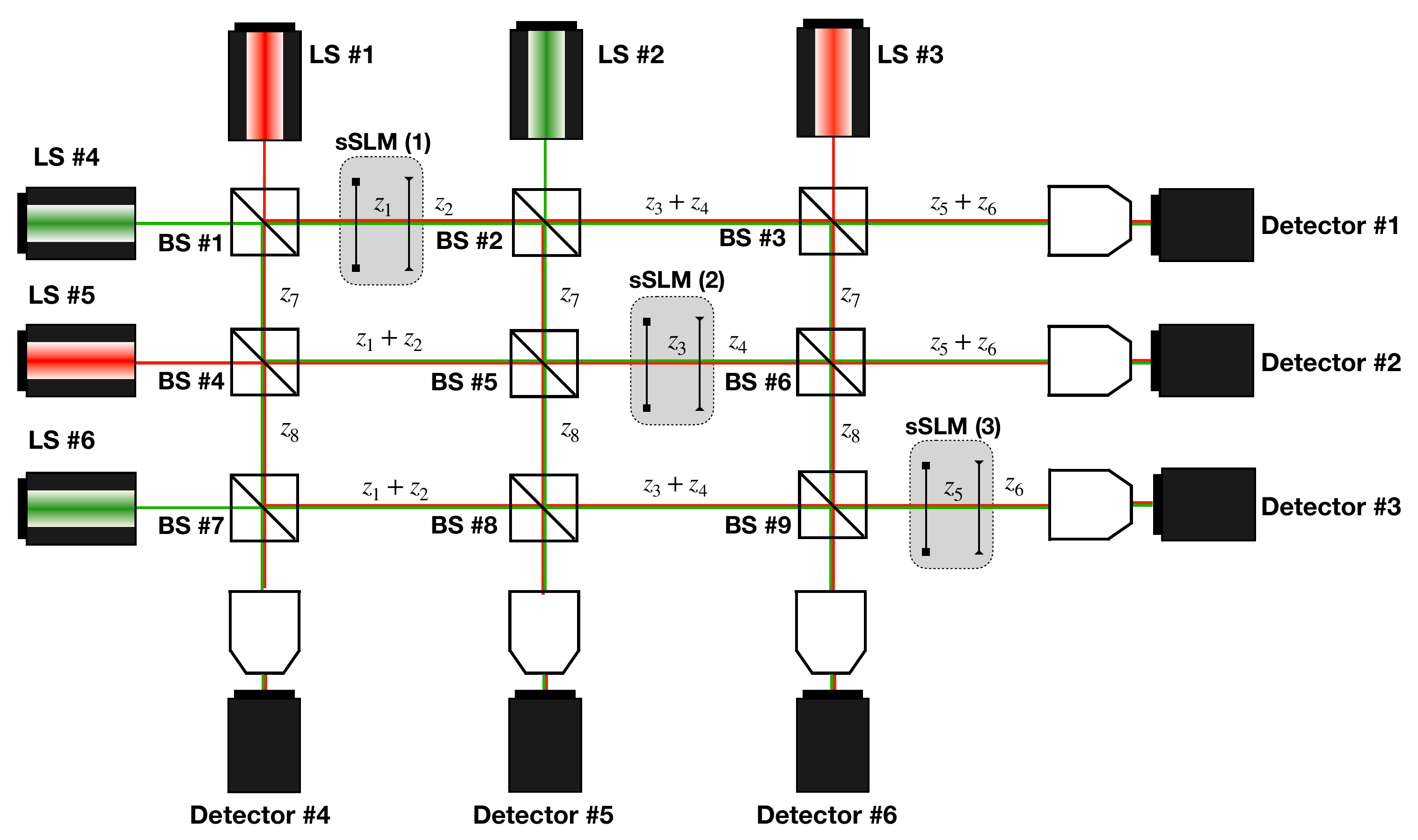}
  \caption{Virtual optical setup utilized for large-scale discovery. It features light sources that emit Gaussian beams with wavelengths of $650$ nm and $532$ nm linearly polarized at $45^o$. Gray boxes, numbered from (1) to (3), represent the building units, each comprising one \textit{super-SLM} (sSLM) and one wave plate (WP). Distances are denoted as $z_i$ where $i=1,...,8$. After interacting with a high NA objective lens (of NA$=0.9$), light gets detected across six detectors ($\#1$ - $\#6$) with $0.05\mu m$ pixel size screen. The parameter space ($\sim$ 4 million optical parameters) contains three \textit{sSLM}, four wave plates (WPs) with variable phase retardance $\eta$ and orientation angle $\theta$, eight distances and nine beam splitter ratios. To discover the new topologies for Dorn, Quabis and Leuchs (2003) in Fig. \ref{fig:XLSetup_topologies}d: (1) all the light sources are set to emit $635$ nm wavelength, (2) the resolution is set to $1024\times1024$ pixels with a pixel size of $4.8$ $\mu m$ and (3) the parameter space is of $\sim$ 6.3 million optical parameters (containing three \textit{sSLM}, four WPs with variable phase retardance $\eta$ and orientation angle $\theta$, eight distances and nine beam splitter ratios).}
  \label{fig:extended_data_init_setup_discovery}
\end{figure*}

\begin{figure*}[ht!]
  \centering
  \includegraphics[width=0.9\linewidth]{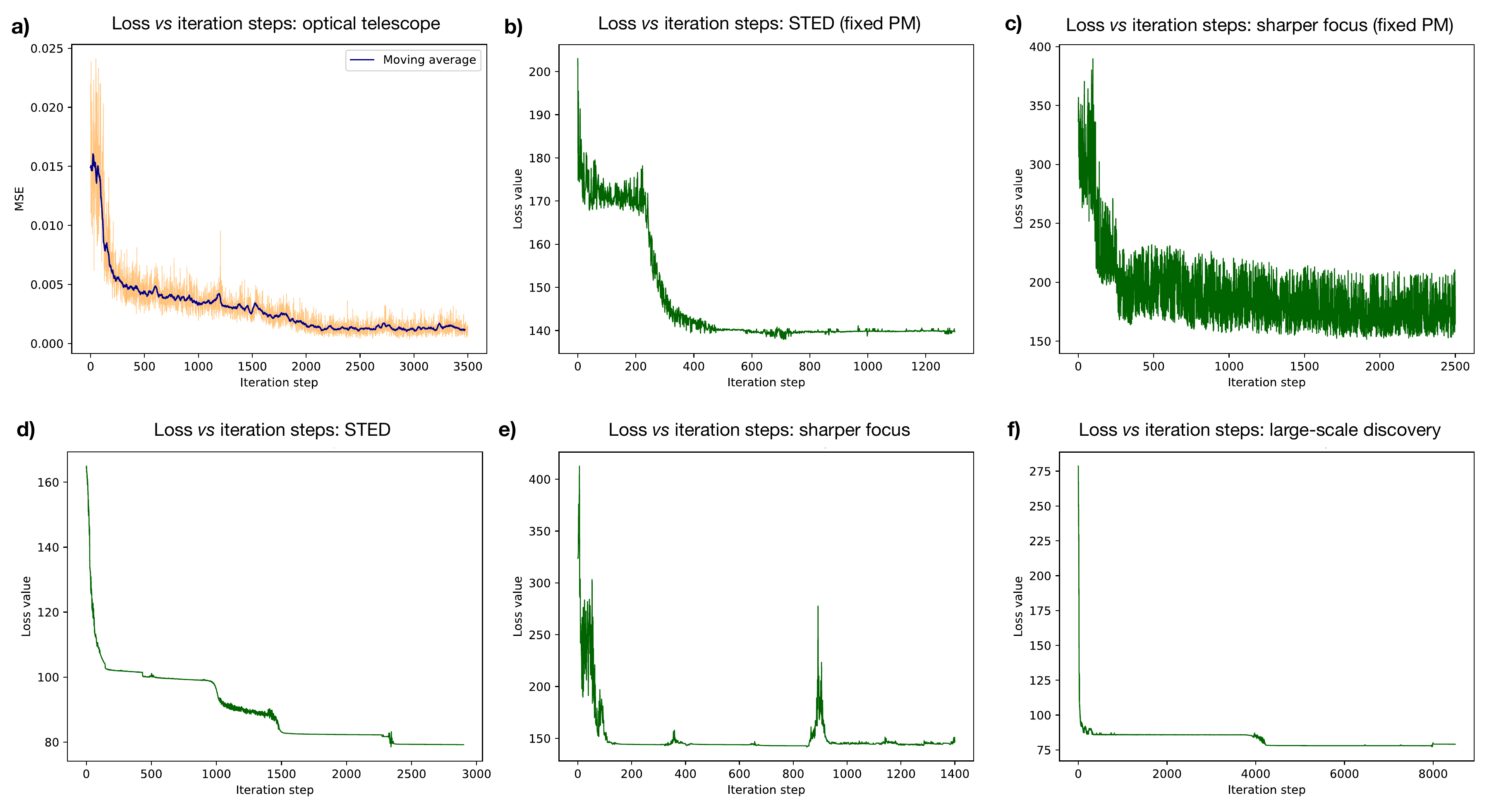}
  \caption{Loss value over iteration steps for the different optical experiments shown in this work. (a) The optical system used to adjust beam and image sizes (i.e., an optical telescope). This solution was identified in roughly 2.4 hours using a GPU. This time includes optics simulations and the data loading processes. (b) The virtual optical setup with fixed phase masks (PM) for the sharper focus employing optical vortices. This solution was identified in roughly 35 minutes using a GPU. (c) The virtual optical setup with fixed phase masks (PM) for STED microscopy. This solution was identified in roughly 1 hour using a GPU. (d) The virtual optical setup for STED microscopy. This solution was identified in roughly 1.3 hours using a GPU. (e) The virtual optical setup for the sharper focus employing optical vortices. This solution was identified in roughly 35 minutes using a GPU. (f) The new experimental blueprint. This solution was identified in roughly 3.8 hours using a GPU.}
  \label{fig:Extended_data_loss}
\end{figure*} 
}
\end{document}